\UseRawInputEncoding
\documentclass[aps,onecolumn,10pt]{revtex4}
\usepackage{amsmath}
\usepackage{amssymb}
\usepackage{hyperref}
\hypersetup{colorlinks=true,linkcolor=red,citecolor=green,urlcolor=blue}
\numberwithin{equation}{section}
\numberwithin{equation}{section}

\begin{document}
\allowdisplaybreaks
\setcounter{equation}{0}

\title{Critique of the use of geodesics in astrophysics and cosmology}

\author{Philip D. Mannheim}
\affiliation{Department of Physics, University of Connecticut, Storrs, CT 06269, USA \\
 philip.mannheim@uconn.edu\\ }

\date{June 28 2022}

\begin{abstract}

Since particles obey wave equations, in general one is not free to postulate that  particles move on the geodesics associated with test particles. Rather, for this to be the case one has to be able to derive such behavior starting from the equations of motion that the particles obey, and to do so one can employ the eikonal approximation. To see what kind of trajectories might occur we explore the domain of support of the propagators associated with the wave equations, and extend the results of some previous propagator studies that have appeared in the literature. For a minimally coupled massless scalar field the domain of support in curved space is not restricted to the light cone, while for a conformally coupled massless scalar field the curved space domain is only restricted to the light cone if the scalar field propagates in a conformal to flat background. Consequently, eikonalization does not in general lead to null geodesics for curved space massless rays even though it does lead to straight line trajectories in flat spacetime. Equal remarks apply to the conformal invariant Maxwell equations. However, for massive particles one does obtain standard geodesic behavior this way, since they do not propagate on the light cone to begin with.  Thus depending on how big the curvature actually is, in principle, even if not necessarily in practice, the standard null-geodesic-based gravitational bending formula and the general behavior of propagating light rays are in need of modification in regions with high enough curvature. We show how to appropriately modify the geodesic equations in such situations. We show that relativistic eikonalization has an intrinsic light-front structure, and show that eikonalization in a theory with local conformal symmetry leads to trajectories that are only globally conformally symmetric. Propagation of massless particles off the light cone is a curved space reflection of the fact that when light travels through a refractive medium in flat  spacetime its velocity is modified from its free flat spacetime value. In the presence of gravity spacetime itself acts as a medium, and this medium can then take light rays off the light cone. This is also manifest in a conformal invariant scalar field theory propagator in two spacetime dimensions. It takes support off the light cone, doing so in fact  even if the geometry is conformal to flat. We show that it is possible to obtain eikonal trajectories that are exact without approximation, and show that normals to advancing wavefronts follow these exact eikonal trajectories, with these trajectories being the trajectories along which energy and momentum are transported. In general then, in going from flat space to curved space one does not generalize flat space geodesics to curved space geodesics. Rather, one generalizes flat space wavefront normals (normals that are geodesic in flat space) to curved space wavefront normals, and in curved space normals to wavefronts do not have to be geodesic.
\end{abstract}

\maketitle

\section{Introduction}
\label{S1}

\subsection{Background}
\label{S1a}

As introduced in Riemannian geometry a geodesic is  the shortest distance between two points in a general curved space. In physics one takes advantage of this fact by introducing the test particle action 
\begin{eqnarray}
I_T=-m\int ds,                                                                               
\label{1.1x}
\end{eqnarray}               
where  $T$ denotes test particle,  $m$ is the particle mass, and $ds=(-g_{\mu\nu}dx^{\mu}dx^{\nu})^{1/2}$ is the proper time (we use the notation and conventions in \cite{Weinberg1972} in which $g_{00}$ is negative). Stationary variation of $I_T$ with respect to the particle coordinate $x^{\lambda}$ leads to the geodesic equation
\begin{eqnarray}
m \left( \frac{D^2x^{\lambda}}{ Ds^2}\right)=m \left( \frac{d^2x^{\lambda} }{ ds^2}
+\Gamma^{\lambda}_{\mu \nu} 
\frac{dx^{\mu}}{ds}\frac{dx^{\nu } }{ ds} \right) 
= 0, 
\label{1.2x}
\end{eqnarray}
where, as introduced here, $D^2x^{\lambda}/Ds^2$ is the total acceleration and $\Gamma^{\lambda}_{\mu \nu}=\tfrac{1}{2}g^{\lambda\sigma}(\partial_{\mu}g_{\nu\sigma} +\partial_{\nu}g_{\mu\sigma}-\partial_{\sigma}g_{\mu\nu})$ is the connection. As such, only the total $D^2x^{\lambda}/Ds^2$ transforms as a general coordinate vector, with neither of its $d^2x^{\lambda}/ds^2$ or $\Gamma^{\lambda}_{\mu \nu}(dx^{\mu}/ds)(dx^{\nu } /ds)$ components separately doing so, and with it being only their sum with the specifically indicated relative weights that is a general coordinate vector. Moreover, precisely because $\Gamma^{\lambda}_{\mu\nu}$ is not a general coordinate tensor we are actually able to remove it from (\ref{1.2x}) at any given point via a general coordinate transformation \cite{footnote1}, and thus establish the equivalence principle between gravity and acceleration. Since having this particular set of relative weights leads to the equality of inertial and gravitational masses it is generally thought that particles must move on geodesics. However the two terms only appear with this particular set of relative weights because any other combination would not lead to the total $D^2x^{\lambda}/Ds^2$  being a general coordinate vector in the first place. To emphasize the point we consider an illustrative, more general,  covariant test particle action
\begin{eqnarray}
I_T=-m\int ds- \kappa\int ds
R^{\alpha}_{\phantom{\alpha}\alpha},                                                               
\label{1.3x}
\end{eqnarray}               
where $R^{\alpha}_{\phantom{\alpha}\alpha}$ is the Ricci scalar and $\kappa$ is a  constant. As such, this action describes a particle that couples to the Ricci scalar as it propagates along a trajectory.
The stationary variation of $I_T$ with respect to $x^{\lambda}$ leads to \cite{Mannheim2006}
\begin{eqnarray}
\left(m +\kappa R^{\alpha}_{\phantom {\alpha} \alpha}\right)\left(
\frac{d^2x^{\lambda} }{ ds^2} +\Gamma^{\lambda}_{\mu \nu} 
\frac{dx^{\mu}}{ds}\frac{dx^{\nu } }{ ds} \right) 
= -\kappa \left( g^{\lambda
\beta}+
\frac{dx^{\lambda}}{ds}                                                      
\frac{dx^{\beta}}{ds}\right)  \nabla_{\beta}R^{\alpha}_{\phantom {\alpha} \alpha}. 
\label{1.4x}
\end{eqnarray}
As we see, again it is the total $D^2x^{\lambda}/Ds^2$ combination that appears, and its ubiquity is due to the fact that covariance would allow no other possibility. Nonetheless,  even though we can still remove the dependence on $\Gamma^{\lambda}_{\mu\nu}$ at any given point, under no coordinate transformation could we also remove the dependence on the Ricci scalar at the same given point precisely because we would simultaneously need both $\Gamma^{\lambda}_{\mu\nu}$ and its derivatives to vanish. Now (\ref{1.4x}) is just as covariant as (\ref{1.2x}), and they both reduce to the flat Cartesian space $d^2x^{\lambda}/ds^2=0$ in the absence of curvature, in exactly the same manner as (\ref{1.3x}) reduces to (\ref{1.1x}) in the same limit. However, only for (\ref{1.2x}) could we simulate the entire effect of a gravitational field at any given point by an acceleration. Since the equivalence principle is commonly understood as the requirement that we can remove all gravitational effects at a given point by a general coordinate transformation, it is commonly assumed that particles move on the purely geodesic (\ref{1.2x}) and not on any other trajectory. However, (\ref{1.4x}) is just as covariant as (\ref{1.2x}), and is allowed by the principle of general coordinate invariance, and in general we should contemplate replacing (\ref{1.2x}) by a more general trajectory such as that given in (\ref{1.4x}). Moreover, in \cite{Weinberg1972} Weinberg also notes that (\ref{1.2x}) could be generalized, and suggested a possible generalization of the form 
\begin{eqnarray}
\frac{d^2x^{\lambda} }{ ds^2} +\Gamma^{\lambda}_{\mu \nu} 
\frac{dx^{\mu}}{ds}\frac{dx^{\nu } }{ ds}  
 =-fR^{\lambda}_{\phantom{\lambda}\mu\nu\kappa}\frac{dx^{\mu}}{ds}\frac{dx^{\nu } }{ ds} S^k
 \label{1.5x}
 \end{eqnarray}
for a particle with spin vector $S^k$ in a geometry with Riemann tensor $R^{\lambda}_{\phantom{\lambda}\mu\nu\kappa}$, with  $f$ being a constant. It is the purpose of this paper to show that  rather than considering generalizations of (\ref{1.2x}) to be just a logical possibility, such generalizations are actually the general rule, with it being (\ref{1.2x}) itself that is the exception. However, since the modifications that we will present from a study of wave equations rather than test particle actions will involve the Ricci tensor and Ricci scalar, such terms are actually absent in the familiar tests of the equivalence principle that are made in the Ricci flat environment of the solar system. In fact it is the very success of these solar system tests that has led to the use (\ref{1.2x}) in other situations, situations that are not Ricci flat.
 
Now procedurally there is nothing  wrong in varying (\ref{1.1x}) to obtain (\ref{1.2x}). However, our concern here is whether real particles as opposed to test particles can actually be described by the test particle action $I_T$ in the first place. That there might actually be a concern is due to the fact that one cannot use (\ref{1.1x}) for massless particles since even if we drop the factor $m$ and only consider $I_T=\int ds$, for particles that propagate on the light cone $ds=(-g_{\mu\nu}dx^{\mu}dx^{\nu})^{1/2}$ is zero. Nonetheless, regardless of whether or not the general coordinate scalar $I_T$ action without any $m$ factor is to be relevant for massless particles, the variation of $I_T$ does show that the left-hand side of (\ref{1.2x}) is a general coordinate vector. Thus for massless particles one ordinarily replaces (\ref{1.2x}) by the null geodesic equation
\begin{eqnarray}
 \frac{d^2x^{\lambda} }{ dq^2}
+\Gamma^{\lambda}_{\mu \nu} 
\frac{dx^{\mu}}{dq}\frac{dx^{\nu } }{ dq} = 0, 
\label{1.6x}
\end{eqnarray}
where $q$ is a general coordinate scalar affine parameter that measures distance along the trajectory associated with (\ref{1.6x}). As such, with its left-hand side being a true general coordinate vector ($dq$ like $ds$ being a general coordinate scalar), (\ref{1.6x}) does covariantly describe the propagation of massless particles. Nonetheless, and despite this, in field theory particles are associated with wave equations rather than with test particle equations, and nowhere in any fundamental Lagrangian description  of the fundamental forces does any test particle action that might lead to the massive particle (\ref{1.2x}) or the massless particle (\ref{1.6x})  actually appear at all. Thus one has to ask whether one could obtain geodesic behavior starting from field equations instead, and whether if in doing so one could instead obtain some modified form of trajectory. This then is the objective of this paper. 

In regard to covariantization, we should note that in and of itself the procedure of replacing  flat Minkowski space metric and derivatives ($\eta_{\mu\nu}$, $\partial_{\mu}$) by covariant metric  and derivatives ($g_{\mu\nu}$ and $\nabla_{\mu}$) is actually independent of the presence or absence of curvature. While the condition $d^2x^{\lambda}/ds^2=0$ is  the special relativistic flat spacetime ($R_{\lambda\mu\nu\tau}=0$) version of Newton's Law of Motion for a free particle, as written it is written in Minkowski coordinates, and as such this equation  is invariant under transformations with uniform velocity. However, it is not left invariant under transformations to accelerating coordinate systems, transformations that will still keep $R_{\lambda\mu\nu\tau}$ zero.  Rather, it is the geodesic equation $d^2x^{\lambda}/ds^2+\Gamma^{\lambda}_{\mu\nu}(dx^{\mu}/ds)(dx^{\nu}/ds)=0$ that is left invariant, and as such it writes Newton's Law of Motion in a form that all flat space observers, accelerating or not accelerating could agree on. However, since all we had done is make a coordinate transformation, if we started in flat spacetime we remain in flat spacetime. Now, as such, covariantization is generally understood as replacing $d^2x^{\lambda}/ds^2=0$ in a space with $R_{\lambda\mu\nu\tau}=0$ by $d^2x^{\lambda}/ds^2+\Gamma^{\lambda}_{\mu\nu}(dx^{\mu}/ds)(dx^{\nu}/ds)=0$ in a space with $R_{\lambda\mu\nu\tau}\neq 0$. As we see, in general this is an incomplete description for curved space trajectories since we also need to allow for the inclusion of curvature-dependent terms, terms that vanish in the flat space limit. 

Moreover, as we will see, not only will we get some modification of geodesic motion per se, for conformally coupled massless particles such as photons we will get trajectories that are not even constrained to lie on the curved space light cone (viz. the covariantized flat spacetime light cone in which $\eta_{\mu\nu}$ is replaced by $g_{\mu\nu}$), unless, however, they are propagating in a geometry that just happens to be conformal to flat.  Such an outcome is actually not without precedent as it already occurs in flat spacetime, where  the velocity of light in a refractive medium is given not by $c$ itself but by $c/n$ where $n$ is the refractive index. The curving of spacetime by gravity causes spacetime to become a medium, and this not only can but actually is able to take light rays off the covariantized version of the light cone that they would otherwise have travelled on in the absence of any curvature. 

To see what can happen in the curved space case we recall that for propagation of electromagnetic waves there are two regimes, geometrical optics and physical optics. If the wave has a wavelength $\lambda$ and propagates through some system with a characteristic scale $L$ the geometrical optics limit is $\lambda \ll L$, while the physical optics regime is where $\lambda$ is of order $L$. In the geometrical optics limit in flat spacetime light rays travel in straight lines that are normal to an advancing wavefront, while in physical optics phenomena such as diffraction occur. It is because these geometrical optics light rays are one-dimensional trajectories that are in the direction of the momentum carried by the wave that one can give a particle interpretation to wave motion in the sense that the ray trajectories can be geodesic. Since just as with flat space rays curved space geodesics are also one-dimensional, one can thus anticipate that for geometrical optics in curved space the normal to the advancing curved wavefront would be null geodesic. Since one can use the eikonal approximation to show that in flat spacetime short wavelength rays do travel in straight lines, we shall thus generalize the optical eikonal approximation presented in \cite{Bruns1895} and \cite{Sommerfeld1911} to curved space. However, as we shall see on doing this, we in general find that the ray trajectories, while still being one-dimensional curves, are not in fact null geodesic. (As we show in Sec. \ref{S1e} the wavefronts themselves and thus the normals to them can go off the light cone.) However, there is an exception if the field obeys a conformal invariant wave equation and is propagating in a geometry that is conformal to flat, since in that case the full content of the theory is the same as that of propagation in flat spacetime, viz. null geodesic.  In Sec. \ref{S8} we will see how this works in a particular conformal to flat case, namely propagation in a  Robertson-Walker geometry. Though even in this case we note (and explore in Sec. \ref{S9}) that fluctuations around a cosmological Robertson-Walker background are not conformal to flat.

\subsection{Flat Spacetime Geometrical Optics}
\label{S1b}

Since a study of the eikonal approximation in the relativistic curved spacetime case is the covariant generalization of the standard treatment of flat spacetime geometrical optics, we quickly review how things work in geometrical optics itself. We follow the treatment given in \cite{Born1959} and \cite{Jackson1998}. We start with the Maxwell equations for electromagnetic fields that oscillate as $e^{-i\omega t}$ in a medium with a spatially-dependent refractive index $n(\textbf{x})=(\epsilon(\textbf{x})/\epsilon_0)^{1/2}$, viz.
\begin{eqnarray}
&&\boldsymbol{\nabla}\cdot(\epsilon(\textbf{x})\boldsymbol{E})=0,\quad \boldsymbol{\nabla}\times\boldsymbol{E}=i\omega\boldsymbol{B},\quad 
\boldsymbol{\nabla}\cdot\boldsymbol{B}=0,\quad \boldsymbol{\nabla}\times\boldsymbol{B}=-i\mu_0\omega \epsilon(\textbf{x})\boldsymbol{E}.
\label{1.7g}
\end{eqnarray}
Manipulation of these equations yields 
\begin{eqnarray}
&&\boldsymbol{\nabla}^2\boldsymbol{E}+\mu_0\omega^2 \epsilon(\textbf{x})\boldsymbol{E}+\boldsymbol{\nabla}\left(\epsilon^{-1}
\boldsymbol{E}\cdot\boldsymbol{\nabla}\epsilon\right)=0,
\nonumber\\
&&\boldsymbol{\nabla}^2\boldsymbol{B}+\mu_0\omega^2 \epsilon(\textbf{x})\boldsymbol{B}-i\mu_0\omega\boldsymbol{\nabla}\epsilon
\times\boldsymbol{E}=0.
\label{1.8g}
\end{eqnarray}
If we assume that the $\boldsymbol{\nabla}\epsilon$ terms are negligible, we can drop the third term in each of the two equations in (\ref{1.8g}) and obtain 
\begin{eqnarray}
&&\left[\boldsymbol{\nabla}^2\boldsymbol+\frac{\omega^2}{c^2} n^2(\textbf{x})\right]\alpha=0,
\label{1.9g}
\end{eqnarray}
where $\alpha$ denotes $\boldsymbol{E}$ or $\boldsymbol{B}$. Setting $\alpha=\exp[i\omega \psi(\textbf{x})/c)]$
we obtain
\begin{eqnarray}
&&\frac{\omega^2}{c^2}\left[n^2(\textbf{x})-\boldsymbol{\nabla}\psi\cdot\boldsymbol{\nabla}\psi\right]+
i\frac{\omega}{c}\boldsymbol{\nabla}^2\psi=0.
\label{1.10g}
\end{eqnarray}
Then on assuming that $\boldsymbol{\nabla}^2\psi$ is much smaller than $\boldsymbol{\nabla}\psi\cdot \boldsymbol{\nabla}\psi$, we obtain
\begin{eqnarray}
&&\boldsymbol{\nabla}\psi\cdot\boldsymbol{\nabla}\psi=n^2(\textbf{x}).
\label{1.11g}
\end{eqnarray}
Eq. (\ref{1.11g})  is  known as the eikonal equation of geometrical optics, with $\psi$ being known as the eikonal function or eikonal phase.

If we set $\boldsymbol{E}=\boldsymbol{E}_0\exp[i\omega \psi(\textbf{x})/c]$, $\boldsymbol{B}=\boldsymbol{B}_0\exp[i\omega \psi(\textbf{x})/c]$, where $\boldsymbol{E}_0$ and $\boldsymbol{B}_0$ are constants,  we obtain
\begin{eqnarray}
\boldsymbol{\nabla}\psi\times\boldsymbol{E}_0=c \boldsymbol{B}_0,\quad 
\boldsymbol{\nabla}\psi\times\boldsymbol{B}_0=- \frac{ n^2(\textbf{x})}{c}\boldsymbol{E}_0.
\label{1.12g}
\end{eqnarray}
From (\ref{1.12g}) it then follows that $\boldsymbol{\nabla}\psi$ is orthogonal to both $\boldsymbol{E}$ and $\boldsymbol{B}$. It thus points in the same direction as the Poynting vector  $\boldsymbol{E}\times\boldsymbol{B}$. With  $\boldsymbol{E}$ and $\boldsymbol{B}$ both lying in the wavefront, $\boldsymbol{\nabla}\psi$ is thus in the direction normal to the wavefront, i.e., in the same direction as the momentum carried by the wave, viz. in the same direction as the  light ray.  The  $\boldsymbol{E}$ and $\boldsymbol{B}$ fields are transverse and $\boldsymbol{E}\times\boldsymbol{B}$ is longitudinal.
If we introduce a unit vector $\hat{\boldsymbol{k}}$ in the direction parallel to the normal to the wave front we can set
\begin{eqnarray}
 \boldsymbol{\nabla}\psi=n(\textbf{x})\hat{\boldsymbol{k}}.
\label{1.13g}
\end{eqnarray}
Now if the wave travels a distance $d\boldsymbol{r}$ in a distance $dq$ as measured along the ray we can set $d\boldsymbol{r}/dq=\hat{\boldsymbol{k}}$. The trajectory of the ray is thus given by 
\begin{eqnarray}
 \frac{d\boldsymbol{r}}{dq}=\frac{\boldsymbol{\nabla}\psi}{n(\textbf{x})}.
\label{1.14g}
\end{eqnarray}
Noting that $d/dq=\hat{\boldsymbol{k}}\cdot\boldsymbol{\nabla}$, differentiation of (\ref{1.14g}) yields
\begin{eqnarray}
 n(\textbf{x})\frac{d^2\boldsymbol{r}}{dq^2}+\frac{d\boldsymbol{r}}{dq}\left(\boldsymbol{\nabla}n(\textbf{x})\cdot\frac{d\boldsymbol{r}}{dq}\right)
 =\boldsymbol{\nabla}n(\textbf{x}).
 \label{1.15g}
\end{eqnarray}
For constant $n(\textbf{x})=n$ (\ref{1.15g}) reduces to $d^2\boldsymbol{r}/dq^2=0$, i.e., light travels on a straight line (with speed $c/n$), while if $n(\textbf{x})$ varies in space the trajectory departs from a straight line according to (\ref{1.15g}). This then is the eikonal approximation in flat spacetime classical electrodynamics

In this regard we note in passing that in the literature it has been noted \cite{Herzberger1936} that (\ref{1.11g}) is analogous to the equation obeyed by Hamilton's characteristic function. In optics as well as eikonalization one also has the least time principle of Fermat. Its status in the gravitational case and its connection to a least action principle has been discussed in \cite{Kovner1990} and \cite{Perlick1990}, with it being shown that on the light cone the associated trajectories are null geodesic. It is the purpose of this paper to recover this result in a procedure that enables us to determine its modification for trajectories that are not restricted to the light cone.

\subsection{Implications for Curved Space Eikonalization}
\label{S1c}

From the above analysis we identify the main ingredients of the eikonal approximation, viz. excluding terms that are negligible (usually achievable at short wavelengths), and identifying  $\boldsymbol{\nabla}\psi$ with the velocity along the ray as in (\ref{1.14g}). Given (\ref{1.11g}), it follows that $d\boldsymbol{r}/dq$ is normalized according to 
\begin{eqnarray}
\frac{d\boldsymbol{r}}{dq}=\frac{\boldsymbol{\nabla}\psi}{(\boldsymbol{\nabla}\psi\cdot\boldsymbol{\nabla}\psi)^{1/2}},\quad 
\frac{d\boldsymbol{r}}{dq}\cdot \frac{d\boldsymbol{r}}{dq}=1.
 \label{1.16g}
\end{eqnarray}
Normalizing $\boldsymbol{\nabla}\psi$, which is standard in optics \cite{Born1959}, will prove central in the following. In fact not only will it prove central, in Sec. \ref{S3} we will see that it is even required for consistency in the covariant case. In generalizing to the covariant Maxwell case we look for covariant generalizations of (\ref{1.16g}) and (\ref{1.15g}), and to this end we first introduce a covariant eikonal four-velocity  $dx^{\mu}/dq$, as defined along a one-dimensional curved space trajectory. For (\ref{1.16g}) there are then two natural generalizations, either $(dx_{\mu}/dq)(dx^{\mu}/dq)=0$ (lightlike) or  $(dx_{\mu}/dq)(dx^{\mu}/dq)=-1$ (timelike in the $ds^2=-g_{\mu\nu}dx^{\mu}dx^{\nu}$ convention with $g_{00}<0$ used in \cite{Weinberg1972}). Similarly, for (\ref{1.15g}) there are also two natural generalizations, either the null geodesics given in (\ref{1.6x}), viz.
\begin{eqnarray}
\frac{d^2x^{\lambda} }{ dq^2} +\Gamma^{\lambda}_{\mu \nu} 
\frac{dx^{\mu}}{dq}\frac{dx^{\nu } }{ dq}=0,
\label{1.17g}
\end{eqnarray}
or Ricci scalar (or Ricci or Riemann  tensor)  dependent ones such as the one given in (\ref{1.4x}) with $m=0$, viz. the equally one-dimensional trajectory
\begin{eqnarray}
R^{\alpha}_{\phantom {\alpha} \alpha}\left(
\frac{d^2x^{\lambda} }{ dq^2} +\Gamma^{\lambda}_{\mu \nu} 
\frac{dx^{\mu}}{dq}\frac{dx^{\nu } }{ dq} \right) 
= - \left( g^{\lambda
\beta}+
\frac{dx^{\lambda}}{dq}                                                      
\frac{dx^{\beta}}{dq}\right)  \nabla_{\beta}R^{\alpha}_{\phantom {\alpha} \alpha}.
\label{1.18g}
\end{eqnarray}

Now initially we would anticipate that the both of these two trajectories would follow from the light cone condition $(dx_{\mu}/dq)(dx^{\mu}/dq)=0$ since neither involves any mass term. And by the same token we would anticipate that the $(dx_{\mu}/dq)(dx^{\mu}/dq)=-1$ option would only be associated with massive particles. What we will actually find is that is not the case, with it being possible to associate $(dx_{\mu}/dq)(dx^{\mu}/dq)=-1$ with the propagation of massless fields. Specifically, in the curved space Maxwell case that we study in Sec. \ref{S4} there is an explicit dependence on the Ricci tensor  in the Maxwell equations themselves, with the curved space vector potential Maxwell equation $\nabla_{\mu}(\nabla^{\mu}A^{\nu}-\nabla^{\nu}A^{\mu})=0$ taking the form
\begin{eqnarray}
\nabla_{\mu}\nabla^{\mu}A^{\nu}-\nabla^{\nu}\nabla_{\mu}A^{\mu}+R^{\nu\alpha}A_{\alpha}=0.
\label{1.19g}
\end{eqnarray}
As we show in Sec. \ref{S4} below, provided the spacetime geometry is not conformal to flat eikonalization of (\ref{1.19g}) in the covariant $\nabla_{\mu}A^{\mu}=0$ gauge where there are three polarization vectors leads to a form (given in (\ref{4.13x})) that is analogous to (\ref{1.18g}), viz.
\begin{eqnarray}
\frac{d^2x^{\lambda} }{ dq^2}+\Gamma^{\lambda}_{\mu \nu} 
\frac{dx^{\mu}}{ dq}\frac{dx^{\nu } }{dq}
=-\frac{1}{2(-(1/3)\sum_{a=1}^3R^{\alpha\beta}\epsilon_{\alpha (a)}\epsilon_{\beta (a)})}
\left[g^{\lambda \mu}+\frac{dx^{\lambda}}{dq}\frac{dx^{\mu}}{dq}\right]\frac{\partial (-(1/3)\sum_{a=1}^3R^{\gamma\delta}\epsilon_{\gamma (a)}\epsilon_{\delta (a)})}{\partial x^{\mu}},
\label{1.20g}
\end{eqnarray}
where the $\epsilon_{\alpha (a)}$ ($a=(1,2,3)$) are three polarization vectors. And not only that, (\ref{1.20g}) is also associated with the non light cone $(dx_{\mu}/dq)(dx^{\mu}/dq)=-1$ condition, with curvature taking photons off the light cone. As we will show in Secs. \ref{S1e} and \ref{S3} however, if the wave equation is conformal invariant and the spacetime geometry in which the wave is propagating is conformal to flat, then the trajectories are on the light cone and described by (\ref{1.17g}).  While photons propagating in a Robertson-Walker geometry fall into this category, fluctuations around this background do not as the fluctuation metric is not conformal to flat, and lead to the modifications presented in Sec. \ref{S9}. Similarly, the metric associated with a spherically symmetric source that is at rest is also not conformal to flat, and thus gravitational lensing off such systems is affected (Secs. \ref{S5}, \ref{S6} and \ref{S7}) by the considerations we present in this paper.

Relating $ d\textbf{r}/dq$ to a derivative function as in (\ref{1.16g}) has a parallel in non-relativistic classical particle mechanics. Specifically, if one introduces an action $I$ and makes an Euler-Lagrange variation in order to obtain the classical equations of motion, one finds that the momentum can be written as $\textbf{p}=\boldsymbol{\nabla}I_{STAT}$ where $I_{STAT}$ is the stationary action, viz. the action as evaluated in the solution to the equations of motion, a solution that is a one-dimensional trajectory. Thus for a free particle for instance, for travel between $(t=0,\textbf{x}=0)$ and $(t=T,\textbf{x}=\textbf{X})$ the action, equation of motion, and its solution are given by 
\begin{eqnarray}
I=\frac{m}{2}\int_0^Tdt\dot{\textbf{x}}^2,\quad m\ddot{\textbf{x}}=0,\quad \textbf{x}(t)=\frac{\textbf{X}t}{T},
\label{1.21h}
\end{eqnarray}
and the stationary action evaluates to
\begin{eqnarray}
I_{STAT}=\frac{m\textbf{X}^2}{2T}. 
\label{1.22h}
\end{eqnarray}
The momentum is then given by varying $I_{STAT}$ with respect to the end point, to thus yield 
\begin{eqnarray}
\textbf{P}=\boldsymbol{\nabla}I_{STAT}=\frac{m\textbf{X}}{T}=m\dot{\textbf{X}}
\label{1.23h}
\end{eqnarray}
along the one-dimensional  trajectory that is the solution to the equation of motion.
For this example  the equation of motion is the geodesic equation, and its solution can be written as a total derivative. Since we can write (\ref{1.16g}) as $(\boldsymbol{\nabla}\psi\cdot\boldsymbol{\nabla}\psi)^{1/2}d\boldsymbol{r}/dq=\boldsymbol{\nabla}\psi$, for eikonalized light rays again we see that the solution to the geodesic equation can be written as a total derivative.  Now initially obtaining a total derivative form is somewhat surprising since there is nothing in the general form of  the geodesic equations given in (\ref{1.2x}) and (\ref{1.6x})  that would suggest that for either of them the solution could be written as a  total derivative, since $dx^{\mu}/dq$ would appear to arbitrary and not restricted in any way \cite{footnote1a}. It is only through the derivation of the geodesic equations in the first place starting from an eikonalized  wave equation or a stationary particle action that such a structure is forced upon us. Thus in the following $dx^{\mu}/dq$ or $dx^{\mu}/ds$ will always be related to a total derivative function.

While our study is motivated by short wavelength geometrical optics, and while  we will study phenomenological implications in such cases, in Sec. \ref{S10} we will show how our formalism can be extended to situations in which the eikonal function term is of the same order as all other terms in the wave equation. By studying some general geometric properties of embeddings we establish that the covariant eikonal velocity generalization $dx^{\mu}/dq$ of the $d\boldsymbol{r}/dq$ eikonal velocity is normal to the wavefront of an advancing wave, with $dx^{\mu}/dq$ being the velocity with which the energy and momentum of the wave are transported. Since the normal to the wavefront and thus equivalently the eikonal velocity $dx^{\mu}/dq$ evolve along one-dimensional curved space trajectories, both can be related to a total derivative of a scalar function. As we will see in Sec. \ref{S2}, that scalar function is the eikonal function $T$, as defined for a scalar field for instance via $S=Ae^{iT}$. As we discuss in Sec. \ref{S10},  it is the relating of $dx^{\mu}/dq$ to the eikonal function that is actually the key step as it can be done without needing to make any short wavelength (viz. large $T$) approximation at all.

To summarize, we see that there are two key steps that go into the eikonal approximation, namely identifying the eikonal velocity $dx^{\mu}/dq$ as the normalized derivative of the eikonal function $T$, and making a short wavelength approximation in order  to determine the equation obeyed by $T$  so as  to obtain the  trajectory that $dx^{\mu}/dq$ follows in the short wavelength case. However, these two steps are independent, and we can make the former without needing to make the latter. Moreover, the eikonal velocity $dx^{\mu}/dq$ is the normal to the wavefront of the associated wave equation, and such normals exist for all wavelengths, short or long. Now while we will discuss the short wavelength approximation in this paper, in Sec. \ref{S10} we will also discuss the case where we make no approximation for the wavelength at all. In flat spacetime the straight line rays of geometrical optics are both normals to plane wave wavefronts and null geodesics. The generalization to curved space is not to replace flat space null geodesics by curved space ones, but to replace normals to flat space wavefronts by normals to curved space wavefronts. For short wavelengths the normals to  curved space wavefronts are null geodesic, but for long wavelengths they need not be null geodesic, and it is then those normals that are the relevant ones, with energy and momentum being transported along them and not along null geodesics. And as we shall see, these normals can take support off the light cone.

\subsection{Comparing Eikonalization with a Harmonic Function Analysis}
\label{S1d}

In flat spacetime the wave equation $(\partial_t^2-\boldsymbol{\nabla}^2)S=0$ with scalar field $S$ has general harmonic solution $S=h(t-|\vec{x}|)$ for arbitrary function $h$. Thus if we repeat this analysis in the curved space situation (the procedure discussed in \cite{Ellis2009}) we would introduce a phase $\chi$, and on setting $S=h(\chi)$ in $\nabla_{\mu}\nabla^{\mu}S=0$  would obtain
\begin{eqnarray}
h^{\prime}\nabla_{\mu}\nabla^{\mu}\chi+h^{\prime\prime}\nabla_{\mu}\chi\nabla^{\mu}\chi=0,
\label{1.24h}
\end{eqnarray}
where the prime denotes the derivative with respect to $\chi$. With $h(\chi)$ being arbitrary, the coefficients of both $h^{\prime}$ and $h^{\prime\prime}$ have to vanish separately. This leads to
\begin{eqnarray}
\nabla_{\mu}\nabla^{\mu}\chi=0,\quad \nabla_{\mu}\chi\nabla^{\mu}\chi=0.
\label{1.25h}
\end{eqnarray}
Then on setting $dx^{\mu}/dq=\nabla^{\mu}\chi$ we obtain
\begin{eqnarray}
\nabla_{\mu}\frac{dx^{\mu}}{dq}=0,\quad \frac{dx_{\mu}}{dq}\frac{dx^{\mu}}{dq}=0,
\label{1.26h}
\end{eqnarray}
from which, and as we explicitly show in Sec. \ref{S2}, the null geodesic equation (\ref{1.17g}) follows.  As such,  this analysis was applied to the Maxwell case in \cite{Ellis2009}, with the presence of the Ricci tensor in (\ref{1.19g}) not affecting the $\nabla_{\mu}\chi\nabla^{\mu}\chi=0$ condition at all, so that the null geodesic (\ref{1.17g}) was recovered.

However, there is another class of solutions to (\ref{1.24h}). Specifically, if $h=e^{\chi}$ then $h^{\prime}$ and $h^{\prime\prime}$ are equal, and from (\ref{1.24h}) we then only obtain
\begin{eqnarray}
\nabla_{\mu}\nabla^{\mu}\chi+\nabla_{\mu}\chi\nabla^{\mu}\chi=0.
\label{1.27h}
\end{eqnarray}
As we show in Sec. \ref{S2}, it is this solution that coincides with eikonalization of  $\nabla_{\mu}\nabla^{\mu}S=0$. And as we show in Sec. \ref{S4}, the Maxwell analog of (\ref{1.27h}) does involve the Ricci tensor directly, leading us not to (\ref{1.17g}) but to (\ref{1.20g}) instead in geometries that are not conformal to flat.

\subsection{Implications from Propagators}
\label{S1e}

Before embarking on a study of eikonalization in curved space it is instructive to study the structure of some curved space propagators, so as  to see what their domains of support might be, since any eikonalized rays would have to lie within those domains.  The analysis we present here extends some earlier studies that have appeared in the literature \cite{DeWitt1960,Hu1998,Mannheim2007,Copi2021}. To begin, we note that for a free massless scalar field in flat spacetime the action and wave equation are of the form 
\begin{eqnarray}
I_{S}=-\int d^4x\frac{1}{2}\partial_{\mu}S\partial^{\mu}S,\quad \left(-\partial_0^2+\vec{\nabla}^2\right)S=0,
\label{1.28h}
\end{eqnarray}
and the retarded propagator obeys
\begin{eqnarray}
\left(-\partial_0^2+\vec{\nabla}^2\right)D_{FLAT}(t,\vec{x};t^{\prime},\vec{x}^{\prime})=-\delta^4(x-x^{\prime}).
\label{1.29h}
\end{eqnarray}
Here $D_{FLAT}(t,\vec{x};t^{\prime},\vec{x}^{\prime})$ is given by
\begin{eqnarray}
D_{FLAT}(t,\vec{x};t^{\prime},\vec{x}^{\prime})=-\frac{1}{(2\pi)^4}\int d^4k \frac{e^{ik\cdot (x-x^{\prime})}}{k_0^2-k_1^2-k_2^2-k_3^2+i\epsilon\theta(k^0)}
=\frac{1}{2\pi}\theta (x_0-x_0^{\prime})\delta[(x-x^{\prime})^2],
\label{1.30h}
\end{eqnarray}
as integrated over the retarded complex $k^0$ plane contour. Massless rays thus travel on $(t-t^{\prime})^2-(\vec{x}-\vec{x}^{\prime})^2=0$. And with $q$ measuring distance along the trajectory the rays obey
\begin{eqnarray}
\frac{d^2x^{\lambda}}{dq^2}=0,
\label{1.31h}
\end{eqnarray}
viz. the flat space null geodesic.

The natural and immediate generalization to curved space would be the minimally coupled
\begin{eqnarray}
I_{S}=-\int d^4x(-g)^{1/2}\frac{1}{2}\nabla_{\mu}S\nabla^{\mu}S,\quad \nabla_{\mu}\nabla^{\mu}S=0,
\label{1.32h}
\end{eqnarray}
with a retarded minimally coupled propagator that obeys
\begin{eqnarray}
\nabla_{\mu}\nabla^{\mu}D_{MIN}(t,\vec{x};t^{\prime},\vec{x}^{\prime})=-(-g)^{-1/2}\delta^4(x-x^{\prime}),
\label{1.33h}
\end{eqnarray}
and geodesics that would be anticipated to be of the null form 
\begin{eqnarray}
\frac{d^2x^{\lambda}}{dq^2}+\Gamma^{\lambda}_{\mu \nu} 
\frac{dx^{\mu}}{dq}\frac{dx^{\nu } }{ dq}=0.
\label{1.34h}
\end{eqnarray}

To determine the domain of support of the minimally coupled $D_{MIN}$ and show that it in fact actually encompasses more than just the light cone we only need to find one example. So consider a geometry that is conformal to flat with a conformal factor $\Omega(p)$ that only depends on the conformal time $p$, viz.
\begin{eqnarray}
ds^2=\Omega^2(p)(dp^2-dx^2-dy^2-dz^2).
\label{1.35h}
\end{eqnarray}
(In terms of comoving Robertson-Walker geometries with expansion radius $a(t)$ the conformal time is $p=\int dt/a(t)$, while $\Omega(p)=a(t)$.) With the $\nabla_{\mu}\nabla^{\mu}S=0$ wave equation being of the form
\begin{eqnarray}
\frac{1}{\Omega^3}\left(-\Omega\partial_0^2-2\dot{\Omega}\partial_0+\Omega\vec{\nabla}^2\right)S=0,
\label{1.36h}
\end{eqnarray}
solutions are of the form
\begin{eqnarray}
S(x)=\frac{g(p)}{\Omega(p)}e^{ik_ix^i},
\label{1.37h}
\end{eqnarray}
where $g(p)$ obeys
\begin{eqnarray}
\Omega\ddot{g}-\ddot{\Omega}g+(k_1^2+k_2^2+k_3^2)\Omega g =0.
\label{1.38h}
\end{eqnarray}
We now choose as special cases $\Omega(p)=\sin(Hp)$ and  $\Omega(p)=e^{Hp}$. ($\Omega(p)=e^{Hp}$ is not a de Sitter geometry as in de Sitter it is the comoving expansion radius $a(t)$ that is equal to $e^{Ht}$, with $p$ being given by $Hp=e^{-Ht}$. To obtain $\Omega(p)=e^{Hp}$ we would need $a(t)=Ht$.) With $\Omega(p)=\sin(Hp)$ or $\Omega(p)=e^{Hp}$ (\ref{1.38h}) respectively reduces to 
\begin{eqnarray}
\ddot{g}\pm H^2g+(k_1^2+k_2^2+k_3^2)g =0.
\label{1.39h}
\end{eqnarray}
With the solution to (\ref{1.39h}) being of the form $g(p)=e^{-ik_0 p}$, where $k_0^2=k_1^2+k_2^2+k_3^2\pm H^2$, we see that the modes behave as massive modes, and are not massless at all. With this solution $D_{MIN}$ is given by 
\begin{eqnarray}
D^+_{MIN}(t,\vec{x};t^{\prime},\vec{x}^{\prime})&=&-\frac{1}{(2\pi)^4}\frac{1}{\Omega(p)\Omega(p^{\prime})}\int d^4k \frac{e^{ik\cdot (x-x^{\prime})}}{k_0^2-k_1^2-k_2^2-k_3^2- H^2+i\epsilon\theta(k^0)},
\nonumber\\
D^-_{MIN}(t,\vec{x};t^{\prime},\vec{x}^{\prime})&=&-\frac{1}{(2\pi)^4}\frac{1}{\Omega(p)\Omega(p^{\prime})}\int d^4k \frac{e^{ik\cdot (x-x^{\prime})}}{k_0^2-k_1^2-k_2^2-k_3^2+H^2+i\epsilon\theta(k^0)}.
\label{1.40h}
\end{eqnarray}
It acts like the propagator of a particle with mass $m^2=H^2$ or $m^2=-H^2$, and is thus not restricted to the light cone \cite{footnote1d}. Moreover, if $D_{MIN}$ is not restricted to the light cone in the special case in which the geometry is conformal to flat, $D_{MIN}$  will not be restricted to the light cone in the more general case in which the geometry is not even conformal to flat. Beyond enabling us to establish that massless field propagators are not automatically restricted to the light cone, since the  model can be solved analytically, we can also show that normals to advancing wavefronts in the model take support off the light cone too. And in addition, these normals follow the same trajectories as the eikonal approximation trajectories that we determine in this paper. This will be discussed further in Sec. \ref{S10} and in the Appendix.

To obtain a propagator that could be restricted to the light cone we instead consider a conformally coupled massless scalar field with action and wave equation 
\begin{eqnarray}
I_{S}=-\int d^4x(-g)^{1/2}\left[\frac{1}{2}\nabla_{\mu}S\nabla^{\mu}S-\frac{1}{12}S^2R^{\alpha}_{\phantom {\alpha} \alpha}\right],\quad \left[\nabla_{\mu}\nabla^{\mu}+\frac{1}{6}R^{\alpha}_{\phantom {\alpha} \alpha}\right]S=0,
\label{1.41h}
\end{eqnarray}
and a retarded conformally coupled propagator that obeys
\begin{eqnarray}
\left[\nabla_{\mu}\nabla^{\mu}+\frac{1}{6}R^{\alpha}_{\phantom {\alpha} \alpha}\right]D_{CON}(t,\vec{x};t^{\prime},\vec{x}^{\prime})=-(-g)^{-1/2}\delta^4(x-x^{\prime}).
\label{1.42h}
\end{eqnarray}
Both the action and the wave equation are left invariant under the local conformal transformation $g_{\mu\nu}(x)\rightarrow \Omega^2(x)g_{\mu\nu}(x)$,  $S(x)\rightarrow \Omega^{-1}(x)S(x)$. Now in the event that the scalar field is propagating in a geometry that is conformal to flat with some arbitrary conformal factor $\Omega(x)$, we can make a conformal transformation that will bring the geometry to flat and the wave equation  to the flat (\ref{1.28h}). Thus the conformally coupled $D_{CON}$ propagator will transform into the $D_{FLAT}$ propagator, and since a propagator is a bilinear function of the fields we obtain \cite{Birrell1982}
\begin{eqnarray}
D_{CON}(t,\vec{x};t^{\prime},\vec{x}^{\prime})=\frac{1}{\Omega(x)}D_{FLAT}(t,\vec{x};t^{\prime},\vec{x}^{\prime})\frac{1}{\Omega(x^{\prime})}.
\label{1.43h}
\end{eqnarray}
This result is of interest for two reasons. First, it shows that the conformally coupled propagator only takes support in the same domain as $D_{FLAT}$, i.e.,  on the light cone \cite{footnote1b}. Second, just as we had noted above, it shows that in generalizing the flat space action given in (\ref{1.28h}) we would miss Riemann-tensor-dependent terms if we were simply to covariantize (\ref{1.28h})   to the minimally coupled action given in (\ref{1.32h}). Rather,  we need to generalize (\ref{1.28h})  to the conformally coupled action given in (\ref{1.41h}), as (\ref{1.28h}) actually is conformal invariant as written in a flat background, and not just general coordinate invariant as written in a flat background. 

However, suppose that instead of propagating in a conformal to flat background the conformally coupled scalar field  is propagating in a geometry that is not conformal to flat, i.e., propagating in a geometry in which the conformal Weyl tensor is nonzero. To determine what can happen then it is instructive to examine the specific way in which the wave equation given in (\ref{1.41h}) is able to be conformal invariant. With the connection transforming as
\begin{eqnarray}
\Gamma^{\lambda}_{\mu\nu}\rightarrow \Gamma^{\lambda}_{\mu\nu}+\Omega^{-1}\left[\delta^{\lambda}_{\mu}\partial_{\nu}\Omega
+\delta^{\lambda}_{\nu}\partial_{\mu}\Omega-g_{\mu\nu}g^{\lambda\sigma}\partial_{\sigma}\Omega\right]
\label{1.44h}
\end{eqnarray}
under a local conformal transformation, we obtain
\begin{eqnarray}
\nabla_{\mu}\nabla^{\mu}S=g^{\mu\nu}\left[\partial_{\mu}\partial_{\nu} -\Gamma^{\lambda}_{\mu\nu}\partial_{\lambda}\right]S\rightarrow
\Omega^{-3}g^{\mu\nu}\left[\partial_{\mu}\partial_{\nu} -\Gamma^{\lambda}_{\mu\nu}\partial_{\lambda}\right]S
-\Omega^{-4}Sg^{\mu\nu}\left[\partial_{\mu}\partial_{\nu} -\Gamma^{\lambda}_{\mu\nu}\partial_{\lambda}\right]\Omega.
\label{1.45h}
\end{eqnarray}
At the same time the Ricci scalar transforms as (see e.g. \cite{Birrell1982})
\begin{eqnarray}
R^{\alpha}_{\phantom {\alpha} \alpha}\rightarrow \bar{R}^{\alpha}_{\phantom {\alpha} \alpha}=\Omega^{-2}R^{\alpha}_{\phantom {\alpha} \alpha}
+6\Omega^{-3}g^{\mu\nu}\left[\partial_{\mu}\partial_{\nu} -\Gamma^{\lambda}_{\mu\nu}\partial_{\lambda}\right]\Omega,
\label{1.46h}
\end{eqnarray}
with (\ref{1.46h}) serving to define  $\bar{R}^{\alpha}_{\phantom {\alpha} \alpha}$. 
Consequently, we find that 
\begin{eqnarray}
\left[\nabla_{\mu}\nabla^{\mu}+\frac{1}{6}R^{\alpha}_{\phantom {\alpha} \alpha}\right]S
\rightarrow\Omega^{-3}\left[\nabla_{\mu}\nabla^{\mu}+\frac{1}{6}R^{\alpha}_{\phantom {\alpha} \alpha}\right]S,
\label{1.47h}
\end{eqnarray}
to thus transform  conformally. If we define $\bar{g}_{\mu\nu}(x)=\Omega^2(x)g_{\mu\nu}(x)$,  $\bar{S}(x)=\Omega^{-1}(x)S(x)$, then from (\ref{1.47h}) we obtain 
\begin{eqnarray}
\left[\nabla_{\mu}\nabla^{\mu}+\frac{1}{6}R^{\alpha}_{\phantom {\alpha} \alpha}\right]S
=\left[\bar{\nabla}_{\mu}\bar{\nabla}^{\mu}+\frac{1}{6}\bar{R}^{\alpha}_{\phantom {\alpha} \alpha}\right]\bar{S}=0.
\label{1.48h}
\end{eqnarray}

If we now consider a conformal invariant wave equation in a background with some general metric $g_{\mu\nu}(x)$ that is not conformal to flat, then on noting that on its own the Ricci scalar is not conformal invariant, we can make a very specific conformal transformation on $R^{\alpha}_{\phantom {\alpha} \alpha}$ chosen so as to make its conformal transform $\bar{R}^{\alpha}_{\phantom {\alpha} \alpha}$ vanish. Thus we chose $\Omega(x)$ to obey
\begin{eqnarray}  
R^{\alpha}_{\phantom {\alpha} \alpha}
+6\Omega^{-1}g^{\mu\nu}\left[\partial_{\mu}\partial_{\nu} -\Gamma^{\lambda}_{\mu\nu}\partial_{\lambda}\right]\Omega=0.
\label{1.49h}
\end{eqnarray}
In making this choice for $\Omega(x)$  we transform to a metric $\bar{g}_{\mu\nu}(x)=\Omega^2(x)g_{\mu\nu}(x)$ that is expressly still not conformal to flat, with the scalar field transforming to the scalar field $\bar{S}(x)=\Omega^{-1}(x)S(x)$, and the wave equation becoming
\begin{eqnarray}
\bar{\nabla}_{\mu}\bar{\nabla}^{\mu}\bar{S}=0,
\label{1.50h}
\end{eqnarray}
i.e., of the minimally coupled form. However, above we showed that $D_{MIN}$ takes support off the light cone. Thus so must the  $D_{CON}$ propagator associated with the conformally coupled wave equation from which we obtained (\ref{1.50h}). Thus the general rule is that $D_{MIN}$ takes support off the light cone in any background geometry, while the domain of $D_{CON}$ is only restricted to the light cone if the Weyl tensor of the associated background geometry is zero \cite{footnote1c}.

To see the transition from off the light cone to on the light cone in more detail, we introduce a non-minimally coupled scalar field that interpolates between the minimal ($\xi=0$) and conformally coupled ($\xi=1$) cases with action and wave equation
\begin{eqnarray}
I_{S}=-\int d^4x(-g)^{1/2}\left[\frac{1}{2}\nabla_{\mu}S\nabla^{\mu}S-\frac{\xi}{12}S^2R^{\alpha}_{\phantom {\alpha} \alpha}\right],\quad \left[\nabla_{\mu}\nabla^{\mu}+\frac{\xi}{6}R^{\alpha}_{\phantom {\alpha} \alpha}\right]S=0.
\label{1.51h}
\end{eqnarray}
For the conformal to flat example given in (\ref{1.35h}) the solution is still given by (\ref{1.37h}). And with the Ricci scalar being given by $R^{\alpha}_{\phantom {\alpha} \alpha}=-6\Omega^{-3}\ddot{\Omega}$, the equation for $g(p)$ is given by 
\begin{eqnarray}
\Omega\ddot{g}-\ddot{\Omega}(1-\xi)g+|\vec{k}|^2\Omega g =0.
\label{1.52h}
\end{eqnarray}
As we see,  even though the metric is conformal to flat, $g(p)$ is not in general given by $e^{-i|\vec{k}|p}$, and it only takes that form if $\xi=1$. Thus only for $\xi=1$ would the domain of support of the propagator be restricted to the light cone if the background geometry is conformal to flat. Though even so, as we have shown above, even if $\xi=1$ the domain would still not be restricted to the light cone if the geometry is not conformal to flat. 

To summarize, from the example given in (\ref{1.52h}) we see that given the fact that the minimally coupled and conformally coupled wave equations presented in (\ref{1.32h}) and (\ref{1.41h}) are different, they could not both be restricted to propagation on the light cone. So at best not more than one could be, and even though it is the conformally coupled one, it too is only restricted to the light cone if it propagates in a geometry that is conformal to flat.

These remarks are not restricted to scalar fields, since as we had shown in (\ref{1.19g}) and as we will discuss below,  conformally invariant wave equations involving the Maxwell field and conformally invariant wave equations involving fermion fields also involve components of the Riemann tensor, the Ricci tensor for Maxwell and the Ricci scalar for fermions. Now neither of these particular components of the Riemann tensor is conformal invariant. The Ricci scalar transforms as given above, while the Ricci tensor transforms as \cite{Birrell1982} 
\begin{eqnarray}
R_{\mu\nu}\rightarrow \Omega^{-2}R_{\mu\nu}+2\Omega^{-3}\nabla_{\mu}\nabla_{\nu}\Omega-4\Omega^{-4}\nabla_{\mu}\Omega\nabla_{\nu}\Omega
+g_{\mu\nu}[\Omega^{-3}\nabla_{\alpha}\nabla^{\alpha}\Omega+\Omega^{-4}\nabla_{\alpha}\Omega\nabla^{\alpha}\Omega].
\label{1.53h}
\end{eqnarray}
(Here the covariant derivatives are taken with respect to the untransformed metric $g_{\mu\nu}$ and not with respect to the transformed $\bar{g}_{\mu\nu}$.) Consequently, in geometries that are not conformal to flat  propagation of Maxwell and fermion fields will not be restricted to the light cone. These remarks equally apply to gravitational waves.

That propagators associated with massless particles are not restricted to the light cone had previously been noted in \cite{DeWitt1960,Hu1998,Mannheim2007,Copi2021}. In \cite{DeWitt1960} minimally coupled scalar field propagators and conformally coupled Maxwell vector field propagators were studied, and both were found to possess both $\delta(-g_{\mu\nu}x^{\mu}x^{\nu})$ and $\theta(-g_{\mu\nu}x^{\mu}x^{\nu})$ type terms (viz. lightlike and timelike). Since the $\theta(-g_{\mu\nu}x^{\mu}x^{\nu})$ term trails the $\delta(-g_{\mu\nu}x^{\mu}x^{\nu})$ term it was referred to as a ``tail"; and in the Maxwell case it was noted that such a tail would be affected by the presence of the Ricci tensor term in the Maxwell equations given in (\ref{1.19g}). (These tails have also been studied in \cite{Copi2021}.) With the Maxwell propagator taking support off the light cone, as noted in Sec. \ref{S1a} above, advancing wavefronts and the normals to them would propagate off the light cone too. In \cite{Hu1998} non conformal to flat stochastic Maxwell field fluctuations around Robertson-Walker and Schwarzschild background geometries were studied, and consistent with the general analysis we have presented here, Maxwell field propagators were found that took support off the light cone. In obtaining this result the authors of \cite{Hu1998} were able to covariantize the flat space statement that in a medium a photon travels with velocity $c/n$ by showing that gravity provides an appropriate refractive index.

Similar propagator structures have also been observed in a conformally coupled scalar field theory in two spacetime dimensions, and in conformal gravity (a conformal invariant, Weyl tensor based theory) in four spacetime dimensions. In two dimensions a free flat space conformal field  obeys
\begin{eqnarray}
\left(-\partial_0^2+\partial_x^2\right)S=0,
\label{1.54h}
\end{eqnarray}
with a propagator that obeys \cite{Mannheim2007}
\begin{eqnarray}
\left(-\partial_0^2+\partial_x^2\right)\frac{1}{2}\theta(t-|x|)=-\delta(t)\delta(x).
\label{1.55h}
\end{eqnarray}
The $\theta(t-|x|)$ propagator takes support inside the light cone.
Now in two dimensions the scalar field is dimensionless and transforms into itself under a conformal transformation, and so now it is the minimally coupled scalar field wave equation that is conformal invariant since 
\begin{eqnarray}
\nabla_{\mu}\nabla^{\mu}S=g^{-1/2}\partial_{\mu}[g^{1/2}g^{\mu\nu}\partial_{\nu}S]\rightarrow g^{-1/2}\Omega^{-2}\partial_{\mu}[\Omega^2g^{1/2}\Omega^{-2}g^{\mu\nu}\partial_{\nu}S]=\Omega^{-2}\nabla_{\mu}\nabla^{\mu}S.
\label{1.56h}
\end{eqnarray}
Thus in conformal to flat geometries we can transform to (\ref{1.54h}) and obtain the $(1/2)\theta(t-|x|)$ propagator. Such propagators are perfectly acceptable in field theory since causality only forbids propagation outside the light cone, but does not forbid propagation inside the light cone. (Covariance alone would allow a solution to (\ref{1.55h}) of the form $-(1/2)\theta(|x|-t)$, a solution that does take support outside the light cone.)

For conformal gravity in four dimensions a linearization around flat space time leads to a $(\nabla_{\mu}\nabla^{\mu})^2$ differential equation and a propagator that obeys \cite{Mannheim2007}
\begin{eqnarray}
\left(-\partial_0^2+\vec{\nabla}^2\right)^2\frac{1}{8\pi}\theta(t-|\vec{x}|)=\delta(t)\delta^3(\vec{x}).
\label{1.57h}
\end{eqnarray}
Being based on the Weyl tensor the wave equation is also conformal invariant, and again we obtain a theta function dependence rather than a delta function dependence for the propagator \cite{footnote1f}.

The reason why we obtain the theta function type propagators in these cases can be traced to dimensional analysis \cite{Mannheim2007}. For (\ref{1.55h})  $(-\partial_0^2+\partial_x^2)$ and $\delta(t)\delta(x)$ both have the same dimension, while for (\ref{1.57h}) 
$(-\partial_0^2+\vec{\nabla}^2)^2$ and $\delta(t)\delta^3(\vec{x})$ both have the same dimension. Thus in both the cases the propagator must be dimensionless, and a delta function is not. Also the propagator must be a Lorentz scalar and must be causal, and thus has to be given by a theta function that depends on $-\eta_{\mu\nu}x^{\mu}x^{\nu}$. In contrast, for the standard (\ref{1.29h})  $(-\partial_0^2+\vec{\nabla}^2)$ and $\delta^4(x-x^{\prime})$ differ by two units of dimension, and thus the associated $D_{FLAT}$ behaves as $\delta(t^2-|\vec{x}|^2)$. Now  $\delta(t^2-|\vec{x}|^2)$ is conformal invariant since if $-\eta_{\mu\nu}x^{\mu}x^{\nu}=0$, then $-\Omega^2\eta_{\mu\nu}x^{\mu}x^{\nu}=0$. However, the theta function propagator is also conformal invariant since if $-\eta_{\mu\nu}x^{\mu}x^{\nu}>0$ then with $\Omega^2$ being positive, $-\Omega^2\eta_{\mu\nu}x^{\mu}x^{\nu}>0$. As we thus see, in general conformal invariance does not force propagation to be restricted to the light cone, and not even in flat or conformal to flat backgrounds (unless dimensional analysis restricts us to a delta function type term).

While it was noted in \cite{DeWitt1960,Hu1998,Mannheim2007,Copi2021} that massless propagation can go off the light cone, there was no discussion as to the effect of the structure of the background geometry on the propagators. Thus in none of these papers was it shown that in four spacetime dimensions there is no tail if the wave equation is conformally invariant and the geometry is conformal to flat, and that there only is a tail for a conformally invariant wave equation if the background geometry is not conformal to flat.  Since the minimally coupled scalar field study of \cite{DeWitt1960} the Maxwell field study of \cite{Hu1998}  and the Maxwell tail study of \cite{Copi2021} were explicitly applied either to wave equations that were not conformal invariant or to geometries that were not conformal to flat, it was indeed found in all of these cases that propagation went off the light cone, to thus be special cases of the general analysis presented here.

To summarize, we note that we have shown that in the curved space case we should expect the presence of curvature-dependent terms in wave equations even though none are obtained by covariantizing flat space expressions. Also we have shown that we should expect propagation off the light cone. Thus in general we can anticipate that the wisdom obtained from flat space geodesics does not automatically carry over to curved space. Unlike the analysis that we present in this paper, in none of \cite{DeWitt1960,Hu1998,Mannheim2007,Copi2021} was there discussion of the explicit way  in which the presence of a Ricci tensor term or a Ricci scalar term in the massless particle equations of motion would modify null geodesics, or any determination of by exactly how much the geodesics would then be modified. Both of our key results below,  viz. the Ricci-scalar-dependent (\ref{3.19qq}) for conformally coupled scalar fields and the Ricci-tensor-dependent  (\ref{4.13x}) for conformally coupled vector fields (essentially (\ref{1.18g}) and (\ref{1.20g}) above, as then generalized without any short wavelength approximation in Sec. \ref{S10}) address this issue, and hold whenever the geometry is not conformal to flat. They are both new to the literature.

In order to investigate how the presence of a Ricci tensor or Ricci scalar term might be expected to affect eikonalization, we note that the general arbitrary $\xi$ scalar field wave equation given in (\ref{1.51h}) can be written as 
\begin{eqnarray}
g^{\mu\nu}\left[\partial_{\mu}\partial_{\nu} -\Gamma^{\lambda}_{\mu\nu}\partial_{\lambda}\right]S+\frac{\xi}{6}R^{\alpha}_{\phantom {\alpha} \alpha}S=0.
\label{1.58h}
\end{eqnarray}
In the limit in which all geometrical factors are negligible (\ref{1.58h})  reduces to
\begin{eqnarray}
g^{\mu\nu}\partial_{\mu}\partial_{\nu}S=0,
\label{1.59h}
\end{eqnarray}
with the wave now being restricted to the light cone. This for instance occurs in the conformal to flat metric case if we set $|\vec{k}|^2\gg \Omega^{-1}(1-\xi)\ddot{\Omega}$ (viz. short wavelength)  in (\ref{1.52h}). However, since the $-g^{\mu\nu}\Gamma^{\lambda}_{\mu\nu}\partial_{\lambda}S$ term in (\ref{1.58h}) would have to be of the same order of magnitude as the $g^{\mu\nu}\partial_{\mu}\partial_{\nu}S$ term if it is to be retained in (\ref{1.58h}), then since the Ricci scalar is built out of derivatives and bilinear products of $\Gamma^{\lambda}_{\mu\nu}$, when $\Gamma^{\lambda}_{\mu\nu}$ is big enough the  $(\xi/6)R^{\alpha}_{\phantom {\alpha} \alpha}S$ term would then be just as important as the $g^{\mu\nu}\partial_{\mu}\partial_{\nu}S$ term. Thus, if we wish to keep the $-g^{\mu\nu}\Gamma^{\lambda}_{\mu\nu}\partial_{\lambda}S$ term we would need to keep the $(\xi/6)R^{\alpha}_{\phantom {\alpha} \alpha}S$ term as well.

Similar arguments apply to trajectories such as the scalar field (\ref{3.19qq}) given below, viz. 
\begin{eqnarray}
\frac{d^2x^{\lambda} }{ dq^2}
+\Gamma^{\lambda}_{\mu \nu} 
\frac{dx^{\mu}}{ dq}\frac{dx^{\nu } }{dq}=
-\frac{1}{2(-R^{\alpha}_{\phantom{\alpha}\alpha}/6)}
\left[g^{\lambda \mu}+\frac{dx^{\lambda}}{dq}\frac{dx^{\mu}}{dq}\right]\frac{\partial (-R^{\alpha}_{\phantom{\alpha}\alpha}/6)}{\partial x^{\mu}},
\label{1.60h}
\end{eqnarray}
an expression that holds if the scalar field wave equation is conformal invariant but the geometry in which it propagates is not conformal to flat. And again, if  $\Gamma^{\lambda}_{\mu\nu}$ is big then so is $R^{\alpha}_{\phantom{\alpha}\alpha}$. Thus our ability to drop the Ricci scalar term requires that $\Gamma^{\lambda}_{\mu\nu}$ be small, and thus we can only come back to null geodesics if  gravity is weak.
In this paper we investigate what happens when the curvature is such that this is not the case. Finally, we note that the right-hand side of (\ref{1.60h}) would not vanish if we let the Ricci scalar go to zero. And yet if we first let the Ricci scalar go to zero in (\ref{1.58h}) and then eikonalize we would be back to trajectories that do not involve the Ricci scalar at all. In Sec. \ref{S3i} we show how to reconcile these two limits.

As constructed, our key (\ref{1.60h}) and its (\ref{4.13x}) vector analog are essentially unique. No matter what trajectories light rays might travel on they travel on trajectories that are one-dimensional. And thus with the left-hand side of (\ref{1.60h}) being parametrized by the one-dimensional affine parameter $q$, the left-hand side of (\ref{1.60h}) must be equal to some function on the right-hand side that is also parameterized by $q$. As noted by us and by \cite{DeWitt1960} this modification must depend on the Ricci tensor or Ricci scalar as they are what appear in the wave equations. By covariance the right-hand side must be a general coordinate vector, and as we show in (\ref{3.25qq}) below the right-hand side must be orthogonal to the velocity four-vector since the covariant four-acceleration on the left-hand side is. The form that appears on the right-hand side of (\ref{1.60h}) meets all of these criteria, and would thus appear to be unique in its structure. 

While short wavelength eikonal approximations can often suffice, we note that we can actually generalize our results so that no short wavelength approximation need be made at all. And thus without approximation  we generalize (\ref{1.60h})  in Secs. \ref{S4} and  \ref{S10}. And again find a similar structure (see (\ref{4.21xy}) and (\ref{10.2x})), viz.
\begin{eqnarray}
\frac{d^2x^{\lambda} }{ dq^2}+\Gamma^{\lambda}_{\mu \nu} 
\frac{dx^{\mu}}{ dq}\frac{dx^{\nu } }{dq}
=-\frac{1}{2X}
\left[g^{\lambda \mu}+\frac{dx^{\lambda}}{dq}\frac{dx^{\mu}}{dq}\right]\frac{\partial X}{\partial x^{\mu}},
\label{1.61h}
\end{eqnarray}
where $X$ is the value of the relevant $-\nabla_{\nu}T\nabla^{\nu}T$, and the eikonal function $T$  is defined as $S=Ae^{iT}$ for scalar fields and $A_{\mu}=A\epsilon_{\mu (a)}e^{iT}$ for vector fields. Specifically, as we show below, $X$ is given by
\begin{eqnarray}
X=-A^{-1}\nabla_{\mu}\nabla^{\mu}A
\label{1.62h}
\end{eqnarray}
for a  scalar field that is minimally coupled, by
\begin{eqnarray}
X=-A^{-1}\nabla_{\mu}\nabla^{\mu}A-\frac{1}{6}R^{\alpha}_{\phantom{\alpha}\alpha}
\label{1.63h}
\end{eqnarray}
for a scalar field that is conformally coupled, and by
\begin{eqnarray}
X=-A^{-1}\nabla_{\mu}\nabla^{\mu}A-\frac{1}{3}\sum_{a=1}^3\epsilon_{\nu(a)}\nabla_{\mu}\nabla^{\mu}\epsilon^{\nu}_{(a)}
-\frac{1}{3} \sum_{a=1}^3\epsilon_{\nu (a)}\epsilon^{\alpha}_{(a)}R^{\nu}_{\phantom{\nu}\alpha}
\label{1.64h}
\end{eqnarray}
for a conformally coupled Maxwell vector field.

In Sec. \ref{S10} we connect our approach to the general theory of geometric embeddings. In this section we additionally show that the normals to advancing wavefronts can not only propagate off the light cone, they follow the same trajectories as the exact  (\ref{1.61h}) eikonal  trajectories that we establish in this paper.  The general prescription that we identify is that regardless of the explicit equation obeyed by $\nabla_{\nu}T\nabla^{\nu}T$,  we can set 
\begin{eqnarray}
n^{\mu}=\frac{dx^{\mu}}{dq}=\frac{\nabla^{\mu}T}{(- \nabla_{\nu}T\nabla^{\nu}T)^{1/2}},\qquad
\frac{dx_{\mu}}{dq}\frac{dx^{\mu}}{dq}=-1
\label{1.65h}
\end{eqnarray}
for timelike normals $n^{\mu}$ ($n_{\mu}n^{\mu}=-1$ in the notation of \cite{Weinberg1972} where $g_{00}$ is negative), with $dx^{\mu}/dq$ being the eikonal velocity.
Since one can consider normals to wavefronts of any wavelength the general expression given in (\ref{1.61h}) holds for any normal and not just for those associated with short wavelength rays.
At any wavelength advancing wavefronts transport energy and momentum along these eikonalized trajectories, doing so with velocity $dx^{\mu}/dq$. To provide further support for our approach, in the Appendix we present an explicitly solvable model in which this is precisely the case.

Also we should recall that, as noted above, for conformally coupled fields the wavefronts can only take support off the light cone if the background geometry is not conformal to flat. If the background is conformal to flat then $dx^{\mu}/dq=\nabla^{\mu}T$, $(dx_{\mu}/dq)(dx^{\mu}/dq)=0$ and the trajectories are the null geodesics
\begin{eqnarray}
\frac{d^2x^{\lambda} }{ dq^2}+\Gamma^{\lambda}_{\mu \nu} 
\frac{dx^{\mu}}{ dq}\frac{dx^{\nu } }{dq}
=0.
\label{1.66h}
\end{eqnarray}

For organizational  purposes we note that in Sec. \ref{S2b} we show  that relativistic eikonalization  must be formulated in light-front coordinates rather than in instant-time coordinates. In Sec. \ref{S3}  we discuss the implications of conformal symmetry for the eikonalization procedure in detail. In Sec. \ref{S3f} we show  that due to its very nature, the eikonal approximation reduces any underlying local conformal symmetry that might be present to a global symmetry alone.  And in Sec. \ref{S3h} we discuss the implications of this result for conformal invariant test particle actions. With the covariant gauge condition $\nabla_{\mu}A^{\mu}=0$ not being locally conformal invariant, in  Sec. \ref{S4} we construct a gauge condition for the vector potential [viz. (\ref{4.15xx})] that is locally conformal invariant. Under local conformal transformations this gauge condition is preserved. In Secs. \ref{S5} to \ref{S9} we provide some phenomenological applications of the approach being developed here, with particular emphasis on the non conformal to flat Schwarzschild geometry and the non conformal to flat cosmological perturbation theory. In Sec. \ref{S10} we present a formalism that involves no short wavelength approximation at all, and provide some applications of this approach in the Appendix.

\section{The Curved Spacetime Eikonal Approximation}
\label{S2}

\subsection{Massless Scalar Fields}
\label{S2a}

If one starts with wave equations one then has both geometrical optics and physical optics, i.e., both ray propagation and diffraction. The ray limit is a short wavelength limit (short on the scale of any system being considered), while the diffraction limit is associated with larger wavelengths that are of the same size as the system under consideration. Now long before the development of general relativity it was known that short wavelength rays travel in straight lines, and as such the use of null geodesics of the form given in (\ref{1.6x}) is just the natural generalization to curved space. However, in curved space one meets a new scale, namely the scale of the curvature rather than the scale of a material system that is being explored. In \cite{Mannheim2010} we explored how such curvature scales can if strong enough cause fluids that are perfect in the absence of gravity to become imperfect in its presence. In this paper we make an analogous analysis for individual light rays. In regard to the covariantization of flat space expressions such as the transition from the flat space $d^2x^{\lambda}/dq^2=0$ to (\ref{1.18g}), we note that the replacing of ordinary derivatives by covariant derivatives is not a sufficient prescription since it ignores the possible presence of intrinsically geometric (i.e., Riemann tensor dependent) terms that have no flat space counterpart. Thus one can go from the curved space (\ref{1.18g}) to the flat space $d^2x^{\lambda}/dq^2=0$, but not vice versa. Thus in studying massless rays in curved space we can anticipate the presence of explicit curvature dependent terms. While for practical purposes such  terms might only be of relevance in regions of high curvature, the point of this paper is only to establish their presence in principle. In regions of low curvature our work in this paper shows how to establish geodesic behavior for both massless and massive particles starting from wave equations without ever encountering any test particle action at all. Some of our results have already been prefigured in \cite{Mannheim2006} and they lead to the more general analysis presented here. The key new ingredient presented here is in the normalization of the eikonal function. 

In developing a relativistic eikonal approximation one starts with a field such as a scalar field $S(x)$ that obeys a minimally coupled, covariant wave equation 
\begin{eqnarray}
\nabla_{\mu}\nabla^{\mu}S=0.
\label{2.1y}
\end{eqnarray}
Following the discussion in \cite{Mannheim2006}, we introduce an eikonal function $T(x)$ via $S(x)=A(x)e^{iT(x)}$ with amplitude $A(x)$. While in general $A$ and $T$ could be complex, modeled on plane waves that behave as $e^{ik\cdot x}$, in the eikonal approximation the phase $T$ is taken to be real. And if we write $A=Be^{iC}$ where $B$ and $C$ are real, then $C$ could be absorbed in $T$. Thus in writing $S=A\exp(iT)$ it is understood that both $T$ and $A$ are real. Insertion of  $S(x)=A(x)e^{iT(x)}$ into (\ref{2.1y})  yields 
\begin{eqnarray}
-A\nabla_{\mu}T\nabla^{\mu}T+\nabla_{\mu}\nabla^{\mu}A+2i\nabla_{\mu}A\nabla^{\mu}T+iA\nabla_{\mu}\nabla^{\mu}T=0.
\label{2.2y}
\end{eqnarray}
With both $A$ and $T$ being real we obtain
\begin{eqnarray}
-A\nabla_{\mu}T\nabla^{\mu}T+\nabla_{\mu}\nabla^{\mu}A=0,
\label{2.3y}
\end{eqnarray}
\begin{eqnarray}
2i\nabla_{\mu}A\nabla^{\mu}T+iA\nabla_{\mu}\nabla^{\mu}T=0.
\label{2.4y}
\end{eqnarray}
In the language of the WKB approximation (\ref{2.3y}) is the analog of the Hamilton-Jacobi equation when $A$ is slowly varying, while (\ref{2.4y}) is the analog of the continuity equation. In the eikonal approximation $A$ is usually taken to be slowly varying  (cf. the constant $\boldsymbol{E}_0$ and $\boldsymbol{B}_0$ in (\ref{1.12g})), while $T$ is taken to be so large that the most important term in (\ref{2.2y}) is the one that is quadratic in $T$ \cite{footnote2}. While we will generally take $A$ to be slowly varying, we will consider some cases (such as in (\ref{3.7z}) below or in Sec. \ref{S10} or in the Appendix) where it expressly is not. With $A$ being slowly varying  or more generally with $\nabla_{\mu}T\nabla^{\mu}T \gg A^{-1}\nabla_{\mu}\nabla^{\mu}A$ (\ref{2.3y}) yields 
\begin{eqnarray}
\nabla_{\mu}T\nabla^{\mu}T=0.
\label{2.5y}
\end{eqnarray}
Applying $\nabla_{\nu}$ to (\ref{2.5y}) and recalling that  since $T$ is a scalar we can set $\nabla_{\mu}\nabla_{\nu}T=\nabla_{\nu}\nabla_{\mu}T$, we obtain 
\begin{eqnarray}
\nabla^{\mu}T\nabla_{\mu}\nabla_{\nu}T=0.
\label{2.6y}
\end{eqnarray}
Now in the geometrical optics regime rays travel in the direction normal to the wavefront and obey the eikonal velocity relation
\begin{eqnarray}
\nabla^{\mu}T=\frac{dx^{\mu}}{dq}=k^{\mu},
\label{2.7y}
\end{eqnarray}
where $q$ is an affine parameter that measures distance along the normal. The vector $k^{\mu}$ is the wave vector of the ray, and according to (\ref{2.5y}) it has to obey $k_{\mu}k^{\mu}=0$ and lie on the momentum space light cone. Similarly, $dx^{\mu}/dq$ has to obey $(dx^{\mu}/dq)(dx_{\mu}/dq)=0$ and lie on the coordinate space light cone. Technically, we should note that given the conventional three-dimensional normalization $d\bar{x}/dq=\bar{\nabla}T/(\bar{\nabla}T\cdot \bar{\nabla}T)^{1/2}$ of the eikonal phase that is used in optics \cite{Born1959}, in the covariant case we should set
\begin{eqnarray}
\frac{dx^{\mu}}{dq}=\frac{\nabla^{\mu}T}{(-\nabla_{\nu}T\nabla^{\nu}T)^{1/2}},\quad \frac{dx_{\mu}}{dq}\frac{dx^{\mu}}{dq}=-1,
\label{2.8y}
\end{eqnarray}
a choice that would then allow $q$ to have the dimension of length.  Below we will meet cases where $\nabla_{\nu}T\nabla^{\nu}T$ is not zero (as would for instance be the case if we were not to drop the $\nabla_{\mu}\nabla^{\mu}A$ term in (\ref{2.3y})), and in those cases we will use the normalization given in (\ref{2.8y}). As we will see, this has  substantive consequences  as it takes $dx^{\mu}/dq$ off the coordinate space light cone, and as we show in Sec. \ref{S3}, would even lead to inconsistencies if not taken into account. Thus on the light cone we use (\ref{2.7y}) and off the light cone we use (\ref{2.8y}).

Now on its own an arbitrary general coordinate vector  $dx^{\mu}/dq$ would have both longitudinal and transverse components \cite{footnote1a}. However, $\nabla^{\mu}T$ is longitudinal, and thus $dx^{\mu}/dq$ has to be longitudinal too. With $dx^{\mu}/dq$ being directed along the normal to the wavefront, we see that the longitudinal component of the eikonal wave is along the normal  (i.e., in the direction of the ray), with the transverse components then being in the wavefront. Since $x^{\mu}$ varies with $q$ along the normal we can therefore use 
\begin{eqnarray}
\frac{d}{dq}=\frac{dx^{\mu}}{dq}\frac{\partial}{\partial x^{\mu}}
\label{2.9y}
\end{eqnarray}
to determine derivatives.  

Thus for the situation in which we do not need to normalize (viz. on the light cone) we can use (\ref{2.6y}) and (\ref{2.7y}), and via (\ref{2.9y}) and (\ref{2.5y}) we obtain
\begin{eqnarray}
\frac{dx^{\mu}}{ dq}\nabla_{\mu}\frac{dx^{\lambda}}{ dq}=\frac{d^2x^{\lambda} }{ dq^2}
+\Gamma^{\lambda}_{\mu \nu} 
\frac{dx^{\mu}}{ dq}\frac{dx^{\nu } }{ dq}=0,\quad \frac{dx_{\mu}}{dq}\frac{dx^{\mu}}{dq}=0.
\label{2.10y}
\end{eqnarray}
We recognize (\ref{2.10y}) as being the massless particle null geodesic equation, with rays then  precisely being found to be geodesic in the large $T$  limit. Thus even though the propagator associated with the $\nabla_{\mu}\nabla^{\mu}$ operator takes support off the light cone, short enough wavelength rays are constrained to be on the light cone.

As such, not only have we derived the massless particle null geodesic equation without needing to make reference to any test particle action, in deriving it we also derive an expression that obeys the equivalence principle. Thus the equivalence principle is seen to follow from the general coordinate invariance of the field wave equation and does not need to be independently postulated. Now historically the equivalence principle was an important guiding principle in Einstein's original development of general relativity. However, once one has general relativity and general coordinate invariance, then on starting with covariant wave equations the equivalence principle is output rather than input to general relativity.

If we momentarily revert back to flat spacetime where $\eta^{\mu\nu}\partial_{\mu}\partial_{\nu}S=0$, solutions to this wave equation are of the form $S(x)=e^{ik\cdot x}$, where $k^{\mu}$ is independent of the spacetime coordinates and obeys $k_{\mu}k^{\mu}=0$. In this case we can set $S(x)=Ae^{iT(x)}$ where $A$ is constant and $T=k\cdot x$, with (\ref{2.3y}) and (\ref{2.4y}) reducing to 
\begin{eqnarray}
-A\nabla_{\mu}T\nabla^{\mu}T=0,\quad iA\nabla_{\mu}\nabla^{\mu}T=0,
\label{2.11y}
\end{eqnarray}
as immediately satisfied by 
\begin{eqnarray}
k_{\mu}k^{\mu}=0,\quad \nabla_{\mu}k^{\mu}=0.
\label{2.12y}
\end{eqnarray}
With $k^{\mu}$ being constant the identification $k^{\mu}=dx^{\mu}/dq$ leads to $dk^{\mu}/dq=d^2x^{\mu}/dq^2=0$, the straight line flat space null geodesic equation. We thus can interpret the null geodesic equation given in (\ref{2.10y}) as the curved space generalization of $d^2x^{\mu}/dq^2=0$. We also see that this interpretation requires that the  $\nabla_{\mu}\nabla^{\mu}A$ term in (\ref{2.3y}) be negligible. However, this interpretation also requires that we will get departures from the null geodesic equation when it is not, something that we will see actually is the case in some specific wave equations that we discuss below. Now we had noted in Sec. \ref{S1e} that there are situations in which propagators take support off the light cone. In such cases such  $\nabla_{\mu}\nabla^{\mu}A$ type terms or whatever it is that causes $\nabla_{\nu}T\nabla^{\nu}T$ to be nonzero (such as a dependence on the Ricci scalar or tensor) cannot be ignored.

\subsection{Relativistic Eikonalization and the Light-Front Approach}
\label{S2b}

We  should note that with $\nabla^{\mu}T=dx^{\mu}/dq=k^{\mu}$, in general the curved space $k^{\mu}$ will be spacetime dependent. Thus it is immediately suggested to set $T=\int^x k_{\mu}dx^{\mu}$ (we could only set $T=k_{\mu}x^{\mu}$ if $k_{\mu}$ is constant, which it could only be in flat space). However, if we do set $T=\int^x k_{\mu}dx^{\mu}$ we would obtain $T=\int^x(dx_{\mu}/dq)dx^{\mu}=\int^x(dx_{\mu}/dq)(dx^{\mu}/dq)dq=\int^xk_{\mu}k^{\mu}dq$, and with $k_{\mu}k^{\mu}=0$, $(dx_{\mu}/dq)(dx^{\mu}/dq)=0$, such a $T$ would vanish identically. To avoid this we note that if we consider a ray propagating in the $z$ direction so that $k^{\mu}=(k,0,0,k)$, and then momentarily do set $T=\int^x k_{\mu}dx^{\mu}$, we would then obtain $\partial_0T=k$, $\partial_3T=-k$, and thus obtain $(\partial_0+\partial_3)T=0$. Now this derivative is in a light-front direction. In light-front quantization (see e.g. \cite{Mannheim2021} and references therein) one introduces metric components, coordinates and derivatives of the form \cite{footnote3}
\begin{eqnarray}
g_{\mu\nu}({\rm front})=
\begin{pmatrix}
0&  \tfrac{1}{2} & 0&0 \cr 
\tfrac{1}{2}& 0&0&0\cr
0&0&-1&0\cr
0&0&0&-1
\end{pmatrix},
\label{2.13y}
\end{eqnarray}
\begin{eqnarray}
x^+=x^0+x^3, \qquad x^-=x^0-x^3, \qquad \partial_+=\frac{1}{2}(\partial_0+\partial_3), \qquad \partial_-=\frac{1}{2}(\partial_0-\partial_3), 
\label{2.14y}
\end{eqnarray}
so that
\begin{eqnarray}
k_{\mu}k^{\mu}=4k_+k_--k_1^2-k_2^2,\qquad x_{\mu}x^{\mu}= g_{\mu\nu}x^{\mu}x^{\nu}=x^+x^--(x^1)^2-(x^2)^2.
\label{2.15y}
\end{eqnarray}
The constraint on $T$ is thus of the form $\partial_+T=0$. This constraint can be satisfied by not actually setting $T=\int^x k_{\mu}dx^{\mu}$ after all, but by instead setting $T$ equal to the non-vanishing $T=\int^x k_{-}dx^{-}$. Then with $k_+=0$, $k_1=0$, $k_2=0$ one still has $k_{\mu}k^{\mu}=0$  even as $T$ is then nonzero (the vanishing of $k_+$, $k_1$ and $k_2$ does not restrict $k_-$ while still keeping $k_{\mu}k^{\mu}=4k_+k_--k_1^2-k_2^2$ zero). Thus while non-relativistic eikonalization occurs with the normal to the wavefront  being in the $x^3$ direction so that a non-vanishing eikonal phase $T$  is given by $T=\int^x k_3dx^3$, $\partial_3T=k_3=dx_3/dq$, in relativistic eikonalization the normal is in the longitudinal $x^-$ direction, with a non-vanishing eikonal phase $T$ being given by $T=\int^xk_-dx^-$, $\partial_+T=0$, $\partial_-T=k_-=dx_-/dq$.

In the light-front approach, and thus in relativistic eikonalization, one considers propagation in $x^+$, while in the instant-time approach one considers propagation in $x^0$. The two approaches thus appear to be different. However, with $g_{\mu\nu}({\rm front})x^{\mu}x^{\nu}=x^+x^--(x^1)^2-(x^2)^2$, we see that for either timelike or lightlike events $x^+x^-$ is positive, so $x^+$ and $x^-$ have the same sign. Now $x^0=(x^++x^-)/2$. Thus if $x^0$ is positive then for timelike or lightlike events $x^+$ is positive too. Hence for timelike or lightlike events forward in $x^+$ is the same as forward in $x^0$.

\subsection{Massive Scalar Fields}
\label{S2c}

If we now give the scalar field a mass the analysis is analogous. For a massive scalar field that obeys
\begin{eqnarray}
\nabla_{\mu}\nabla^{\mu}S-m^2S=0,
\label{2.16y}
\end{eqnarray}
we again set $S=A\exp(iT)$ and obtain 
\begin{eqnarray}
-A\nabla_{\mu}T\nabla^{\mu}T+\nabla_{\mu}\nabla^{\mu}A+2i\nabla_{\mu}A\nabla^{\mu}T+iA\nabla_{\mu}\nabla^{\mu}T-Am^2=0.
\label{2.17y}
\end{eqnarray}
We again  drop the $\nabla_{\mu}\nabla^{\mu}A$, $2i\nabla_{\mu}A\nabla^{\mu}T$ and $iA\nabla_{\mu}\nabla^{\mu}T$ terms, though in the presence of the mass term we could achieve this by having them be much smaller in magnitude than $m^2$ rather than much smaller than $\nabla_{\mu}T\nabla^{\mu}T$. On dropping these terms for whichever one of these reasons might be relevant we obtain  
\begin{eqnarray}
\nabla_{\mu}T\nabla^{\mu}T+m^2=0.
\label{2.18y}
\end{eqnarray}
Then, since $ds$ is nonzero for massive particles this time we can identify 
\begin{eqnarray}
\frac{\nabla^{\mu}T}{m}=\frac{dx^{\mu}}{ds}=\frac{P^{\mu}}{m},
\label{2.19y}
\end{eqnarray}
where $P^{\mu}$ is the momentum. Since $(-\nabla_{\mu}T\nabla^{\mu}T)^{1/2}$ is equal to $m$, (\ref{2.19y}) satisfies the normalization condition given in (\ref{2.8y}), just as it can since massive particles propagate off the light cone.
As introduced, $P^{\mu}$ and $dx^{\mu}/ds$ obey  
\begin{eqnarray}
P_{\mu}P^{\mu}+m^2=0,\quad \frac{dx_{\mu}}{ds}\frac{dx^{\mu}}{ds}=-1.
\label{2.20y}
\end{eqnarray}
(With $g_{00}<0$ in our notation the $(dx_{\mu}/ds)(dx^{\mu}/ds)$ interval is timelike.) 
With $P_{\mu}=\nabla_{\mu}T$ we recognize (\ref{2.20y}) as the Hamilton-Jacobi equation, an equation whose solution is precisely $S=\exp(i\int^x P_{\mu}dx^{\mu})$, $T=\int^x P_{\mu}dx^{\mu}$. This is just as we would want, with the Hamilton-Jacobi equation describing eikonalization in the massive particle case. Unlike in the massless case, this time 
there is no difficulty in setting  $T=\int^x P_{\mu}dx^{\mu}$  since the modes are not propagating on the light cone. Then, with $m^2$ being a constant we find that on taking the $\nabla^{\nu}$ derivative of (\ref{2.18y}) we again obtain (\ref{2.6y}), viz. $\nabla^{\mu}T\nabla_{\mu}\nabla^{\nu}T=0$. 
Finally, with $(dx^{\mu}/ds)(\partial/\partial x^{\mu})=d/ds$ (\ref{1.2x}) then follows, viz.
\begin{eqnarray}
m^2 \left( \frac{D^2x^{\lambda}}{ Ds^2}\right)=m^2 \left( \frac{d^2x^{\lambda} }{ ds^2}
+\Gamma^{\lambda}_{\mu \nu} 
\frac{dx^{\mu}}{ds}\frac{dx^{\nu } }{ ds} \right) 
= 0. 
\label{2.21y}
\end{eqnarray}
We thus derive the massive particle geodesic equation starting from a massive field wave equation. Also we note that as well as satisfy (\ref{2.21y}), on integrating the solution has to obey the timelike constraint $(dx_{\mu}/ds)(dx^{\mu}/ds)=-1$ appropriate to a massive particle. Moreover, below (in e.g. Secs. \ref{S5} and \ref{S9}) we shall see further examples of the value of  $(dx_{\mu}/ds)(dx^{\mu}/ds)$ being fixed as an integration constant, and this will prove to  actually be quite central to the general analysis that we make in this paper. Moreover, as we see, at no point in the calculation that leads us to (\ref{2.21y}) did we utilize the test particle action $I_T$ given in (\ref{1.1x}). So now that we have divorced the geodesic equation from the test particle action, we can consider what happens when we use the more complicated, conformal invariant wave equations of relevance to the case of particular concern to us in this paper, namely the propagation of massless particles such as photons.

\section{Conformal Invariant Scalar Field Wave Equation}
\label{S3}

\subsection{Conformal Scalar Field with no Mass Term}
\label{S3a}

While we had initially taken the scalar field to be massless, we were subsequently able to add on a mass term since no principle had been invoked that might have prevented us from doing so. However, there is such a principle, namely local conformal invariance, an invariance possessed by  massless fermion and massless gauge boson wave equations, an invariance that is thus of relevance to the gravitational bending and lensing of light. We shall explore the gauge boson case below, but first we explore the illustrative conformal scalar field case. Under a local conformal transformation  $g_{\mu\nu}(x)$ and $S(x)$ transform as $g_{\mu\nu}(x)\rightarrow \Omega^2(x)g_{\mu\nu}(x)$, $S(x)\rightarrow \Omega^{-1}(x)S(x)$, where $\Omega(x)$ is a spacetime dependent conformal factor. Since $\nabla_{\mu}\nabla^{\mu}S(x)$ would not be left invariant if $\Omega(x)$ is spacetime dependent, one has to introduce a coupling to the geometry, with the conformally coupled wave equation
\begin{eqnarray}
\nabla_{\mu}\nabla^{\mu}S
+\frac{1}{6}SR^{\alpha}_{\phantom{\alpha}\alpha}=0
\label{3.1x}
\end{eqnarray}
then being locally conformal invariant,  and with this same local conformal invariance forbidding the presence of any $-m^2S$ term. For (\ref{3.1x})  the substitution $S=A\exp(iT)$ leads to 
\begin{eqnarray}
-A\nabla_{\mu}T\nabla^{\mu}T+\nabla_{\mu}\nabla^{\mu}A+2i\nabla_{\mu}A\nabla^{\mu}T+iA\nabla_{\mu}\nabla^{\mu}T+\frac{1}{6}AR^{\alpha}_{\phantom{\alpha}\alpha}=0,
\label{3.2x}
\end{eqnarray}
and with $A$ and $T$ real  (\ref{3.2x}) breaks up into  
\begin{eqnarray}
-A\nabla_{\mu}T\nabla^{\mu}T+\nabla_{\mu}\nabla^{\mu}A+\frac{1}{6}AR^{\alpha}_{\phantom{\alpha}\alpha}=0,
\label{3.3z}
\end{eqnarray}
\begin{eqnarray}
2i\nabla_{\mu}A\nabla^{\mu}T+iA\nabla_{\mu}\nabla^{\mu}T=0.
\label{3.4z}
\end{eqnarray}
In light of our discussion of the domains of support of propagators in Sec. \ref{S1e}, the domains of support of eikonalized solutions to (\ref{3.3z}) will depend on the nature of the background geometry in which the scalar field is propagating.

\subsection{Conformal to Flat Background Geometry}
\label{S3b}

With (\ref{3.1x}) being locally conformal invariant, if the background geometry is conformal to flat with some general conformal factor $\Omega^2(x)$ so that the metric is given by $g_{\mu\nu}=\Omega^2(x)\eta_{\mu\nu}$, then, on making a conformal transformation $\bar{S}(x)=\Omega^{-1}(x)S(x)$, we find that $\bar{S}(x)$ obeys the flat space wave equation $\eta^{\mu\nu}\partial_{\mu}\partial_{\nu}\bar{S}=0$. Geodesics are thus of the flat space null $d^2x^{\mu}/dq^2=0$ form given in (\ref{1.31h}). With both the flat background and the conformal to flat background propagators only taking support on the light cone, it must be the case that the trajectories associated with $g_{\mu\nu}=\Omega^2(x)\eta_{\mu\nu}$ must be null geodesic too. However, the (\ref{1.31h}) null  geodesics are determined with the metric $\eta_{\mu\nu}$, while for the case of interest in (\ref{3.1x}) null geodesics are determined with respect to $g_{\mu\nu}=\Omega^2(x)\eta_{\mu\nu}$, and take the form given in (\ref{1.6x}). In Sec. \ref{S3g} we will show that with a redefinition of the affine parameter $q$ the (\ref{1.31h}) and (\ref{1.6x}) null geodesics conformally transform into each other, and are thus completely equivalent. Since $\Omega^2(x)\eta_{\mu\nu}dx^{\mu}dx^{\nu}$ is zero if $\eta_{\mu\nu}dx^{\mu}dx^{\nu}$ is, the  (\ref{1.31h}) and  (\ref{1.6x}) geodesics are both associated with propagation on the light cone. We shall exhibit this equivalence of null geodesics in a specific case in Sec. \ref{S8}, where we study propagation in a conformal to flat background Robertson-Walker geometry.

\subsection{Background Geometry not Conformal to Flat}
\label{S3c}

However, the analysis changes if the background geometry is not conformal to flat and at the same time the Ricci scalar contribution to (\ref{3.3z}) is comparable in magnitude with the $\nabla_{\mu}T\nabla^{\mu}T$ term. (In the  conformal to flat case it did not matter how big the Ricci scalar was, as we could always conformally transform (\ref{3.1x}) to flat.) If the background is not conformal to flat, then as noted in (\ref{1.49h}), under a very specific local conformal transformation that makes the transformed $\bar{R}^{\alpha}_{\phantom {\alpha} \alpha}$ vanish, viz.
\begin{eqnarray}  
\bar{R}^{\alpha}_{\phantom {\alpha} \alpha}=\Omega^{-2}\left[R^{\alpha}_{\phantom {\alpha} \alpha}
+6\Omega^{-1}g^{\mu\nu}\left[\partial_{\mu}\partial_{\nu} -\Gamma^{\lambda}_{\mu\nu}\partial_{\lambda}\right]\Omega\right]=
\Omega^{-2}\left[R^{\alpha}_{\phantom {\alpha} \alpha}
+6\Omega^{-1}\nabla_{\mu}\nabla^{\mu} \Omega\right]=0, 
\label{3.5z}
\end{eqnarray}
we can bring (\ref{3.1x}) to the form 
\begin{eqnarray}
\bar{\nabla}_{\mu}\bar{\nabla}^{\mu}\bar{S}(x)=0,
\label{3.6z}
\end{eqnarray}
where $\bar{g}_{\mu\nu}=\Omega^2(x)g_{\mu\nu}$,  $\bar{S}(x)=\Omega^{-1}S(x)$, as evaluated with the very specific $\Omega(x)$ that satisfies (\ref{3.5z}). To eikonalize we  set $\bar{S}=\bar{A}\exp(i\bar{T})$ with real $\bar{A}$ and $\bar{T}$. And with $\bar{S}=\Omega^{-1}S=\Omega^{-1}A\exp(iT)$ we identify $\bar{A}=\Omega^{-1}A$, $\bar{T}=T$.  With $\bar{S}=\bar{A}\exp(i\bar{T})$ the general (\ref{3.3z}) takes the form
\begin{eqnarray}
-\bar{A}\bar{\nabla}_{\mu}\bar{T}\bar{\nabla}^{\mu}\bar{T}+\bar{\nabla}_{\mu}\bar{\nabla}^{\mu}\bar{A}=0.
\label{3.7z}
\end{eqnarray}
However, since according to (\ref{3.5z}) $\Omega^{-1}\nabla_{\mu}\nabla^{\mu}\Omega$ is to be of the same order as the initial Ricci scalar $R^{\alpha}_{\phantom {\alpha} \alpha}$, the $\Omega^{-1}(x)$ prefactor in $\bar{S}(x)=\Omega^{-1}S(x)$ has to be taken into account. Thus the $\bar{\nabla}_{\mu}\bar{\nabla}^{\mu}\bar{A}$ term in (\ref{3.7z}) cannot be ignored, and thus $\bar{\nabla}_{\mu}\bar{T}\bar{\nabla}^{\mu}\bar{T}$ cannot be on the light cone. This is just as we found with the  propagator associated with the $\bar{\nabla}_{\mu}\bar{\nabla}^{\mu}$ operator in Sec. \ref{S1e}, as it takes support off the light cone. 

With $\bar{\nabla}_{\mu}\bar{T}\bar{\nabla}^{\mu}\bar{T}$ not being zero,  to eikonalize (\ref{3.7z}) we follow (\ref{2.8y}), and with affine parameter $\bar{q}$ normalize according to 
\begin{eqnarray}
\frac{dx^{\mu}}{d\bar{q}}=\frac{\bar{\nabla}^{\mu}\bar{T}}{(-\bar{\nabla}_{\mu}\bar{T}\bar{\nabla}^{\mu}\bar{T})^{1/2}},
=\frac{\bar{\nabla}^{\mu}\bar{T}}{(-\bar{A}^{-1}\bar{\nabla}_{\mu}\bar{\nabla}^{\mu}\bar{A})^{1/2}}, \qquad \bar{g}_{\mu\nu}\frac{dx^{\mu}}{d\bar{q}}\frac{dx^{\nu}}{d\bar{q}}=-1.
\label{3.8z}
\end{eqnarray}
We note that  even though (\ref{3.1x}) has no mass term, $dx^{\mu}/d\bar{q}$ is  not a lightlike vector but a unit timelike four-vector instead. (In contrast we should note that for massive particles the constraint given in (\ref{2.20y})  is  the equally timelike $(dx_{\mu}/ds)(dx^{\mu}/ds)=-1$, but with $s$ being the proper time and not the  affine parameter $\bar{q}$). With (\ref{3.7z}) leading to 
\begin{eqnarray}
\bar{\nabla}^{\mu}\bar{T}\bar{\nabla}_{\mu}\bar{\nabla}^{\lambda}\bar{T}=\frac{1}{2}\bar{\nabla}^{\lambda}\left[\bar{A}^{-1}\bar{\nabla}_{\mu}\bar{\nabla}^{\mu}\bar{A}\right],
\label{3.9z}
\end{eqnarray}
the substitution into (\ref{3.9z}) of the eikonal relation given in (\ref{3.8z}) leads to
\begin{eqnarray}
X^{1/2}\frac{dx^{\mu}}{d\bar{q}}\left[X^{1/2}\frac{\partial}{\partial x^{\mu}}\left(\frac{dx^{\lambda}}{d\bar{q}}\right)
+\frac{dx^{\lambda}}{d\bar{q}}\frac{1}{2X^{1/2}}\frac{\partial X}{\partial x^{\mu}}
+\bar{\Gamma}^{\lambda}_{\mu\nu}X^{1/2}\frac{dx^{\nu}}{d\bar{q}}\right]=-\frac{1}{2}\bar{\nabla}^{\lambda}X,
\label{3.10z}
\end{eqnarray}
where $X=-\bar{A}^{-1}\bar{\nabla}_{\mu}\bar{\nabla}^{\mu}\bar{A}$. On recalling (\ref{2.9y}), we can rewrite  (\ref{3.10z}) as 
\begin{eqnarray}
\frac{d^2x^{\lambda} }{ d\bar{q}^2}
+\bar{\Gamma}^{\lambda}_{\mu \nu} 
\frac{dx^{\mu}}{ d\bar{q}}\frac{dx^{\nu } }{d\bar{q}}=
-\frac{1}{2X}
\left[\bar{g}^{\lambda \mu}+\frac{dx^{\lambda}}{d\bar{q}}\frac{dx^{\mu}}{d\bar{q}}\right]\frac{\partial X}{\partial x^{\mu}},
\label{3.11z}
\end{eqnarray}
and with  $X=-\bar{A}^{-1}\bar{\nabla}_{\mu}\bar{\nabla}^{\mu}\bar{A}$ we finally obtain 
\begin{eqnarray}
\frac{d^2x^{\lambda} }{ d\bar{q}^2}
+\bar{\Gamma}^{\lambda}_{\mu \nu} 
\frac{dx^{\mu}}{ d\bar{q}}\frac{dx^{\nu } }{d\bar{q}}=
-\frac{1}{2(-\bar{A}^{-1}\bar{\nabla}_{\mu}\bar{\nabla}^{\mu}\bar{A})}
\left[\bar{g}^{\lambda \mu}+\frac{dx^{\lambda}}{d\bar{q}}\frac{dx^{\mu}}{d\bar{q}}\right]\frac{\partial (-\bar{A}^{-1}\bar{\nabla}_{\mu}\bar{\nabla}^{\mu}\bar{A})}{\partial x^{\mu}}.
\label{3.12z}
\end{eqnarray}
This is our key result  and it shows that in a background that is not conformal to flat  the trajectories are not null geodesics even though the starting wave equation given in (\ref{3.1x}) is conformal invariant,

As written (\ref{3.12z}) refers to quantities in the barred system. We thus need to transform (\ref{3.12z}) to the unbarred system to see how the trajectory looks in the unbarred starting basis. To relate the bases we note that in (\ref{3.3z}) we had taken $\nabla_{\mu}\nabla^{\mu}A$ term to be slowly varying, We now set $\bar{A}=\Omega^{-1}A$ in  $X=-\bar{A}^{-1}\bar{\nabla}_{\mu}\bar{\nabla}^{\mu}\bar{A}$, and with slowly varying $A$ obtain
\begin{eqnarray}
X=-A^{-1}\Omega\bar{\nabla}_{\mu}\bar{\nabla}^{\mu}(\Omega^{-1}A)=-\Omega\bar{\nabla}_{\mu}\bar{\nabla}^{\mu}\Omega^{-1}=\Omega^{-1}\bar{\nabla}_{\mu}\bar{\nabla}^{\mu}\Omega
-2\Omega^{-2}\bar{\nabla}_{\mu}\Omega\bar{\nabla}^{\mu}\Omega.
\label{3.13x}
\end{eqnarray}
However the $\bar{\nabla}_{\mu}$ derivative operators refer to $\bar{g}_{\mu\nu}$ and we need to convert them to $\nabla_{\mu}$ derivative operators that refer to $g_{\mu\nu}$. Thus from (\ref{1.44h}) we obtain  
\begin{eqnarray}
X&=&\Omega^{-3}g^{\mu\nu}\left[\partial_{\mu}\partial_{\nu}-\Gamma^{\lambda}_{\mu\nu}\partial_{\lambda}-
\Omega^{-1}\left(\delta^{\lambda}_{\mu}\partial_{\nu}\Omega
+\delta^{\lambda}_{\nu}\partial_{\mu}\Omega-g_{\mu\nu}g^{\lambda\sigma}\partial_{\sigma}\Omega\right)\partial_{\lambda}\right]\Omega
-2\Omega^{-4}g^{\mu\nu}\partial_{\mu}\Omega\partial_{\nu}\Omega
\nonumber\\
&=&\Omega^{-3}\nabla_{\mu}\nabla^{\mu}\Omega+2\Omega^{-4}g^{\mu\nu}\partial_{\mu}\Omega\partial_{\nu}\Omega-2\Omega^{-4}g^{\mu\nu}\partial_{\mu}\Omega\partial_{\nu}\Omega
\nonumber\\
&=&\Omega^{-3}\nabla_{\mu}\nabla^{\mu}\Omega
\label{3.14x}
\end{eqnarray}
Then, since this particular $\Omega(x)$ was chosen to make $\bar{R}^{\alpha}_{\phantom {\alpha} \alpha}$ vanish, from (\ref{3.5z}) we obtain
\begin{eqnarray}
X=-\frac{1}{6}\Omega^{-2}R^{\alpha}_{\phantom{\alpha}\alpha}.
\label{3.15x}
\end{eqnarray}
In order to maintain the same normalization in going from the barred basis  to the unbarred basis, viz.
\begin{eqnarray}
\bar{g}_{\mu\nu}\frac{dx^{\mu}}{d\bar{q}}\frac{dx^{\nu}}{d\bar{q}}=g_{\mu\nu}\frac{dx^{\mu}}{dq}\frac{dx^{\nu}}{dq}=-1,
\label{3.16x}
\end{eqnarray}
we must set 
\begin{eqnarray}
\frac{d}{d\bar{q}}=\Omega^{-1}\frac{d}{dq},
\label{3.17x}
\end{eqnarray}
a transformation that we note differs from the $d/d\bar{q}=\Omega^{-2}d/dq$ transformation that we will use in Sec. \ref{S3g} when we make an analogous transformation in the null geodesic case. With the transformation given in (\ref{3.17x}) and with $X$ given in (\ref{3.15x}), the two sides of (\ref{3.11z}) take the form
\begin{eqnarray}
\textrm{LHS}&=&\Omega^{-2}\left[\frac{d^2x^{\lambda} }{ dq^2}
+\Gamma^{\lambda}_{\mu \nu} 
\frac{dx^{\mu}}{ dq}\frac{dx^{\nu } }{dq}\right]+\Omega^{-3}\frac{dx^{\lambda}}{dq}\frac{dx^{\mu}}{dq}\partial_{\mu}\Omega
+\Omega^{-3}g^{\lambda\sigma}\partial_{\sigma}\Omega,
\nonumber\\
\textrm{RHS}&=&-\frac{1}{2}\frac{1}{(-R^{\alpha}_{\phantom{\alpha}\alpha}/6)}\left[g^{\lambda\mu}+\frac{dx^{\lambda}}{dq}\frac{dx^{\mu}}{dq}\right]
\left[-\frac{1}{6}\Omega^{-2}\partial_{\mu}R^{\alpha}_{\phantom{\alpha}\alpha}+\frac{1}{3}R^{\alpha}_{\phantom{\alpha}\alpha}\Omega^{-3}\partial_{\mu}\Omega\right].
\label{3.18x}
\end{eqnarray}
Finally then, in the unbarred basis (\ref{3.12z}) and (\ref{3.16x}) take the form
\begin{eqnarray}
\frac{d^2x^{\lambda} }{ dq^2}
+\Gamma^{\lambda}_{\mu \nu} 
\frac{dx^{\mu}}{ dq}\frac{dx^{\nu } }{dq}=
-\frac{1}{2(-R^{\alpha}_{\phantom{\alpha}\alpha}/6)}
\left[g^{\lambda \mu}+\frac{dx^{\lambda}}{dq}\frac{dx^{\mu}}{dq}\right]\frac{\partial (-R^{\alpha}_{\phantom{\alpha}\alpha}/6)}{\partial x^{\mu}},
\quad g_{\mu\nu}\frac{dx^{\mu}}{d\bar{q}}\frac{dx^{\nu}}{d\bar{q}}=-1.
\label{3.19qq}
\end{eqnarray}

It is  of interest to derive this result  in the original basis where  $S=A\exp(iT)$, With $A$ being slowly varying (\ref{3.3z}) takes the form  
\begin{eqnarray}
-\nabla_{\mu}T\nabla^{\mu}T+\frac{1}{6}R^{\alpha}_{\phantom{\alpha}\alpha}=0,
\label{3.20qq}
\end{eqnarray}
so that
\begin{equation}
\nabla^{\mu}T\nabla_{\mu}\nabla^{\lambda}T=\frac{1}{12}\nabla^{\lambda}R^{\alpha}_{\phantom{\alpha}\alpha}.
\label{3.21qq}
\end{equation}
We now eikonalize (\ref{3.20qq}) by setting 
\begin{eqnarray}
\frac{dx^{\mu}}{dq}=\frac{\nabla^{\mu}T}{(-\nabla_{\nu}T\nabla^{\nu}T)^{1/2}}=\frac{\nabla^{\mu}T}{(-R^{\alpha}_{\phantom{\alpha}\alpha}/6)^{1/2}}, \qquad \frac{dx_{\mu}}{dq}\frac{dx^{\mu}}{dq}=-1.
\label{3.22qq}
\end{eqnarray}
Since the steps leading to (\ref{3.11z}) are generic, (\ref{3.11z}) also holds in the unbarred basis where now  $X= -R^{\alpha}_{\phantom{\alpha}\alpha}/6$. Consequently, from (\ref{3.22qq}) we obtain none other than  (\ref{3.19qq}), just as we should. The scalar field (\ref{3.19qq}), its Maxwell analog given in (\ref{4.13x}) and their  generalizations given in (\ref{4.21xy}) and (\ref{10.2x}) below constitute our main result for conformally coupled scalar or Maxwell fields propagating in background geometries that are not conformal to flat \cite{footnote3b}. Though, as we had noted before, if the background geometry is conformal to flat then eikonalized conformally coupled scalar and Maxwell field modes follow null geodesics.

As we see, despite obtaining the geodesic operator on the left-hand side of (\ref{3.19qq}), we find that there is a nonzero geometric term on the right-hand side. Moreover, we quite remarkably find that  (\ref{3.19qq}) is of precisely the same generic form as  (\ref{1.18g}). In fact it would have been identical in form if instead of (\ref{1.3x}) we had varied the test particle action \cite{footnote4}
\begin{eqnarray}I_T=- \kappa\int ds
(-R^{\alpha}_{\phantom{\alpha}\alpha})^{1/2}.
\label{3.23qq}
\end{eqnarray}
Now in \cite{Mannheim2006} we had introduced (\ref{1.3x}) purely for illustrative purposes. Its emergence as being associated with a dynamical wave equation is both surprising and intriguing \cite{footnote5}. 

To check that there is no obvious error in (\ref{3.19qq}), we note that on taking the derivative of $(dx_{\mu}/dq)(dx^{\mu}/dq)=-1$ with respect to $q$ we obtain
\begin{eqnarray}
\frac{1}{2}\frac{d}{dq}\left(g_{\lambda\alpha}\frac{dx^{\lambda}}{ dq}\frac{dx^{\alpha} }{dq}\right)
=\frac{dx^{\alpha}}{dq}\left(g_{\lambda\alpha}\frac{d^2x^{\lambda } }{dq^2}
+\frac{1}{2} \frac{dx^{\lambda}}{dq}\frac{dx^{\nu}}{dq}\frac{\partial g_{\lambda\alpha}}{\partial x^{\nu}}\right)=0.
\label{3.24qq}
\end{eqnarray}
We recognize (\ref{3.24qq}) as being of the form 
\begin{eqnarray}
g_{\lambda \alpha}\frac{dx^{\alpha} }{dq}\left(\frac{d^2x^{\lambda} }{ dq^2}
+\Gamma^{\lambda}_{\mu \nu} 
\frac{dx^{\mu}}{ dq}\frac{dx^{\nu } }{dq}\right)=0,
\label{3.25qq}
\end{eqnarray}
with the covariant four-velocity being orthogonal to the covariant four-acceleration. Consequently, the right-hand side of (\ref{3.19qq}) must be orthogonal to the covariant four-velocity as well. And with it  precisely having the form that it does have, one can readily check that this is in fact the case.

It is also of interest to note that had we set $dx^{\mu}/dq=\nabla^{\mu}T$ without the normalization factor given in (\ref{3.22qq}), we would have obtained \cite{Mannheim2006}
\begin{eqnarray}
\frac{d^2x^{\lambda} }{ dq^2}
+\Gamma^{\lambda}_{\mu \nu} 
\frac{dx^{\mu}}{ dq}\frac{dx^{\nu } }{dq}=
\frac{1}{12}
g^{\lambda \mu}\frac{\partial R^{\alpha}_{\phantom{\alpha}\alpha}}{\partial x^{\mu}}.
\label{3.26qq}
\end{eqnarray}
Since (\ref{3.25qq}) is a mathematical identity, the right-hand side of (\ref{3.26qq}) would have to obey 
\begin{eqnarray}
\frac{dx^{\mu}}{dq}\frac{\partial R^{\alpha}_{\phantom{\alpha}\alpha}}{\partial x^{\mu}}=\frac{d R^{\alpha}_{\phantom{\alpha}\alpha}}{d q}=0,
\label{3.27qq}
\end{eqnarray}
a relation that could only hold for specific values of $ R^{\alpha}_{\phantom{\alpha}\alpha}$ such as a constant or zero 
\cite{footnote6}. We thus see that with the normalization of the eikonal that we have used  there is no constraint on $ R^{\alpha}_{\phantom{\alpha}\alpha}$, with (\ref{3.19qq}) holding for any 
nonzero $R^{\alpha}_{\phantom{\alpha}\alpha}$. (For $ R^{\alpha}_{\phantom{\alpha}\alpha}=0$ we would be back to (\ref{2.1y}).) Consequently, we see that the very covariance of the theory requires that we must use the normalized form of the eikonal whenever $\nabla_{\mu}T\nabla^{\mu}T$ is nonzero. Moreover, if we do use the normalized form then no matter how small $R^{\alpha}_{\phantom{\alpha}\alpha}$ might be, because of the appearance of $ R^{\alpha}_{\phantom{\alpha}\alpha}$ in the denominator of the right-hand side of (\ref{3.19qq}) the contribution of $ R^{\alpha}_{\phantom{\alpha}\alpha}$ will still be non-trivial. However, when  $R^{\alpha}_{\phantom{\alpha}\alpha}$ is zero we must be able to come back to the lightlike (\ref{2.10y}). The $R^{\alpha}_{\phantom{\alpha}\alpha}\rightarrow 0$ limit is thus delicate, and we discuss it in more detail in Sec. \ref{S3i} below.

In addition we note that while we normalized $\nabla_{\mu}T$ in (\ref{3.22qq}), we are still free to rescale $q$ by a numerical constant in (\ref{3.22qq}) as it would still measure distance along the trajectory and would not affect (\ref{3.19qq}). Thus we can introduce a constant $E$ by setting
\begin{eqnarray}
q^{\prime}= \frac{q}{E^{1/2}},
\label{3.28qq}
\end{eqnarray}
so that
\begin{eqnarray}
\frac{1}{E^{1/2}}\frac{dx^{\mu}}{dq^{\prime}}=\frac{\nabla^{\mu}T}{(-R^{\alpha}_{\phantom{\alpha}\alpha}/6)^{1/2}}, \qquad -g_{\mu\nu}\frac{dx^{\mu}}{dq^{\prime}}\frac{dx^{\nu}}{dq^{\prime}}=E.
\label{3.29qq}
\end{eqnarray}
We will see below that this will allow us to make contact with the standard analysis of solar system geodesics in the Schwarzschild de Sitter case, where $E$ serves as an energy. 

Moreover, we note that even though we had started with the locally conformal invariant equation of motion given in (\ref{3.1x}), and even though both it and the light cone $ds^2=-g_{\mu\nu}dx^{\mu}dx^{\nu}=0$ are left invariant under $g_{\mu\nu}\rightarrow \Omega^2(x)g_{\mu\nu}(x)$, nonetheless according to (\ref{3.22qq}) $-g_{\mu\nu}dx^{\mu}dx^{\nu}$ is not zero \cite{footnote7}. Since it is nonzero it would need to be timelike if it is to describe physical events, and thus $E$ has to be positive, just as needed in (\ref{3.28qq}) so as to have $q^{\prime}$ be real.  However, since the standard discussion of propagation of light sets $E=0$, below we will find modifications of the standard discussion of gravitational bending of light once $E$ is nonzero. We comment more on the fact that $-g_{\mu\nu}dx^{\mu}dx^{\nu}$ is not zero in our discussion of the implications of conformal invariance and in our discussion the Maxwell field that we give below. 

While the left-hand side of (\ref{3.19qq}) is the standard term that appears in the geodesic equation, the term on the right-hand side is not. It leads to a departure from strict geodesic behavior and has to be taken into consideration if the derivative of the Ricci scalar is not zero. That this term appears at all is a reflection of our remark in Sec. \ref{S1} that covariantization of a flat space quantity can miss explicit geometric dependent terms. Thus if we start in flat spacetime with $\partial_{\mu}\partial^{\mu}\exp(iT)=0$, eikonalization would lead us to $\partial_{\mu}T\partial^{\mu}T=0$. Covariantization would then lead us to $\nabla_{\mu}T\nabla^{\mu}T=0$, and not to $\nabla_{\mu}T\nabla^{\mu}T-R^{\alpha}_{\phantom{\alpha}\alpha}/6=0$. Thus by covariantization we cannot reach $\nabla_{\mu}T\nabla^{\mu}T-R^{\alpha}_{\phantom{\alpha}\alpha}/6=0$ starting from $\partial_{\mu}T\partial^{\mu}T=0$, even though starting $\nabla_{\mu}T\nabla^{\mu}T-R^{\alpha}_{\phantom{\alpha}\alpha}/6=0$ we can reach $\partial_{\mu}T\partial^{\mu}T=0$ by taking the flat space limit.

Now while the right-hand side  of (\ref{3.19qq}) contains the derivative of the Ricci scalar, in the event that the derivative is zero there are actually two possibilities: the Ricci scalar is either zero or  equal to a nonzero constant. In either case (\ref{3.19qq}) reduces to the standard massless geodesic equation given in (\ref{2.10y}). However, if the Ricci scalar is a nonzero constant, (\ref{3.20qq}) does not reduce to (\ref{2.5y}). In this case (\ref{3.20qq}) can be considered to be the integral of  (\ref{3.21qq})  in the event of a vanishing $\nabla^{\lambda}R^{\alpha}_{\phantom{\alpha}\alpha}$, with a then constant $\tfrac{1}{6}R^{\alpha}_{\phantom{\alpha}\alpha}$ term being a nonzero integration constant. In the exterior Schwarzschild situation the Ricci scalar is zero and the standard geodesic discussion of the gravitational bending of massless geodesics by the Sun remains untouched. However, in the Schwarzschild de Sitter situation the Ricci scalar is a nonzero constant and (\ref{3.21qq}) (with $\nabla^{\lambda}R^{\alpha}_{\phantom{\alpha}\alpha}=0$) has to be integrated under the constraint provided by (\ref{3.20qq}). We discuss the implications of these remarks in more detail in Secs. \ref{S5}, \ref{S6} and \ref{S7}.

\subsection{Conformal Scalar Field with Dynamical Mass}
\label{S3d}

While conformal invariance would exclude any mass term at the level of the Lagrangian, mass could still be generated through the vacuum by spontaneous symmetry breakdown. For the conformally coupled scalar field $S$ that we are considering here we would need to introduce a second scalar field  $Q$ that would provide the symmetry breaking by acquiring a non-vanishing expectation value. Thus for a conformal invariant action of the form
\begin{eqnarray}
I=-\int d^4x(-g)^{1/2}\left[\frac{1}{2}\nabla_{\mu}S\nabla^{\mu}S -\frac{1}{12}S^2R^{\alpha}_{\phantom{\alpha}\alpha}+\frac{1}{2}S^2Q^2+V(Q)\right],
\label{3.30qq}
\end{eqnarray}
where $V(Q)$ is to generate a possibly spacetime-dependent expectation value $Q_0$ for $Q$, the equation of motion for the scalar field takes the form
\begin{eqnarray}
\nabla_{\mu}\nabla^{\mu}S
+\frac{1}{6}SR^{\alpha}_{\phantom{\alpha}\alpha}-Q_0^2S=0,
\label{3.31qq}
\end{eqnarray}
while the eikonal phase defined by $S=A\exp(iT)$ obeys 
\begin{eqnarray}
-A\nabla_{\mu}T\nabla^{\mu}T+\nabla_{\mu}\nabla^{\mu}A+2i\nabla_{\mu}A\nabla^{\mu}T+iA\nabla_{\mu}\nabla^{\mu}T+\frac{1}{6}AR^{\alpha}_{\phantom{\alpha}\alpha}-AQ_0^2=0.
\label{3.32qq}
\end{eqnarray}
On dropping the $\nabla_{\mu}\nabla^{\mu}A$, $2i\nabla_{\mu}A\nabla^{\mu}T$ and $iA\nabla_{\mu}\nabla^{\mu}T$ terms, and on setting 
\begin{eqnarray}
\frac{dx^{\mu}}{dq}=\frac{\nabla^{\mu}T}{(Q_0^2-R^{\alpha}_{\phantom{\alpha}\alpha}/6)^{1/2}}, \qquad \frac{dx_{\mu}}{dq}\frac{dx^{\mu}}{dq}=-1,
\label{3.33qq}
\end{eqnarray}
the discussion now follows as above, with (\ref{3.19qq}) being replaced by
\begin{eqnarray}
\frac{d^2x^{\lambda} }{ dq^2}
+\Gamma^{\lambda}_{\mu \nu} 
\frac{dx^{\mu}}{ dq}\frac{dx^{\nu } }{dq}=
-\frac{1}{2}\frac{1}{(Q_0^2-R^{\alpha}_{\phantom{\alpha}\alpha}/6)}
\left[g^{\lambda \mu}+\frac{dx^{\lambda}}{dq}\frac{dx^{\mu}}{dq}\right]\frac{\partial(Q_0^2-R^{\alpha}_{\phantom{\alpha}\alpha}/6)}{\partial x^{\mu}}.
\label{3.34qq}
\end{eqnarray}
When $Q_0$ is a constant $m$, (\ref{3.34qq}) reduces to  
\begin{eqnarray}
\frac{d^2x^{\lambda} }{ dq^2}
+\Gamma^{\lambda}_{\mu \nu} 
\frac{dx^{\mu}}{ dq}\frac{dx^{\nu } }{dq}=
-\frac{1}{2}\frac{1}{(m^2-R^{\alpha}_{\phantom{\alpha}\alpha}/6)}
\left[g^{\lambda \mu}+\frac{dx^{\lambda}}{dq}\frac{dx^{\mu}}{dq}\right]\frac{\partial(-R^{\alpha}_{\phantom{\alpha}\alpha}/6)}{\partial x^{\mu}}.
\label{3.35qq}
\end{eqnarray}
Interestingly, the scalar field equation based (\ref{3.35qq}) is a direct analog of the test particle (\ref{1.4x}), though such an analog of (\ref{1.4x}) is output here not input. Finally, if  $Q_0^2=m^2 \gg R^{\alpha}_{\phantom{\alpha}\alpha}/6$ and $q^{\prime}=q/E^{1/2}$, (\ref{3.33qq}) and (\ref{3.35qq}) reduce to 
\begin{eqnarray}
\frac{1}{E^{1/2}}\frac{dx^{\mu}}{dq^{\prime}}=\frac{\nabla^{\mu}T}{m}, \qquad \frac{dx_{\mu}}{dq^{\prime}}\frac{dx^{\mu}}{dq^{\prime}}=-E, \qquad \frac{d^2x^{\lambda} }{ dq^{\prime 2}}
+\Gamma^{\lambda}_{\mu \nu} 
\frac{dx^{\mu}}{ dq^{\prime}}\frac{dx^{\nu } }{dq^{\prime}}=0,
\label{3.36qq}
\end{eqnarray}
the standard massive particle geodesic equation with energy $E$. To ascertain whether $m$ might be much larger than any curvature scale, we note that the factor that should appear in the scalar field wave equation is not $m$ but something with the dimension of an inverse length. In quantum field theory the natural length is $mc/\hbar$, the inverse of the Compton wavelength of $m$ (viz. $10^{13}~cm^{-1}$ for a proton). For this choice we would need to compare the square of the inverse of the Compton wavelength of a typical solar mass star (order $10^{140}~cm^{-2}$) with a typical scale that appears in the Ricci scalar. Since the Ricci scalar vanishes identically in the pure Schwarzschild case, we look instead at a de Sitter geometry where $R^{\alpha}_{\phantom{\alpha}\alpha}=-12K$. The natural expectation for $K$ is that it would be of order the inverse square of the Hubble radius, i.e., of order $10^{-56}~cm^{-2}$. Alternatively, if the $m$ that appears in the wave equation is to be associated with a classical length scale, the natural length scale is the Schwarzschild radius $mG/c^2$ (viz. $10^5~cm$ for a solar mass star). The square of the inverse of this quantity is $10^{-10}~cm^{-2}$. Thus for either choice the inverse length term dominates over the Ricci scalar, and so for massive particles we are back to standard geodesics.  

Effects similar to the ones found here for scalars also occur for fermions, either massless \cite{footnote8} or massive \cite{footnote9}, where again there is a dependence on the Ricci scalar. However, while we could generate a mass dynamically in the conformal scalar or conformal fermion field cases, and while such masses could dominate over curvature, below we discuss a case where there is no mass generation, namely the Maxwell field, an equally conformal case where now there is no mass term that is able to dominate over any curvature contribution in the first place. But before doing so,  in order to show that the departure from geodesic behavior exhibited in (\ref {3.19qq}) is not actually restricted to fields with an underlying conformal structure, we now show that it also occurs with perfect fluids, and in particular for perfect fluid energy-momentum tensors that have no need to be of the traceless form that would be required of conformal perfect fluids.

\subsection{Perfect Fluids}
\label{S3e}

For a perfect fluid the energy-momentum tensor is of the form
\begin{eqnarray}
T^{\mu \nu}=(\rho+p)U^{\mu}U^{\nu}+pg^{\mu\nu},
\label{3.37qq}
\end{eqnarray}                                 
with energy density $\rho$, pressure $p$ and a unit timelike fluid
four-vector $U^{\mu}$ that is normalized to $U_{\mu}U^{\mu}=-1$.  We will have no need here to require the perfect fluid to obey the $g_{\mu\nu}T^{\mu\nu}=3p-\rho=0$ condition that massless radiation fluids obey. For any general perfect fluid covariant conservation leads to 
\begin{eqnarray}
\nabla_{\nu}\left[ (\rho + p)U^{\mu}U^{\nu} +pg^{\mu \nu}\right]=
U^{\mu} \nabla_{\nu}[(\rho+p)U^{\nu}]                                             
+(\rho+p)U^{\nu}\nabla_{\nu}U^{\mu} +\nabla^{\mu}p=0,
\label{3.38qq}
\end{eqnarray}                                 
and thus to
\begin{eqnarray}
- \nabla_{\nu}[(\rho+p)U^{\nu}]
+U_{\mu}\nabla^{\mu}p=-                  
(\rho+p)U_{\mu}U^{\nu}\nabla_{\nu}U^{\mu}.
\label{3.39qq}
\end{eqnarray}                                 
Then since $\nabla_{\nu}[U_{\mu}U^{\mu}]=2[U_{\mu}\nabla_{\nu}U^{\mu}]=0$ we obtain
\begin{eqnarray}
- \nabla_{\nu}[(\rho+p)U^{\nu}]
+U_{\mu}\nabla^{\mu}p=0.
\label{3.40qq}
\end{eqnarray}                                 
The insertion of (\ref{3.40qq}) into (\ref{3.38qq}) then yields
\begin{eqnarray}
(\rho+p)U^{\nu}\nabla_{\nu}U^{\mu}                                   
+[g^{\mu\nu} + U^{\mu}U^{\nu}]\nabla_{\nu}p=0. 
\label{3.41qq}
\end{eqnarray}                                 
On setting $U^{\mu}=dx^{\mu}/ds$ (we do not need to set  $U^{\mu}=dx^{\mu}/dq$ since $U^{\mu}$ is timelike, not lightlike), we thus impose the eikonal normalization condition 
\begin{eqnarray}                                                                              
\frac{dx^{\mu}}{ds}=\frac{U^{\mu}}{(-U_{\nu}U^{\nu})^{1/2}} =U^{\mu}
\label{3.42qq}
\end{eqnarray}                                 
on using  $U_{\nu}U^{\nu}=-1$, with the direction of the velocity of the fluid being  in the normal, viz. longitudinal, direction. Thus from (\ref{3.41qq})   we obtain 
\begin{eqnarray}
\frac{d^2x^{\lambda} }{ ds^2}
+\Gamma^{\lambda}_{\mu \nu} 
\frac{dx^{\mu}}{ ds}\frac{dx^{\nu } }{ds}=                                   
-\frac{1}{(\rho+p)}\left[g^{\lambda\nu} + \frac{dx^{\lambda}}{ ds}\frac{dx^{\nu } }{ds}\right]\nabla_{\nu}p. 
\label{3.43qq}
\end{eqnarray}                                 
Thus in general not only does the fluid velocity vector not move on a geodesic, the structure of (\ref{3.43qq}) is  identical to that of the  scalar field (\ref{3.19qq}), as it would essentially have to be since with the left-hand side of (\ref{3.43qq})  being orthogonal to $dx^{\lambda}/dq$, the  right-hand side would have to be orthogonal to $dx^{\lambda}/dq$ too.

While the right-hand side of (\ref{3.43qq}) would vanish in the pressureless dust situation in which $p$ is zero, it would not vanish otherwise. However, since our derivation of (\ref{3.43qq}) holds no matter what the relation between $\rho$ and $p$ might be, it therefore holds for radiation fluids in which $\rho=3p$. Thus for a radiation fluid the right-hand side of (\ref{3.43qq}) is necessarily nonzero, and the motion of the fluid velocity is necessarily neither geodesic nor on the light cone. Now this behavior will hold for any background be it conformal to flat or not conformal to flat, and yet for any radiation mode propagating in a conformal to flat background propagation is restricted to null geodesics. The distinction here is that a radiation fluid is not a single light cone mode but an incoherent statistical averaging of a whole ensemble of light cone modes, and as shown  in \cite{Mannheim2006}, this averaging produces a fluid whose velocity vector is timelike not lightlike. 

\subsection{Implications for Conformal Invariance}
\label{S3f}

While we started with a locally conformal invariant wave equation in  (\ref{3.1x}), the act of dropping the $-i\nabla_{\mu}\nabla^{\mu}T$ term in (\ref{3.2x}) leaves us with an eikonal equation  in (\ref{3.20qq}) that is no longer locally conformal invariant. The dropping of the $\nabla_{\mu}\nabla^{\mu}A+2i\nabla_{\mu}A\nabla^{\mu}T$ term does not affect conformal invariance as this particular linear combination is locally conformal invariant on its own. To see these specific features  explicitly we note that under $S\rightarrow e^{-\alpha}S$ the eikonal phase changes according to $T\rightarrow T+i\alpha$, with $\Gamma^{\lambda}_{\mu\nu}$ changing by $\delta^{\lambda}_{\mu}\partial_{\nu}\alpha+\delta^{\lambda}_{\nu}\partial_{\mu}\alpha -g^{\lambda\sigma}g_{\mu\nu}\partial_{\sigma}\alpha$. With $A$ not changing,  the five terms in (\ref{3.2x}) transform as 
\begin{eqnarray}
 A\nabla_{\mu}T\nabla^{\mu}T &\rightarrow&Ae^{-2\alpha}\left[\nabla_{\mu}T\nabla^{\mu}T +2i\nabla_{\mu}T\nabla^{\mu}\alpha-\nabla_{\mu}\alpha\nabla^{\mu}\alpha\right],  
 \nonumber\\
 -iA\nabla_{\mu}\nabla^{\mu}T &\rightarrow&-iAe^{-2\alpha}\left[\nabla_{\mu}\nabla^{\mu}T+i\nabla_{\mu}\nabla^{\mu}\alpha+2\nabla_{\mu}T\nabla^{\mu}\alpha+2i\nabla_{\mu}\alpha\nabla^{\mu}\alpha\right],
 \nonumber\\
 -\frac{1}{6}AR^{\alpha}_{\phantom{\alpha}\alpha}&\rightarrow&-e^{-2\alpha}A\left[\frac{1}{6}R^{\alpha}_{\phantom{\alpha}\alpha}+\nabla_{\mu}\nabla^{\mu}\alpha+\nabla_{\mu}\alpha\nabla^{\mu}\alpha\right],
 \nonumber\\
 \nabla_{\mu}\nabla^{\mu}A&\rightarrow&e^{-2\alpha}\left[\nabla_{\mu}\nabla^{\mu}A+2\nabla_{\mu}A\nabla^{\mu}\alpha\right],
 \nonumber\\
 2i\nabla_{\mu}A\nabla^{\mu}T&\rightarrow& e^{-2\alpha}\left[2i\nabla_{\mu}A\nabla^{\mu}T- 2\nabla_{\mu}A\nabla^{\mu}\alpha\right].
  \label{3.44qq}
 \end{eqnarray}
 As we see, the $ -i\nabla_{\mu}\nabla^{\mu}T$ term is needed to maintain the local conformal invariance of $ \nabla_{\mu}T\nabla^{\mu}T -i\nabla_{\mu}\nabla^{\mu}T-(1/6)R^{\alpha}_{\phantom{\alpha}\alpha}$, with $\nabla_{\mu}\nabla^{\mu}A+2i\nabla_{\mu}A\nabla^{\mu}T$ being  locally conformal invariant on its own. In the absence of the $ -i\nabla_{\mu}\nabla^{\mu}T$ term we could only maintain local conformal invariance for those particular $\alpha$ that obey $i\nabla_{\mu}\nabla^{\mu}\alpha+2\nabla_{\mu}T\nabla^{\mu}\alpha+2i\nabla_{\mu}\alpha\nabla^{\mu}\alpha=0$. However, as we show momentarily, we should actually satisfy this requirement by taking $\alpha$ to be a constant. Now this does not mean that we have lost all conformal symmetry, only that we have lost local conformal symmetry, as we still have global conformal invariance, an invariance that  is often also referred to as global scale invariance \cite{footnote10a}.  Global conformal invariance is still sufficient to forbid any kinematical mass terms such as the one exhibited in (\ref{2.16y}). Moreover, if we were not to eikonalize, i.e., in the diffraction regime, we would still need to keep the $-i\nabla_{\mu}\nabla^{\mu}T$ term and would still have full local conformal symmetry. However, for eikonalization purposes global  conformal symmetry will suffice.
 
In Sec. \ref{S1a} we had noted that the eikonal approximation required that the wavelength of a wave obey $\lambda \ll L$, where $L$ is a characteristic scale of a system through which the massless wave propagates. Now this condition is not locally conformal invariant. However, it is globally conformal invariant since on scaling  with any positive parameter $c$ the condition becomes $c\lambda \ll cL$. In addition, we note that the  test particle action of the form given in (\ref{1.3x}), viz.  $I_T=-\kappa\int ds R^{\alpha}_{\phantom{\alpha}\alpha}$, is neither locally nor globally conformal invariant. However, while not locally conformal invariant the test particle action given in (\ref{3.23qq}), viz. $I_T=-\kappa\int ds (R^{\alpha}_{\phantom{\alpha}\alpha})^{1/2}$, is globally conformal invariant. Consequently, the Ricci-scalar-dependent trajectory given in (\ref{3.19qq}) is globally conformal invariant too.
 
That we would want to take $\alpha$ to be a constant when we eikonalize can be seen from the eikonal condition given in (\ref{2.8y}). Specifically, we note that while the fields and the metric transform non-trivially under a local conformal transformation, the coordinates do not.  We thus could not enforce the eikonal relation $dx^{\mu}/dq=\nabla^{\mu}T/(-\nabla_{\nu}T\nabla^{\nu}T)^{1/2}$ if one side of this equation is left untouched while the other side acquires derivatives of $\alpha$. Thus the only option is to keep $\alpha$ constant, since then $\nabla^{\mu}T\rightarrow \nabla^{\mu}T+i\nabla^{\mu}\alpha=\nabla^{\mu}T$. We should also note that even if we had not normalized the eikonal relation at all and simply set $dx^{\mu}/dq=\nabla^{\mu}T$, it still would not be invariant under $T\rightarrow T+i\alpha$ unless $\alpha $ is constant. Moreover, we note that even if we did not take $T$ to dominate and did not approximate the equations of motion at all (we discuss this possibility in Sec. \ref{S10}), the eikonal condition $dx^{\mu}/dq=\nabla^{\mu}T/(-\nabla_{\nu}T\nabla^{\nu}T)^{1/2}$ would in and of itself already reduce the symmetry from local to global conformal invariance.

\subsection{Conformal Invariance and Null Geodesics}
\label{S3g}

With the transition from local conformal invariance to global conformal invariance the modified geodesic equation given in (\ref{3.19qq}) is only globally conformal invariant. However, we should note that the standard null geodesic equation  given in (\ref{1.6x}) does itself actually admit of a local conformal invariance. In order to maintain it we not only need to set $(dx_{\mu}/dq)(dx^{\mu}/dq)=0$, we, in addition, need to introduce an affine parameter according to $d/d\bar{q}=\Omega^{-2}(x)d/dq$, and define geodesics with respect to this $\bar{q}$. Specifically,  under a local conformal transformation we obtain
\begin{eqnarray}
 \frac{d^2x^{\lambda} }{ d\bar{q}^2}
+\Gamma^{\lambda}_{\mu \nu} 
\frac{dx^{\mu}}{d\bar{q}}\frac{dx^{\nu } }{ d\bar{q}}&\rightarrow&  \frac{1}{\Omega^2}\frac{d}{dq}\left[\frac{1}{\Omega^2}\frac{dx^{\lambda} }{ dq}\right]
+\frac{1}{\Omega^4}\left[\Gamma^{\lambda}_{\mu \nu} 
+\frac{1}{\Omega}\left(\delta^{\lambda}_{\nu}\partial_{\mu}+\delta^{\lambda}_{\mu}\partial_{\nu}-g_{\mu\nu}\partial^{\lambda}\right)\Omega
\right]
\frac{dx^{\mu}}{dq}\frac{dx^{\nu } }{ dq}
 \nonumber\\
 &=&\frac{1}{\Omega^4}\left[\frac{d^2x^{\lambda} }{ dq^2}
+\Gamma^{\lambda}_{\mu \nu} 
\frac{dx^{\mu}}{dq}\frac{dx^{\nu } }{ dq}\right]-\frac{\partial^{\lambda}\Omega}{\Omega^5}g_{\mu\nu}\frac{dx^{\mu}}{dq}\frac{dx^{\nu } }{ dq}.
  \label{3.45qq}
 \end{eqnarray}
 Then,  if $(dx_{\mu}/dq)(dx^{\mu}/dq)=0$  (\ref{3.45qq}) reduces to
\begin{eqnarray}
 \frac{d^2x^{\lambda} }{ d\bar{q}^2}
+\Gamma^{\lambda}_{\mu \nu} 
\frac{dx^{\mu}}{d\bar{q}}\frac{dx^{\nu } }{ d\bar{q}}&\rightarrow&  \frac{1}{\Omega^4}\left[\frac{d^2x^{\lambda} }{ dq^2}
+\Gamma^{\lambda}_{\mu \nu} 
\frac{dx^{\mu}}{dq}\frac{dx^{\nu } }{ dq}\right].
  \label{3.46qq}
 \end{eqnarray}
Null geodesics on the light cone can thus transform into each other provided we locally rescale the affine parameter along each individual trajectory according to $d/d\bar{q}=\Omega^{-2}(x)d/dq$.

We now recall that in Secs. \ref{S1e} and \ref{S3b} we had noted that for a conformally coupled  scalar field propagating  in a conformal to flat background geometry with metric $g_{\mu\nu}=\Omega^2(x)\eta_{\mu\nu}$, the theory is completely equivalent to a free scalar field propagating in a flat background with metric $\eta_{\mu\nu}$. While the trajectories in both cases would then be null geodesics the question was raised as to whether the null geodesics should be determined with respect to $g_{\mu\nu}$ or $\eta_{\mu\nu}$. From (\ref{3.46qq}) we see that the two null geodesics are equivalent. In Sec. \ref{S8} we will see this explicitly in  a completely solvable case, namely null geodesics in a conformal to flat Robertson-Walker geometry.

\subsection{Conformal Invariant Massive Test Particle Action}
\label{S3h}

While trajectories have to be derived from wave equations and while massive particles do not propagate on the light cone, nonetheless,  it is of interest to note that there exists a massive test particle action that is locally conformal invariant. It is of the form

\begin{eqnarray}
I_T=-\int dsQ_0(x),
\label{3.47qq}
\end{eqnarray}
where the scalar field $Q_0(x)$ varies along the trajectory and $ds=[-g_{\mu\nu}dx^{\mu}dx^{\nu}]^{1/2}$ is the proper time. Here local conformal invariance is due to the fact that the local conformal transformation $Q_0(x)\rightarrow \Omega^{-1}(x)Q_0(x)$, $ds\rightarrow \Omega(x)ds$ leaves this action invariant. It is crucial that the proper time be involved as $\int dq Q_0(x)$ is not locally conformal invariant since the affine parameter $q$ does not transform under a conformal transformation. Since the proper time vanishes for massless particles it follows that (\ref{3.47qq}) can only apply to massive particles, with (\ref{3.47qq}) becoming the test particle action $I_T=-m\int ds$ given in (\ref{1.1x}) if $Q_0$ is equal to a constant $m$. Variation of (\ref{3.47qq}) with respect to $x^{\lambda}$ yields the locally conformal invariant \cite{Mannheim1993}
\begin{eqnarray}
\frac{d^2x^{\lambda} }{ ds^2}
+\Gamma^{\lambda}_{\mu \nu} 
\frac{dx^{\mu}}{ ds}\frac{dx^{\nu } }{ds}=
-\frac{1}{Q_0}
\left[g^{\lambda \mu}+\frac{dx^{\lambda}}{ds}\frac{dx^{\mu}}{ds}\right]\frac{\partial Q_0}{\partial x^{\mu}},
\label{3.48qq}
\end{eqnarray}
when $Q_0$ is not constant. Then, when $Q_0$ is equal to a constant $m$, (\ref{3.48qq}) reduces to  
\begin{eqnarray}
\frac{d^2x^{\lambda} }{ ds^2}
+\Gamma^{\lambda}_{\mu \nu} 
\frac{dx^{\mu}}{ ds}\frac{dx^{\nu } }{ds}=0.
\label{3.49qq}
\end{eqnarray}
The $I_T$ action given in (\ref{3.47qq}) allows us to construct conformal invariant trajectories for massive particles, with the mass parameter transforming under a local conformal transformation since it is actually a classical scalar field (i.e., the expectation value of a quantum scalar field) and not a kinematic or mechanical parameter in a test particle action. In other words, if mass is generated dynamically in a theory with an underlying local conformal symmetry, then if that procedure were to generate (\ref{3.48qq}) the particle trajectory would be locally conformal invariant. However, we now show that the generation of (\ref{3.48qq}) cannot in fact occur.

Specifically, noting the similarity between (\ref{3.48qq})  and  (\ref{3.34qq}), we see that if set $dq=ds$ in (\ref{3.34qq}), which we can do since for massive modes $ds^2\neq 0$, we can rewrite  (\ref{3.34qq}) as 
\begin{eqnarray}
\frac{d^2x^{\lambda} }{ ds^2}
+\Gamma^{\lambda}_{\mu \nu} 
\frac{dx^{\mu}}{ ds}\frac{dx^{\nu } }{ds}=
-\frac{1}{2}\frac{1}{(Q_0^2-R^{\alpha}_{\phantom{\alpha}\alpha}/6)}
\left[g^{\lambda \mu}+\frac{dx^{\lambda}}{ds}\frac{dx^{\mu}}{ds}\right]\frac{\partial(Q_0^2-R^{\alpha}_{\phantom{\alpha}\alpha}/6)}{\partial x^{\mu}}.
\label{3.50qq}
\end{eqnarray}
Moreover, if $Q_0^2\gg R^{\alpha}_{\phantom{\alpha}\alpha}$, (\ref{3.50qq}) reduces to
\begin{eqnarray}
\frac{d^2x^{\lambda} }{ ds^2}
+\Gamma^{\lambda}_{\mu \nu} 
\frac{dx^{\mu}}{ ds}\frac{dx^{\nu } }{ds}=
-\frac{1}{Q_0}
\left[g^{\lambda \mu}+\frac{dx^{\lambda}}{ds}\frac{dx^{\mu}}{ds}\right]\frac{\partial Q_0}{\partial x^{\mu}}.
\label{3.51qq}
\end{eqnarray}
In recovering (\ref{3.51qq}) again we see, just as with (\ref{3.19qq}) and (the variation of) (\ref{3.23qq}), that eikonalized trajectory solutions to wave equations can coincide with trajectories obtained by varying test particle actions.Though we should note that in this particular case this is only possible if mass is generated dynamically in a locally conformal invariant theory, while not being possible if mass is purely mechanical.

However, there is a key distinction this time. While both (\ref{3.47qq}) and (\ref{3.48qq}) are locally conformal invariant, the very derivation of (\ref{3.50qq}) and (\ref{3.51qq}) involves going from the locally conformal invariant (\ref{3.31qq}) to the eikonal condition given in (\ref{3.33qq}), a condition that is only globally conformal invariant. Consequently, (\ref{3.50qq}) is only globally conformal invariant. Moreover, this is also the case for (\ref{3.51qq})  since if we set  $Q_0^2\gg R^{\alpha}_{\phantom{\alpha}\alpha}$ the eikonal condition that would then lead to (\ref{3.51qq}) would not be (\ref{3.33qq}) but would be 
\begin{eqnarray}
\frac{dx^{\mu}}{dq}=\frac{\nabla^{\mu}T}{Q_0}
\label{3.52qq}
\end{eqnarray}
instead. And (\ref{3.52qq}) is only invariant under  $T\rightarrow T+i\alpha$ if $\alpha$ is constant. (While at the same $Q_0$ would transform into $\Omega^{-1}Q_0$, this can be accommodated by setting $d/d\bar{q}=\Omega^{-1}d/dq$ just as (\ref{3.17x})). Moreover, since (\ref{3.51qq}) has been generated starting from the  scalar field theory given in (\ref{3.30qq}),
if we did implement a local conformal transformation we would not obtain a transform of (\ref{3.48qq}) but would instead generate a new Ricci scalar contribution, just as in (\ref{3.5z}). We thus can use (\ref{3.48qq}) and (\ref{3.51qq}) provided that we only subject them to global but not to local conformal transformations. (In \cite{Horne2016} and \cite{Hobson2021}  (\ref{3.48qq}) was subjected to  a radially-dependent local conformal transformation and in consequence various difficulties for the candidate alternate conformal gravity, itself a locally conformal invariant theory, were identified. While these difficulties have been resolved in  \cite{Mannheim2021a}, we add here that it is not valid to make local conformal transformations on eikonalized trajectories in the first place, with only global transformations on them being permitted.) 

\subsection{The Zero Ricci Scalar Limit}
\label{S3i}
With $A$ slowly varying the conformally coupled wave equation given in (\ref{3.1x}) leads to (\ref{3.20qq}), viz. 
\begin{eqnarray}
-\nabla_{\mu}T\nabla^{\mu}T+\frac{1}{6}R^{\alpha}_{\phantom{\alpha}\alpha}=0,
\label{3.53qq}
\end{eqnarray}
and to the timelike eikonal condition given in (\ref{3.22qq}), viz. 
\begin{eqnarray}
\frac{dx^{\mu}}{dq}=\frac{\nabla^{\mu}T}{(-\nabla_{\nu}T\nabla^{\nu}T)^{1/2}}=\frac{\nabla^{\mu}T}{(-R^{\alpha}_{\phantom{\alpha}\alpha}/6)^{1/2}}, \qquad \frac{dx_{\mu}}{dq}\frac{dx^{\mu}}{dq}=-1.
\label{3.54qq}
\end{eqnarray}
In analog to (\ref{3.10z}) and (\ref{3.11z}) this leads to 
\begin{eqnarray}
X^{1/2}\frac{dx^{\mu}}{dq}\left[X^{1/2}\frac{\partial}{\partial x^{\mu}}\left(\frac{dx^{\lambda}}{dq}\right)
+\frac{dx^{\lambda}}{dq}\frac{1}{2X^{1/2}}\frac{\partial X}{\partial x^{\mu}}
+\Gamma^{\lambda}_{\mu\nu}X^{1/2}\frac{dx^{\nu}}{dq}\right]=-\frac{1}{2}\nabla^{\lambda}X,
\label{3.55qq}
\end{eqnarray}
and
\begin{eqnarray}
\frac{d^2x^{\lambda} }{ dq^2}
+\Gamma^{\lambda}_{\mu \nu} 
\frac{dx^{\mu}}{ dq}\frac{dx^{\nu } }{dq}=
-\frac{1}{2X}
\left[g^{\lambda \mu}+\frac{dx^{\lambda}}{dq}\frac{dx^{\mu}}{dq}\right]\frac{\partial X}{\partial x^{\mu}},
\label{3.56qq}
\end{eqnarray}
where $X=-R^{\alpha}_{\phantom{\alpha}\alpha}/6$.

On the other hand, if we set $R^{\alpha}_{\phantom{\alpha}\alpha}=0$ in (\ref{3.53qq}) these equations are replaced by the lightlike (\ref{2.5y}), (\ref{2.7y})  and (\ref{2.10y}), viz.
\begin{eqnarray}
-\nabla_{\mu}T\nabla^{\mu}T=0,
\label{3.57qq}
\end{eqnarray}
\begin{eqnarray}
\frac{dx^{\mu}}{dq}=\nabla^{\mu}T, \qquad \frac{dx_{\mu}}{dq}\frac{dx^{\mu}}{dq}=0,
\label{3.58qq}
\end{eqnarray}
and
\begin{eqnarray}
\frac{d^2x^{\lambda} }{ dq^2}
+\Gamma^{\lambda}_{\mu \nu} 
\frac{dx^{\mu}}{ dq}\frac{dx^{\nu } }{dq}=0.
\label{3.59qq}
\end{eqnarray}

Since the $R^{\alpha}_{\phantom{\alpha}\alpha}\rightarrow 0$ limit of (\ref{3.53qq}) leads continuously to (\ref{3.57qq}),  the $R^{\alpha}_{\phantom{\alpha}\alpha}\rightarrow 0$ limit of (\ref{3.56qq}) must continuously lead to (\ref{3.59qq}), and yet it does not appear to do so. To reconcile (\ref{3.56qq}) with (\ref{3.59qq}) we set
\begin{eqnarray}
X^{1/2}\frac{dx^{\mu}}{dq}=\frac{dx^{\mu}}{d\bar{q}}.
\label{3.60qq}
\end{eqnarray}
With this change in the affine parameter (\ref{3.55qq}) takes the form 
\begin{eqnarray}
\frac{dx^{\mu}}{d\bar{q}}\frac{\partial}{\partial x^{\mu}}\left(\frac{dx^{\lambda}}{d\bar{q}}\right)
-\frac{dx^{\mu}}{d\bar{q}}\frac{dx^{\lambda}}{d\bar{q}}\frac{1}{2X}\frac{\partial X}{\partial x^{\mu}}
+\frac{dx^{\mu}}{d\bar{q}}\frac{dx^{\lambda}}{d\bar{q}}\frac{1}{2X}\frac{\partial X}{\partial x^{\mu}}
+\frac{dx^{\mu}}{d\bar{q}}\Gamma^{\lambda}_{\mu\nu}\frac{dx^{\nu}}{d\bar{q}}=-\frac{1}{2}\nabla^{\lambda}X.
\label{3.61qq}
\end{eqnarray}
Thus from (\ref{3.61qq}) and (\ref{3.54qq}) we obtain
\begin{eqnarray}
\frac{d^2x^{\lambda} }{ d\bar{q}^2}
+\Gamma^{\lambda}_{\mu \nu} 
\frac{dx^{\mu}}{ d\bar{q}}\frac{dx^{\nu } }{d\bar{q}}
=-\frac{1}{2}\nabla^{\lambda}X,\quad \frac{dx_{\mu}}{ d\bar{q}}\frac{dx^{\mu } }{d\bar{q}}=-X.
\label{3.62qq}
\end{eqnarray}
And now the limit $X=-R^{\alpha}_{\phantom{\alpha}\alpha}/6\rightarrow 0$ is continuous, and in the limit (\ref{3.62qq}) recovers (\ref{3.58qq}) and (\ref{3.59qq}) in the form
\begin{eqnarray}
\frac{d^2x^{\lambda} }{ d\bar{q}^2}
+\Gamma^{\lambda}_{\mu \nu} 
\frac{dx^{\mu}}{ d\bar{q}}\frac{dx^{\nu } }{d\bar{q}}=0, \quad \frac{dx_{\mu}}{ d\bar{q}}\frac{dx^{\mu } }{d\bar{q}}=0.
\label{3.63qq}
\end{eqnarray}
The utility of (\ref{3.62qq}) is that it shows in exactly what specific way lightlike null geodesics begin to be modified as the contribution of the Ricci scalar begins to increase from zero. Moreover, the fact that we can recover the $\nabla_{\nu}T\nabla^{\nu}T\rightarrow 0$ limit continuously supports our use of $dx^{\mu}/dq=\nabla^{\mu}T/(-\nabla_{\nu}T\nabla^{\nu}T)^{1/2}$ as the eikonal condition when $\nabla_{\nu}T\nabla^{\nu}T$ is not zero.

\section{Conformal Invariant Maxwell  Field Wave Equation}
\label{S4}

In curved space the electromagnetic field is associated with the Maxwell action $I=-\tfrac{1}{4}\int d^4x (-g)^{1/2}F_{\mu\nu}F^{\mu\nu}$. On setting $F_{\mu\nu}=\nabla_{\mu}A_{\nu}-\nabla_{\nu}A_{\mu}$ this action is left invariant under the local conformal transformation  $g_{\mu\nu}(x)\rightarrow \Omega^2(x)g_{\mu\nu}(x)$, $A_{\mu}(x)\rightarrow A_{\mu}(x)$ \cite{footnote10}. Consequently, the wave equation $\nabla_{\mu}F^{\mu\nu}=0$, viz.
\begin{eqnarray}
\nabla_{\mu}(\nabla^{\mu}A^{\nu}-\nabla^{\nu}A^{\mu})=0,
\label{4.1x}
\end{eqnarray}
is locally conformal invariant. Thus just as with the locally conformal invariant scalar field case, if the background geometry is conformal to flat we can conformally transform to a flat background, and thus the eikonalized Maxwell trajectories are null geodesics.

However, if the background is not conformal to flat, it is convenient to rewrite the Maxwell equations in a form analogous to the conformally coupled scalar field wave equation. Thus, recalling the geometric identity
\begin{eqnarray}
\nabla_{\kappa}\nabla_{\nu}V_{\lambda}-\nabla_{\nu}\nabla_{\kappa}V_{\lambda}=V^{\sigma}R_{\lambda\sigma\nu\kappa}
\label{4.2x}
\end{eqnarray}
that holds for any general coordinate vector $V_{\lambda}$, 
we can rewrite (\ref{4.1x}) in the convenient form
\begin{eqnarray}
\nabla_{\mu}\nabla^{\mu}A^{\nu}-\nabla^{\nu}\nabla_{\mu}A^{\mu}+R^{\nu\alpha}A_{\alpha}=0.
\label{4.3x}
\end{eqnarray}
As we see, just as in the scalar field case discussed above and in fermion field case discussed in \cite{footnote8} and \cite{footnote9}, there is an intrinsic dependence on some of the components of the Riemann tensor.

The utility of rewriting (\ref{4.1x}) in this form is that it greatly simplifies in the covariant Lorentz gauge. Thus if set $\nabla_{\mu}A^{\mu}=0$ (\ref{4.3x}) reduces to
\begin{eqnarray}
\nabla_{\mu}\nabla^{\mu}A^{\nu}+R^{\nu\alpha}A_{\alpha}=0.
\label{4.4x}
\end{eqnarray}
To eikonalize (\ref{4.4x}) we initially introduce a set of four spacetime-dependent polarization four-vectors $\epsilon_{(a)}^{\mu}(x)$ ($a=1,2,3,4$) that  are covariantly normalized according to  $\sum_{a=1}^4\epsilon_{(a)}^{\mu}\epsilon^{\nu}_{ (a)}=g^{\mu\nu}$ as summed over the four $a$.  With this normalization we can then use a scalar amplitude to fix the overall scale of $A_{\mu}$ according to $A^{\mu}_{(a)}(x)=\epsilon^{\mu}_{(a)}(x)A(x)\exp(iT(x))$. And with this form the gauge condition takes the form 
\begin{eqnarray}
A\nabla_{\mu}\epsilon^{\mu}_{(a)}+\epsilon^{\mu}_{(a)}\nabla_{\mu}A+i A\epsilon^{\mu}_{(a)}\nabla_{\mu}T=0,
\label{4.5x}
\end{eqnarray}
while (\ref{4.4x}) takes the form
\begin{eqnarray}
&&-A\epsilon^{\nu}_{(a)}\nabla_{\mu}T\nabla^{\mu}T+\epsilon^{\nu}_{(a)}\nabla_{\mu}\nabla^{\mu}A
+A\nabla_{\mu}\nabla^{\mu}\epsilon^{\nu}_{(a)}+2\nabla_{\mu}A\nabla^{\mu}\epsilon^{\nu}_{(a)}
+AR^{\nu}_{\phantom{\nu}\alpha}\epsilon^{\alpha}_{(a)}
\nonumber\\
&&+2i\epsilon^{\nu}_{(a)}\nabla_{\mu}A\nabla^{\mu}T+iA\epsilon^{\nu}_{(a)}\nabla_{\mu}\nabla^{\mu}T+2iA\nabla_{\mu}\epsilon^{\nu}_{(a)}\nabla^{\mu}T
=0.
\label{4.6x}
\end{eqnarray}

However with $A$, $\epsilon^{\mu}_{(a)}$ and $T$ all being real the gauge condition would yield $\epsilon_{\mu (a)}\nabla^{\mu}T=0$. Then on contracting with $\epsilon^{\sigma}_{ (a)}$ we would obtain $\nabla^{\sigma}T=0$, and the eikonal function would vanish. To avoid this we instead restrict to three polarization vectors, all of which must lie on the surface orthogonal to  $\nabla^{\mu}T$, with the gauge condition $\nabla_{\mu}A^{\mu}=0$ only allowing for three independent polarization vectors. As we discuss in more detail in Sec. \ref{S10} below, as long as $\nabla_{\nu}T\nabla^{\nu}T\neq 0$ we can introduce either a timelike or a spacelike normal $n^{\mu}=\nabla^{\mu}T/(\mp \nabla_{\nu}T\nabla^{\nu}T)^{1/2}$, and this normal will be along the direction in which energy and momentum are transported normal to a wavefront. If $\nabla_{\nu}T\nabla^{\nu}T=0$ the direction of energy and momentum transport would be lightlike with $n^{\mu}=\nabla^{\mu}T$, and as had been discussed earlier, the trajectories would be null geodesics. Thus as long as $\nabla_{\nu}T\nabla^{\nu}T\neq 0$ trajectories must take support off the light cone, and will do so in a form that we will identify momentarily.

In the $\nabla_{\nu}T\nabla^{\nu}T\neq 0$ case with $n^{\mu}=\nabla^{\mu}T/(\mp \nabla_{\nu}T\nabla^{\nu}T)^{1/2}$, each of the three polarization vectors obeys $\epsilon_{\mu (a)}n^{\mu}=0$. Thus in same way as the flat space  $\boldsymbol{E}$ and $\boldsymbol{B}$ fields are orthogonal to the Poynting vector (see  Sec. \ref{S1b}), each of the three  $\epsilon_{\mu (a)}$ lies in the tangent plane normal to $n^{\mu}$. Moreover, in Sec. \ref{S10} we will identify this normal with a velocity vector according to $n^{\mu}=dx^{\mu}/dq$, a velocity that we can refer to as the eikonal velocity, with $dx^{\mu}/dq$ being the velocity with which energy and momentum are transported  (in analog to the group velocity of optics). We also note that in making this identification we have no need to restrict to large $T$ as it does not require any specification as to what $\nabla_{\nu}T\nabla^{\nu}T$ might be equal to or approximated by, other than that it be nonzero. Following the discussion given in Sec. \ref{S10} we introduce an induced  metric whose form depends on whether $n^{\mu}$ is timelike or spacelike.  For timelike $n_{\mu}n^{\mu}=-1$ we define $I^{\mu\nu}(-)=g^{\mu\nu}+n^{\mu}n^{\nu}$, while for spacelike $n_{\mu}n^{\mu}=+1$ we define $I^{\mu\nu}(+)=g^{\mu\nu}-n^{\mu}n^{\nu}$. In both cases $n_{\mu}I^{\mu\nu}(\mp)=0$, so $I^{\mu\nu}(\mp)$ also lies in the tangent plane orthogonal to the normal. With there only being three polarization vectors all of which are in the tangent plane, we  set $\sum_{a=1}^3\epsilon_{(a)}^{\mu}\epsilon^{\nu}_{(a)}=I^{\mu\nu}(\mp)$ as summed over three values for $a$. From this it follows that $\sum_{a=1}^3\epsilon_{\nu(a)}\epsilon^{\nu}_{(a)}=\sum_{a=1}^3g_{\sigma\nu}\epsilon^{\sigma}_{(a)}\epsilon^{\nu}_{(a)}=g_{\sigma \nu}I^{\sigma \nu}(\mp)=g_{\sigma\nu}(g^{\sigma\nu}\pm n^{\sigma}n^{\nu})=3$, and thus that $\sum_{a=1}^3\epsilon_{\nu(a)}\nabla^{\mu}\epsilon^{\nu}_{(a)}=0$. Thus while we cannot impose $\sum_{a=1}^4\epsilon_{(a)}^{\mu}\epsilon^{\nu}_{ (a)}=g^{\mu\nu}$ as summed over four $a$, we can impose $\sum_{a=1}^3\epsilon_{(a)}^{\mu}\epsilon^{\nu}_{ (a)}\mp n^{\mu}n^{\nu}=g^{\mu\nu}$ as summed over three $a$.

On now contracting the real part of (\ref{4.6x}) with $\epsilon_{\nu (a)}$ we obtain 
\begin{eqnarray}
&&-\nabla_{\mu}T\nabla^{\mu}T=-A^{-1}\nabla_{\mu}\nabla^{\mu}A-\frac{1}{3}\epsilon_{\nu(a)}\nabla_{\mu}\nabla^{\mu}\epsilon^{\nu}_{(a)}
-\frac{1}{3} I_{\nu}^{\phantom{\nu}\alpha}(\mp)R^{\nu}_{\phantom{\nu}\alpha}.
\label{4.7xx}
\end{eqnarray}
So far everything is exact. In the eikonal limit we take $T$ to be very large, and in particular we take $T$ to be much larger than any of the components of $A\epsilon_{\mu (a)}$ or their derivatives  \cite{footnote9a}.  Thus since the Ricci tensor term involves no $\nabla_{\mu}$  derivative, in the short wavelength limit the real part of (\ref{4.6x}) reduces to 
\begin{eqnarray}
-\epsilon^{\nu}_{(a)}\nabla_{\mu}T\nabla^{\mu}T=-R^{\nu}_{\phantom{\nu}\alpha}\epsilon^{\alpha}_{(a)},
\label{4.8x}
\end{eqnarray}
provided the Ricci tensor term is competitive with the $\nabla_{\mu}T\nabla^{\mu}T$ term. 
If we set $T=\int^xk_{\mu}dx^{\mu}$, $\nabla_{\mu}T=k_{\mu}$, then from the real parts of (\ref{4.5x}) and (\ref{4.6x}) we obtain 
\begin{eqnarray}
k_{\mu}\epsilon^{\mu}_{(a)}=0,
\label{4.9x}
\end{eqnarray}
\begin{eqnarray}
\epsilon^{\nu}_{(a)}k_{\mu}k^{\mu}
-R^{\nu\alpha}\epsilon_{\alpha (a)}=0.
\label{4.10x}
\end{eqnarray}

In passing, we note that if we were to work in light-front coordinates, then as had been discussed in Sec. \ref{S2b} we would set $T=\int k_-dx^-$, i.e., in light-front coordinates we would set $k_{\mu}=(0,k_-,0,0)$ and thus set $n_{\mu}=k_{\mu}/(\mp k_{\nu}k^{\nu})^{1/2}$. From (\ref{4.9x}) this leads to $\epsilon^{-}_{(a)}=0$. However, whether we use standard $(x^0,x^1,x^2,x^3)$ coordinates or light-front coordinates the three polarization vectors obey  $\sum_{a=1}^3\epsilon^{\mu}_{ (a)}\epsilon^{\nu}_{(a)}=I^{\mu\nu}(\mp)$. Thus  (\ref{4.8x}) yields
\begin{eqnarray}
-\nabla_{\mu}T\nabla^{\mu}T
=-\frac{1}{3}\sum_{a=1}^3R^{\alpha\beta}\epsilon_{\alpha (a)}\epsilon_{\beta (a)}=-\frac{1}{3} I_{\nu}^{\phantom{\nu}\alpha}(\mp)R^{\nu}_{\phantom{\nu}\alpha},
\label{4.11x}
\end{eqnarray}
as summed over the three polarizations.

We can now proceed as in (\ref{3.22qq}), and in the timelike case (viz. $I^{\mu\nu}(-)$) we set 
\begin{eqnarray}
\frac{dx^{\mu}}{dq}=\frac{\nabla^{\mu}T}{(-\nabla_{\nu}T\nabla^{\nu}T)^{1/2}}=\frac{\nabla^{\mu}T}{(-(1/3)\sum_{a=1}^3R^{\alpha\beta}\epsilon_{\alpha (a)}\epsilon_{\beta (a)})^{1/2}},\qquad \frac{dx_{\mu}}{dq}\frac{dx^{\mu}}{dq}=-1,
\label{4.12x}
\end{eqnarray}
again as summed over the three polarizations.
From this point the calculation follows the conformal scalar field case, and leads to 
\begin{eqnarray}
\frac{d^2x^{\lambda} }{ dq^2}+\Gamma^{\lambda}_{\mu \nu} 
\frac{dx^{\mu}}{ dq}\frac{dx^{\nu } }{dq}
=-\frac{1}{2(-(1/3)\sum_{a=1}^3R^{\alpha\beta}\epsilon_{\alpha (a)}\epsilon_{\beta (a)})}
\left[g^{\lambda \mu}+\frac{dx^{\lambda}}{dq}\frac{dx^{\mu}}{dq}\right]\frac{\partial (-(1/3)\sum_{a=1}^3R^{\gamma\delta}\epsilon_{\gamma (a)}\epsilon_{\delta (a)})}{\partial x^{\mu}}.
\label{4.13x}
\end{eqnarray}
This relation and its (\ref{4.21xy}) generalization constitute our main result in the Maxwell case. And just as in the conformally coupled scalar field case, in the Maxwell case there again are departures from pure geodesic behavior in the presence of curvature provided that, as noted above, the background geometry is not conformal to flat. In general, (\ref{4.13x}) would hold in any background geometry that is not conformal to flat, with the right-hand side of (\ref{4.13x}) not vanishing. However in a Schwarzschild de Sitter geometry,  viz. a particular geometry that is not conformal to flat, the Ricci tensor  $R_{\mu\nu}$ is equal to $-3Kg_{\mu\nu}$. Thus  with the polarization vectors normalized so that   $\sum_{a=1}^3\epsilon_{\mu (a)}\epsilon_{\nu (a)}=I_{\mu\nu}(\mp)$, the quantity $(1/3)\sum_{a=1}^3R^{\alpha\beta}\epsilon_{\alpha (a)}\epsilon_{\beta (a)}$ would then be equal to the constant $-3K$, i.e.,  not $-4K$ since there are only three polarization states not four. In that case even though the right-hand side of (\ref{4.13x}) would then vanish, nonetheless since $K$ is nonzero the second equation in (\ref{4.12x}) would provide  what would then be an integration constant for (\ref{4.13x}). Thus for Schwarzschild de Sitter (\ref{4.12x}) can provide an integration constant, though it would not provide one for pure Schwarzschild  itself \cite{footnote9b}. Consequently, gravitational bending of light by the Sun is not modified from the standard pure geodesic-based Ricci flat discussion, but when the $\sum_{a=1}^3R^{\alpha\beta}\epsilon_{\alpha (a)}\epsilon_{\beta (a)}$ term is not negligible there in principle are modifications. We discuss further aspects of Schwarzschild de Sitter modifications below.

It is also of interest to note that covariantizing the flat space Maxwell equations would miss the $R^{\nu\alpha}A_{\alpha}$ term that appears in (\ref{4.3x}). Specifically, in flat space we can write the wave equation in the form 
\begin{eqnarray}
\partial_{\mu}\partial^{\mu}A^{\nu}-\partial_{\mu}\partial^{\nu}A^{\mu}=\partial_{\mu}\partial^{\mu}A^{\nu}-\partial^{\nu}\partial_{\mu}A^{\mu}
+(\partial^{\nu}\partial^{\mu}-\partial^{\mu}\partial^{\nu})A_{\mu}=0.
\label{4.14x}
\end{eqnarray}
Now while we can drop the $(\partial^{\nu}\partial^{\mu}-\partial^{\mu}\partial^{\nu})A_{\mu}$ term in flat space, in curved space covariant derivatives do not commute, and thus we have to generalize $(\partial^{\nu}\partial^{\mu}-\partial^{\mu}\partial^{\nu})A_{\mu}$ to $(\nabla^{\nu}\nabla^{\mu}-\nabla^{\mu}\nabla^{\nu})A_{\mu}$ and this quantity is not zero, being instead equal to $R^{\nu\mu}A_{\mu}$. Thus in covariantizing a flat space expression the order in which derivatives are applied cannot be ignored.

As well as needing to covariantize correctly, we need to maintain local conformal invariance. However, while the gauge condition $\nabla_{\mu}A^{\mu}=0$ that we use is general coordinate invariant it is not locally conformal invariant. Alternatively,  the gauge condition
\begin{eqnarray}
\nabla_{\mu}A^{\mu}-\frac{1}{4}A^{\mu}g^{\alpha\beta}\partial_{\mu}g_{\alpha\beta}=0
\label{4.15xx}
\end{eqnarray}
is locally conformal invariant. However, it is not general coordinate invariant. Given this latter gauge condition we replace (\ref{4.3x}) by 
\begin{eqnarray}
\nabla_{\mu}\nabla^{\mu}A^{\nu}-\nabla^{\nu}\nabla_{\mu}A^{\mu}+\frac{1}{4}\nabla^{\nu}\left[A^{\mu}g^{\alpha\beta}\partial_{\mu}g_{\alpha\beta}\right]
-\frac{1}{4}\nabla^{\nu}\left[A^{\mu}g^{\alpha\beta}\partial_{\mu}g_{\alpha\beta}\right]+R^{\nu\alpha}A_{\alpha}=0,
\label{4.16xx}
\end{eqnarray}
which then reduces to 
\begin{eqnarray}
&&\nabla_{\mu}\nabla^{\mu}A^{\nu}-\frac{1}{4}\nabla^{\nu}\left[A^{\mu}g^{\alpha\beta}\partial_{\mu}g_{\alpha\beta}\right]+R^{\nu\alpha}A_{\alpha}=0
\label{4.17xx}
\end{eqnarray}
when (\ref{4.15xx}) is imposed. The utility of this gauge is that it allows us to conformally transform the Maxwell equations using 
\begin{eqnarray}
R_{\mu\nu}\rightarrow \Omega^{-2}R_{\mu\nu}+2\Omega^{-3}\nabla_{\mu}\nabla_{\nu}\Omega-4\Omega^{-4}\nabla_{\mu}\Omega\nabla_{\nu}\Omega
+g_{\mu\nu}[\Omega^{-3}\nabla_{\alpha}\nabla^{\alpha}\Omega+\Omega^{-4}\nabla_{\alpha}\Omega\nabla^{\alpha}\Omega]
\label{4.18xx}
\end{eqnarray}
without needing to introduce a new gauge condition. Thus if the background geometry is conformal to flat we can conformally transform  (\ref{4.17xx}) to the flat space $\partial_{\mu}\partial^{\mu}A^{\nu}=0$, with (\ref{4.15xx}) becoming $\partial_{\mu}A^{\mu}=0$, viz. the standard flat space Maxwell equations in the flat space Lorentz gauge.

We should also note that it is possible to satisfy (\ref{4.8x}) without needing to constrain the polarization vectors beyond the fact that they have to be slowly varying in order for us to have been able to derive (\ref{4.8x}) in the first place. Specifically, if $R^{\mu}_{\phantom{\mu}\nu}$ is diagonal in its indices and $R^{\theta}_{\phantom{\theta}\theta} =   R^{\phi}_{\phantom{\phi}\phi}$ (this being the case for the Schwarzschild metric for instance), it follows that (\ref{4.8x}) reduces to 
\begin{eqnarray}
-\epsilon^{\theta}_{(a)}\nabla_{\mu}T\nabla^{\mu}T
+R^{\theta}_{\phantom{\theta}\theta}\epsilon^{\theta}_{(a)}=0, \quad -\epsilon^{\phi}_{(a)}\nabla_{\mu}T\nabla^{\mu}T
+R^{\phi}_{\phantom{\phi}\phi}\epsilon^{\phi}_{(a)}=0,\quad
\nabla_{\mu}T\nabla^{\mu}T
=R^{\theta}_{\phantom{\theta}\theta}=R^{\phi}_{\phantom{\phi}\phi},
\label{4.19xx}
\end{eqnarray}
with both $\epsilon^{\theta}_{(a)}$ and $\epsilon^{\phi}_{(a)}$ being unconstrained. In the event that (\ref{4.8x}) might overconstrain the polarization vectors in some specific  background geometries [(\ref{4.8x}) could be viewed as an eigenvalue problem for $R^{\nu}_{\phantom{\nu}\alpha}/\nabla_{\mu}T\nabla^{\mu}T$] we would then have to take into account a spatial variation of $\epsilon_{\nu (a)}$, as this is allowed by the $-(1/3)\epsilon_{\nu(a)}\nabla_{\mu}\nabla^{\mu}\epsilon^{\nu}_{(a)}$ term that is present in the full unapproximated (\ref{4.7xx}). In fact we could take the full (\ref{4.7xx})  into account anyway by proceeding without any short wavelength approximation for $T$ at all. Specifically, if we denote the right-hand side of (\ref{4.7xx}) by 
\begin{eqnarray}
X=-A^{-1}\nabla_{\mu}\nabla^{\mu}A-\frac{1}{3}\epsilon_{\nu(a)}\nabla_{\mu}\nabla^{\mu}\epsilon^{\nu}_{(a)}
-\frac{1}{3} I_{\nu}^{\phantom{\nu}\alpha}(\mp)R^{\nu}_{\phantom{\nu}\alpha},
\label{4.20xy}
\end{eqnarray}
then instead of focussing solely on the contribution of  the Ricci tensor, (\ref{4.13x}) would generalize to
\begin{eqnarray}
\frac{d^2x^{\lambda} }{ dq^2}+\Gamma^{\lambda}_{\mu \nu} 
\frac{dx^{\mu}}{ dq}\frac{dx^{\nu } }{dq}
=-\frac{1}{2X}
\left[g^{\lambda \mu}+\frac{dx^{\lambda}}{dq}\frac{dx^{\mu}}{dq}\right]\frac{\partial X}{\partial x^{\mu}}.
\label{4.21xy}
\end{eqnarray}
As such, (\ref{4.21xy}) is exact without approximation, provided only that we identify  $dx^{\mu}/dq$ with a  normal $n^{\mu}=\nabla^{\mu}T/(\mp \nabla_{\nu}T\nabla^{\nu}T)^{1/2}$ to a propagating wavefront. We shall return to this general solution in Sec. \ref{S10}.

While we have shown that there can be departures from null geodesic behavior if the background geometry is not conformal to flat, in a sense more interesting in such a case is that  (\ref{3.22qq}) and (\ref{4.12x})  are both of the timelike form
\begin{eqnarray}
\frac{dx_{\mu}}{dq}\frac{dx^{\mu}}{dq}=-1.
\label{4.22xy}
\end{eqnarray}
Thus in the massless particle curved space eikonal approximation it is possible for modes to propagate off the light cone  in a conformal invariant theory in which there is no intrinsic mass.  With $ds^2=-g_{\mu\nu}dx^{\mu}dx^{\nu}$, (\ref{4.22xy}) leads to timelike $ds^2$, and thus modes that obey (\ref{4.22xy}) will nicely be causally inside the $ds^2=0$ light cone \cite{footnote11}. In addition, we note that the modification to geodesic behavior in the light wave case depends on the polarization of the light wave. As such, (\ref{4.13x}) is somewhat reminiscent of (\ref{1.5x}) as they both involve both spin and geometric factors.

Propagation of light off the light cone is actually a quite general phenomenon, and already occurs in flat space electromagnetism. Specifically,  in the presence of a current source  the Maxwell equations take the form $\partial_{\mu}\partial^{\mu}A^{\nu}=J^{\nu}$ in the Lorentz gauge $\partial_{\mu}A^{\mu}=0$, and admit of no solution of the  form $A^{\nu}=\epsilon^{\nu}\exp(ik\cdot x)$ where $k_{\mu}k^{\mu}=0$. However, if the source is localized one then does get such lightlike solutions far from the source. But if the ray enters a refractive medium with refractive index $n$, then in the medium the ray travels with a velocity $c/n$ and thus not at velocity $c$. As discussed earlier,  the refractive index is often spatially dependent, and in optics the eikonal approximation with a spatially-dependent eikonal phase was developed for this very purpose \cite{Born1959}. In curved space there is also an effective refractive index, one produced by gravity itself as curvature causes spacetime to become a medium. This gravitational medium will then take a light ray off (the covariantized version of) the light cone that it otherwise would have travelled on in the absence of gravity. The role of the $\sum_{a=1}^3R^{\alpha\beta}\epsilon_{\alpha}\epsilon_{\beta}$ and $R^{\alpha}_{\phantom{\alpha}\alpha}$ terms in the vector and scalar cases is thus akin to that of a refractive medium, and in this paper we are exploring what effects such a medium could  in principle, even if not necessarily in practice, provide.

\section{Departures from  Geodesic Behavior in the Schwarzschild de Sitter Case}
\label{S5}

We consider the geometry exterior to a static, spherically symmetric system with line element 
\begin{eqnarray}
ds^2=B(r)dt^2-A(r)dr^2 -r^2d\theta^2-r^2\sin^2\theta d\phi^2. 
\label{5.1x}
\end{eqnarray}
For the Schwarzschild case where $R_{\mu\nu}=0$ the exterior metric is given by $B(r)=1/A(r)=1-2MG/r$, while for the Schwarzschild de Sitter case where $R_{\mu\nu}=-3Kg_{\mu\nu}$, $R^{\alpha}_{\phantom{\alpha}\alpha}=-12K$, the exterior metric is given by $B(r)=1/A(r)=1-2MG/r-Kr^2$. For a general $B(r)=1/A(r)$ metric all of the non-trivial components of the Weyl  tensor are proportional to $B^{\prime\prime}-2B^{\prime}/r+2(B-1)/r^2$, and  this quantity takes the nonzero value $-12MG/r^3$ when $B(r)=1/A(r)=1-2MG/r-Kr^2$. Consequently, the geometry is not conformal to flat and for the propagation of massless modes (\ref{3.19qq}) and (\ref{4.13x}) apply. In both of the Schwarzschild and Schwarzschild de Sitter  geometries the right-hand side of  the massless scalar field (\ref{3.19qq}) vanishes, while for an appropriate choice of polarization vectors (viz. $(1/3)\sum_{a=1}^3R^{\alpha\beta}\epsilon_{\alpha (a)}\epsilon_{\beta (a)}=-3K$) the right-hand side of the massless vector field  (\ref{4.13x}) vanishes too. However, while both (\ref{3.19qq}) and (\ref{4.13x}) then reduce to the null geodesic equation given in (\ref{1.6x}), we would have to integrate the null geodesic equation subject to  the $(dx_{\mu}/dq)(dx^{\mu}/dq)=-1$ constraint given in (\ref{3.22qq}) and (\ref{4.12x}). While for massive particles we should use (\ref{3.35qq}), for  $m^2 \gg R^{\alpha}_{\phantom{\alpha}\alpha}/6$ (\ref{3.35qq}) reduces to the standard massive particle (\ref{1.6x}) as written with a nonzero proper time. We can thus treat massless and massive cases simultaneously since they both involve a non-vanishing $(dx_{\mu}/dq)(dx^{\mu}/dq)$, whose value will serve as an integration constant.

Starting with the null geodesic equation given in (\ref{1.6x}), or also with the massive geodesic equation given in (\ref{1.2x}) with $q$ replacing $s$ in it we obtain \cite{Weinberg1972}
\begin{eqnarray}
&&\frac{d^2r}{dq^2}+\frac{A^{\prime}}{2A}\left(\frac{dr}{dq}\right)^2-\frac{r}{A}\left(\frac{d\theta}{dq}\right)^2 -\frac{r\sin^2\theta}{A}\left(\frac{d\phi}{dq}\right)^2+\frac{B^{\prime}}{2A}\left(\frac{dt}{dq}\right)^2=0,
\nonumber\\
&&\frac{d^2\theta}{dq^2}+\frac{2}{r}\frac{d\theta}{dq}\frac{dr}{dq}-\sin\theta\cos\theta\left(\frac{d\phi}{dq}\right)^2=0,
\nonumber\\
&&\frac{d^2\phi}{dq^2}+\frac{2}{r}\frac{d\phi}{dq}\frac{dr}{dq}+2\cot\theta\frac{d\phi}{dq}\frac{d\theta}{dq}=0,
\nonumber\\
&&\frac{d^2t}{dq^2}+\frac{B^{\prime}}{B}\frac{dt}{dq}\frac{dr}{dq}=0.
\label{5.2x}
\end{eqnarray}
On setting $\theta=\pi/2$ these equations integrate to
\begin{align}
&\frac{dt}{dq}=\frac{1}{B},\qquad r^2\frac{d\phi}{dq}=J,\qquad \left(\frac{dr}{dq}\right)^2=\frac{1}{AB}-\frac{J^2}{Ar^2}-\frac{E}{A},
\label{5.3x}
\end{align}
where the angular momentum $J$ and the energy $E$ are integration constants. (The integration constant for $dt/dq$ has been set to one.)
From these relations we obtain
\begin{eqnarray}
ds^2=-g_{\mu\nu}\frac{dx^{\mu}}{dq}\frac{dx^{\nu}}{dq}dq^2=Edq^2,
\label{5.4x}
\end{eqnarray}
and on comparing with (\ref{3.29qq}) we reinterpret $q$ as $q^{\prime}$.
Now while for material particles with nonzero mass we take the integration constant $E$ to be nonzero and positive, for massless particles one ordinarily  sets $E=0$ so as to put them on the light cone. Now while one can do this in the Schwarzschild case, one cannot do this in the Schwarzschild de Sitter case since according to (\ref{3.22qq}) $-(dx_{\mu}/dq)(dx^{\mu}/dq)$ is greater than zero (and analogously for (\ref{4.12x})). In the Schwarzschild de Sitter case then geodesic behavior for massless particles is not compatible with the  particles being on the light cone. While these remarks would appear to pertain to pure de Sitter as well since for it $R^{\alpha}_{\phantom{\alpha}\alpha}=-12K$ is nonzero, unlike Schwarzschild de Sitter a pure de Sitter geometry is conformal to flat. Thus for pure de Sitter we can set $E=0$ for massless particles, and can take $E$ to be nonzero for massive ones.

For a particle of energy $E$ coming in from an asymptotically Minkowski geometry the gravitational bending of the trajectory of a particle in the solution given in (\ref{5.3x})  is of the form \cite{Weinberg1972}
\begin{eqnarray}
\left(\frac{d\phi}{dr}\right)^2=\frac{ABJ^2}{r^2(r^2-J^2B-r^2EB)},\quad 
\phi(r)-\phi(\infty)=\int _r^{\infty} dr \frac{A^{1/2}(r)B^{1/2}(r)r_0(1-EB(r_0))^{1/2}}{r[r^2B(r_0)-r_0^2B(r)+EB(r)B(r_0)(r_0^2-r^2)]^{1/2}},
\label{5.5x}
\end{eqnarray}
where $r_0$ is the distance of closest approach to the gravitational source located at $r=0$. When $A=1/B$ and $E=0$ (\ref{5.5x}) reduces to
\begin{eqnarray}
\left(\frac{d\phi}{dr}\right)^2=\frac{J^2}{r^2(r^2-J^2B)},\quad \phi(r)-\phi(\infty)=\int _r^{\infty} dr \frac{r_0}{r[r^2B(r_0)-r_0^2B(r)]^{1/2}},
\label{5.6x}
\end{eqnarray}
the standard geodesic-based bending formula in a curved space.

As written, (\ref{5.6x}) has a well-known shortcoming. If we set $B(r)=1-2MG/r-Kr^2$ we find that the $Kr^2$ term drops out of $\phi(r)-\phi(\infty)$ identically \cite{footnote11a}. However, it is not actually valid to use (\ref{5.5x}) in the Schwarzschild de Sitter or pure de Sitter cases \cite{Rindler2009,Ishak2010} since (\ref{5.5x}) was derived on the assumption that far from the source the geometry is asymptotically flat, and for Schwarzschild de Sitter or pure de Sitter this is not the case. Moreover, in  \cite{Rindler2009,Ishak2010} Rindler and Ishak  provided an alternate prescription for determining gravitational bending that does take the non-asymptotic flatness of the Schwarzschild de Sitter and pure de Sitter geometries into account. While these remarks are valid, to them we should add that for Schwarzschild de Sitter one should anyway not be using (\ref{5.6x}) in the first place. Rather, with $R_{\mu\nu}=-3Kg_{\mu\nu}$ and  $(1/3)\sum_{a=1}^3R^{\alpha\beta}\epsilon_{\alpha (a)}\epsilon_{\beta(a)}=-3K$ both being nonzero, for Schwarzschild de Sitter one should use (\ref{5.5x}) with $E=1$ for massless scalar rays and massless vector rays, and now the dependence on $K$ does not drop out. However, even so one still has to incorporate the asymptotic non-flatness considerations discussed in  \cite{Rindler2009,Ishak2010} as well. The discussion of gravitational bending is thus instructive since while one does not need to depart from $(dx_{\mu}/dq)(dx^{\mu}/dq)=0$ in the pure Schwarzschild case, one does have to do so in Schwarzschild de Sitter. 

In passing we should note that even without the asymptotic considerations of \cite{Rindler2009} and \cite{Ishak2010},  as noted in\cite{Kraniotis2011,Sultana2013,Kraniotis2014,Arakida2021} the de Sitter contribution to gravitational bending does not drop out of the rotating source generalization of (\ref{5.6x}) \cite{footnote11b}.

For particles with mass we can discuss bound orbits, and for circular bound orbits the analysis is different from the one just given for the gravitational bending of massless light rays, since with $dr/dq$ set to zero in a circular orbit we do not need to integrate the $d^2r/dq^2$ equation given in (\ref{5.2x}) or address the appropriate value for its integration constant $E$. Specifically, if we take $\theta=\pi/2$ and set $dr/dq=0$ in (\ref{5.2x}) we obtain \cite{footnote12}
\begin{eqnarray}
&&-\frac{r}{A}\left(\frac{d\phi}{dq}\right)^2+\frac{B^{\prime}}{2A}\left(\frac{dt}{dq}\right)^2=0,\qquad \frac{d^2\phi}{dq^2}=0,\qquad \frac{d^2t}{dq^2}=0.
\label{5.7x}
\end{eqnarray}
In an orbit of fixed radius $R$ we have
\begin{eqnarray}
\left(\frac{d\phi}{dt}\right)^2=\frac{B^{\prime}(R)}{2R},
\qquad \frac{d\phi}{dq}=C(R),\qquad \frac{dt}{dq}=D(R),
\qquad \left(\frac{d\phi}{dt}\right)^2=\frac{C^2(R)}{D^2(R)}=\frac{B^{\prime}(R)}{2R},
\label{5.8x}
\end{eqnarray}
where $C(R)$ and $D(R)$ are integration constants.  From (\ref{5.8x}) the orbital velocity $v$ is given by
\begin{eqnarray}
v=R\frac{d\phi}{dt}=\left(\frac{RB^{\prime}(R)}{2}\right)^{1/2}.
\label{5.9x}
\end{eqnarray}
We recognize (\ref{5.9x}) as the standard expression for massive particle circular orbits with $v^2<c^2$, i.e., $v<1$.  

Now while we did not need to utilize the condition $(dx_{\mu}/dq)(dx^{\mu}/dq)=-1$ given in (\ref{3.22qq}) in order to derive the massive particle (\ref{5.9x}), nonetheless (\ref{3.22qq}) still holds. Moreover, since for massive particles there is a nonzero integration constant $E$ to begin with, and since we can absorb its magnitude  via (\ref{3.28qq}), the discussion is actually the standard one.  Thus on absorbing $E$, for the metric given in (\ref{5.1x})  the $(dx_{\mu}/dq)(dx^{\mu}/dq)=-1$ condition  takes the form
\begin{eqnarray}
-B\left(\frac{dt}{dq}\right)^2+A\left(\frac{dr}{dq}\right)^2+r^2\left(\frac{d\phi}{dq}\right)^2=-1
\label{5.10x}
\end{eqnarray}
when $\theta=\pi/2$. In a circular orbit this requires that
\begin{eqnarray}
B(R)D^2(R)-R^2C^2(R)=1,
\label{5.11x}
\end{eqnarray}
to thus constrain the integration constants given in (\ref{5.8x}) according to
\begin{eqnarray}
D^2(R)=\frac{2}{2B(R)-RB^{\prime}(R)}=\frac{1}{B(R)-v^2},\qquad C^2(R)=\frac{B^{\prime}(R)}{2RB(R)-R^2B^{\prime}(R)}=\frac{v^2}{R^2(B(R)-v^2)}.
\label{5.12x}
\end{eqnarray}
With both $D^2(R)$ and $C^2(R)$ being positive, $v^2$ is constrained to be less than $B(R)$. If $B(R)$ is itself less than one then $v^2<1$. With $B(R)=1-2MG/R-KR^2$ this can readily be achieved with positive $MG$ and $K$.

\section{Departures from Geodesic Behavior in the General Static Spherically Symmetric Massless Case}
\label{S6}

For the conformally coupled massless scalar field and massless vector field the trajectories are given by (\ref{3.19qq}) and (\ref{4.13x}) as constrained by (\ref{3.22qq}) no matter what the values of the Ricci scalar or tensor, provided the background geometry is not conformal to flat.  With the general static, spherically symmetric  metric  given in (\ref{5.1x}) not being conformal to flat, the equations that replace (\ref{5.2x}) and (\ref{5.4x}) are of the form
\begin{align}
\frac{d^2r}{dq^2}+\frac{A^{\prime}}{2A}\left(\frac{dr}{dq}\right)^2-\frac{r}{A}\left(\frac{d\theta}{dq}\right)^2 -\frac{r\sin^2\theta}{A}\left(\frac{d\phi}{dq}\right)^2+\frac{B^{\prime}}{2A}\left(\frac{dt}{dq}\right)^2&=-\frac{1}{2X}\left(\frac{1}{A(r)}+\left(\frac{dr}{dq}\right)^2\right)\frac{d X}{dr},
\nonumber\\
\frac{d^2\theta}{dq^2}+\frac{2}{r}\frac{d\theta}{dq}\frac{dr}{dq}-\sin\theta\cos\theta\left(\frac{d\phi}{dq}\right)^2&=
-\frac{1}{2X}\frac{d\theta}{dq}\frac{dr}{dq}\frac{d X}{dr},
\nonumber\\
\frac{d^2\phi}{dq^2}+\frac{2}{r}\frac{d\phi}{dq}\frac{dr}{dq}+2\cot\theta\frac{d\phi}{dq}\frac{d\theta}{dq}
&=-\frac{1}{2X}\frac{d\phi}{dq}\frac{dr}{dq}\frac{d X}{dr},
\nonumber\\
\frac{d^2t}{dq^2}+\frac{B^{\prime}}{B}\frac{dt}{dq}\frac{dr}{dq}
&=-\frac{1}{2X}\frac{dt}{dq}\frac{dr}{dq}\frac{d X}{dr},
\label{6.1x}
\end{align}
\begin{eqnarray}
g_{\mu\nu}\frac{dx^{\mu}}{dq}\frac{dx^{\nu}}{dq}=-B(r)\left(\frac{dt}{dq}\right)^2+A(r)\left(\frac{dr}{dq}\right)^2+r^2\left(\frac{d\theta}{dq}\right)^2
+r^2\sin^2\theta\left(\frac{d\phi}{dq}\right)^2=-1,
\label{6.2x}
\end{eqnarray}
where $X=-R^{\alpha}_{\phantom{\alpha}\alpha}/6$ or $X=-(1/3)\sum_{a=1}^3R^{\alpha\beta}\epsilon_{\alpha (a)}\epsilon_{\beta (a)}$.
With $\theta$ being fixed at $\pi/2$, the equations for $d\phi/dq$ and $dt/dq$ can readily be integrated, and with $dr/dq$ then being determinable  from (\ref{6.2x}) the general solution is of the form 
\begin{align}
&\left(\frac{dt}{dq}\right)^2=\frac{d_1}{B^2X},\qquad \left(\frac{d\phi}{dq}\right)^2=\frac{d_2}{r^4X},\qquad \left(\frac{dr}{dq}\right)^2=\frac{d_1r^2-d_2B-r^2BX}{r^2ABX},
\label{6.3x}
\end{align}
where $d_1$ and $d_2$ are integration constants.
In the solution given in (\ref{6.3x}) gravitational bending is  determined from
\begin{align}
&\left(\frac{d\phi}{dr}\right)^2=\frac{d_2AB}{r^2(d_1r^2-d_2B-r^2BX)}.
\label{6.4x}
\end{align}
With $A$ and $B$ both being positive, the reality of the solution requires that
\begin{align}
\frac{d_1}{X}>0,\qquad \frac{d_2}{X}>0,\qquad \frac{d_1r^2-d_2B-r^2BX}{X}>0.
\label{6.5x}
\end{align}
If these conditions are not satisfied (conditions that could restrict the range of allowed values of $r$), eikonalization is not possible. And when these conditions are satisfied one should in general then determine the dependence of $\phi$ on $r$  for massless rays from (\ref{6.4x}), and not from the value for $d\phi/dr$ given in (\ref{5.5x}) when evaluated at $E=0$, viz.
\begin{eqnarray}
\left(\frac{d\phi}{dr}\right)^2=\frac{ABJ^2}{r^2(r^2-J^2B)}.
\label{6.6x}
\end{eqnarray}

\section{Departures from  Geodesic Behavior in the Conformal Gravity Case}
\label{S7}

In static, spherically symmetric geometries there is another situation in which the above analysis is of relevance, namely gravitational bending in the conformal gravity theory. Conformal gravity is based on the action $I_{\rm W}=-\alpha_g\int d^4x(-g)^{1/2}C_{\lambda\mu\nu\kappa}
C^{\lambda\mu\nu\kappa}$, where $C_{\lambda\mu\nu\kappa}$ is the Weyl tensor and $\alpha_g$ is a dimensionless  gravitational coupling constant. As with the other conformal invariant theories discussed above, $I_{\rm W}$ is invariant under $g_{\mu\nu}\rightarrow \Omega^2(x)g_{\mu\nu}(x)$ with arbitrary local $\Omega(x)$. And if we are going to impose local conformal invariance for all fields  we must then also use conformal invariant scalar or vector field wave equations for the fields coupled to conformal gravity. In the static, spherically symmetric case we must thus use (\ref{6.3x}) and (\ref{6.4x}), with the gravitational equations of the conformal gravity theory then supplying us with explicit expressions for $A$, $B$ and $X$.

To this end the conformal gravity gravitational equations of motion can be written in the compact form \cite{Mannheim2006}
\begin{eqnarray}                                                                               
4\alpha_g[2\nabla_{\kappa}\nabla_{\lambda}C^{\mu\lambda\nu\kappa}-R_{\lambda \kappa}C^{\mu\lambda\nu\kappa}]=T^{\mu\nu}.
\label{7.1z}
\end{eqnarray}      
In the static, spherically symmetric situation (\ref{7.1z}) can be solved exactly \cite{Mannheim1989,Mannheim1994}, and since such geometries are not conformal to flat, in them the Weyl tensor does not vanish. For such  geometries the imposition of conformal invariance on (\ref{5.1x}) allows us to set $A=1/B$ as a kinematic condition \cite{Mannheim1989}, with all nonzero components of the Weyl tensor then being proportional to $B^{\prime\prime}-2B^{\prime}/r+2(B-1)/r^2$. With $A=1/B$ the gravitational equations of motion associated with the metric given in (\ref{5.1x}) reduce exactly without any approximation at all to the remarkably compact 
\begin{eqnarray}                                                                               
&&B^{\prime\prime\prime\prime}+\frac{4}{r}B^{\prime\prime\prime}= \nabla^4 B(r) = \frac{3}{4\alpha_g B(r)}\left(T^0_{{\phantom 0} 0}
-T^r_{{\phantom r} r}\right) =f(r),
\nonumber\\
&&
\frac{1}{3r^4}(1+y^3y^{\prime\prime})=\frac{T^r_{{\phantom r} r}}{4\alpha_g}.
\label{7.2z}
\end{eqnarray}      
Here the first equation in (\ref{7.2z}) serves to define $f(r)$, and in the second we have set $y^2=r^2B^{\prime}-2rB$. 

The solution to the first equation in (\ref{7.2z}) can be determined in closed form and is given by  \cite{Mannheim1994} 
\begin{eqnarray}
B(r)&=&w-Kr^2-\frac{r}{2}\int_0^r
dr^{\prime}r^{\prime 2}f(r^{\prime})
-\frac{1}{6r}\int_0^r
dr^{\prime}r^{\prime 4}f(r^{\prime})
-\frac{1}{2}\int_r^{\infty}
dr^{\prime}r^{\prime 3}f(r^{\prime})
-\frac{r^2}{6}\int_r^{\infty}
dr^{\prime}r^{\prime }f(r^{\prime}),
\nonumber\\
B^{\prime}(r)&=&-2Kr-\frac{1}{2}\int_0^r
dr^{\prime}r^{\prime 2}f(r^{\prime})
+\frac{1}{6r^2}\int_0^rdr^{\prime}r^{\prime 4}f(r^{\prime})
-\frac{r}{3}\int_r^{\infty}
dr^{\prime}r^{\prime }f(r^{\prime}),
\nonumber\\
B^{\prime\prime}(r)&=&-2K
-\frac{1}{3r^3}\int_0^rdr^{\prime}r^{\prime 4}f(r^{\prime})
-\frac{1}{3}\int_r^{\infty}
dr^{\prime}r^{\prime }f(r^{\prime}),
\nonumber\\
B^{\prime\prime\prime}(r)&=&\frac{1}{r^4}\int_0^rdr^{\prime}r^{\prime 4}f(r^{\prime}),
\nonumber\\
B^{\prime\prime\prime\prime}(r)&=&-\frac{4}{r^5}\int_0^rdr^{\prime}r^{\prime 4}f(r^{\prime})+f(r),
\label{7.3z}
\end{eqnarray}                                 
from which it follows that  (see e.g. \cite{Horne2016})                      
\begin{eqnarray}                                 
\frac{1}{12r^4}\left[4+\int_0^r
dr^{\prime}r^{\prime 2}f(r^{\prime})\int_0^r
dr^{\prime}r^{\prime 4}f(r^{\prime}) -\left(\int_r^{\infty}
dr^{\prime}r^{\prime 3}f(r^{\prime})-2w\right)^2\right]=\frac{T^r_{{\phantom r} r}}{4\alpha_g}.
\label{7.4x}
\end{eqnarray}                                 
In (\ref{7.3z}) we have given derivatives of $B(r)$ so that it can readily be checked that the $f(r)$ equation in (\ref{7.2z}) is indeed obeyed. Also we have included a $B=w-Kr^2$ contribution as it satisfies $\nabla^4 B(r)=0$ identically. On evaluating the non-vanishing components of the Weyl tensor for the $B(r)$ given in (\ref{7.3z}) we obtain
\begin{eqnarray}
B^{\prime\prime}-\frac{2}{r}B^{\prime}+\frac{2}{r^2}(B-1)=-\frac{1}{r^3}\int_0^rdr^{\prime}r^{\prime 4}f(r^{\prime})-\frac{1}{r^2}\int_r^{\infty}
dr^{\prime}r^{\prime 3}f(r^{\prime})+\frac{2}{r^2}(w-1),
\label{7.5x}
\end{eqnarray}
and confirm that the $-Kr^2$ contribution cancels identically, just as it should since a de Sitter geometry is conformal to flat. We note the  absence of any $\int_0^rdr^{\prime}r^{\prime 2}f(r^{\prime})$ term in (\ref{7.5x}), a point we will return to below.

For the metric given in (\ref{5.1x}) the various components of the Ricci tensor take the form
\begin{align}
R_{rr}&=\frac{B^{\prime\prime}}{2B}+\frac{B^{\prime}}{rB},\qquad R_{\theta\theta}=-1+rB^{\prime}+B,
\nonumber\\
R_{\phi\phi}&=\sin^2\theta R_{\theta\theta},\qquad
R_{tt}=-\frac{B^{\prime\prime}B}{2}-\frac{B^{\prime}B}{r},
\nonumber\\
R^{\alpha}_{\phantom{\alpha}\alpha}&=B^{\prime\prime}+\frac{4B^{\prime}}{r}+\frac{2B}{r^2}-\frac{2}{r^2}
\label{7.6x}
\end{align}
when $A=1/B$. From (\ref{7.3z}) and (\ref{7.6x}) we could determine the form of $R^{\alpha\beta}\epsilon_{\alpha (a)}\epsilon_{\beta (a)}$ in any specified polarization state, but we content ourselves here by only determining $R^{\alpha}_{\phantom{\alpha}\alpha}$, and find that it is given by
\begin{align}
R^{\alpha}_{\phantom{\alpha}\alpha}&=-\frac{3}{r}\int_0^r
dr^{\prime}r^{\prime 2}f(r^{\prime})-\frac{1}{r^2}\int_r^{\infty}
dr^{\prime}r^{\prime 3}f(r^{\prime})-2\int_r^{\infty}
dr^{\prime}r^{\prime }f(r^{\prime})-12K+\frac{2}{r^2}(w-1).
\label{7.7x}
\end{align}
Thus for a given $f(r)$, to determine the gravitational bending of massless rays, in the general (\ref{6.4x}) one should, modulo a modification that is given  in (\ref{7.15x}) and (\ref{7.16x}) below, use the relation given in (\ref{7.3z}) for $B(r)$ and $A(r)=1/B(r)$, and use the $R^{\alpha}_{\phantom{\alpha}\alpha}$ given in (\ref{7.7x}) for $-6X$.

Some simplification of the general $B(r)$ is possible depending on how the matter sources are distributed. For an isolated system such as a star of radius $R_0$ with $f(r)$ only being nonzero in the range $(0,R_0)$, the metric in $r>R_0$ takes the form
\begin{eqnarray}
B(r>R_0)&=&w-Kr^2-\frac{1}{6r}\int_0^{R_0}
dr^{\prime}r^{\prime 4}f(r^{\prime})-\frac{r}{2}\int_0^{R_0}
dr^{\prime}r^{\prime 2}f(r^{\prime})
=w-Kr^2-\frac{2\beta^*}{r}+\gamma^* r,
\label{7.8x}
\end{eqnarray}                                 
where
\begin{eqnarray}
\beta^*&=& \frac{1}{12}\int_0^{R_0}
dr^{\prime}r^{\prime 4}f(r^{\prime}),\qquad
\gamma^*=-\frac{1}{2}\int_0^{R_0}
dr^{\prime}r^{\prime 2}f(r^{\prime}).
\label{7.9x}
\end{eqnarray}                                 
It was in the form 
\begin{eqnarray}
B(r>R_0)=w-\frac{2\beta^*}{r}+\gamma^* r -K r^2
\label{7.10x}
\end{eqnarray}
that this solution was presented in \cite{Mannheim1994}. (For a solar mass star $\beta^*=M_{\odot}G/c^2$.) With $T_{rr}(r>R_0)=0$, insertion of (\ref{7.10x}) into (\ref{7.4x}) yields the constraint $w^2=1-6\beta^*\gamma^*$, to then give the solution presented in  \cite{Riegert1984b} and \cite{Mannheim1989}.

However, driven by attempts to fit the systematics of galactic rotation curve data it was gradually realized that this was not the complete solution. Specifically, since sources are putting out potentials that grow with distance (viz. the $\gamma r$ type term), one cannot ignore the contribution of material outside of any system of interest. There are two forms of such external contributions, one due to the background cosmological Hubble flow and the other due to the inhomogeneities in it. Since the cosmological background is described by a Robertson-Walker geometry and since such a geometry is conformal to flat, in it the Weyl tensor vanishes. However, as seen for instance from (\ref{7.5x}), for the inhomogeneities in the Hubble flow the Weyl tensor does not vanish. In \cite{Mannheim1989,Mannheim1997} it was realized that when transformed by a coordinate transformation to any static, spherically symmetric rest frame coordinate system, a conformally transformed  conformal to flat comoving global Robertson-Walker metric took the form of a universal linear potential term $\gamma_0r$ contribution to $B(r)$, where $\gamma_0$ is fixed by the spatial curvature $k$ of the Universe according to $\gamma_0=(-4k)^{1/2}$. With the $\int_0^rdr^{\prime}r^{\prime 2}f(r^{\prime})$ term not contributing to the Weyl tensor in (\ref{7.5x}) and with this same integral being related to $\gamma^*$ in (\ref{7.9x}), a contribution of the form $B(r)=\gamma_0r$ would not couple to the Weyl tensor either, just as it should not since it originates from cosmology where the Weyl tensor vanishes. Thus just like its cosmological $-Kr^2$ partner, the $\gamma_0r$ term can also be added on to the solution to $\nabla^4B(r)=f(r)$. Thus in \cite{Mannheim1997} the exterior metric 
\begin{eqnarray}
B(r>R_0)=w-\frac{2\beta^*}{r}+\gamma^* r+\gamma_0r-Kr^2
\label{7.11x}
\end{eqnarray}
and its associated massive particle geodesic equation were used to successfully fit the  galactic rotation curve bound orbit data for a then available set of 11  galaxies for which there were at the time  both good radio data (HI) and good photometry (HII), and good fits were found (the contribution of the $-Kr^2$ term was not needed) without the need to introduce any dark matter.

For inhomogeneities, it was noted in \cite{Mannheim2011} that they would contribute to the $r^2\int_r^{\infty}
dr^{\prime}r^{\prime }f(r^{\prime })$ and $\int_r^{\infty}dr^{\prime}r^{\prime 3}f(r^{\prime })$ terms in $B(r)$ that appear in (\ref{7.3z}), as those integrals continue all the way to infinity, to thus encompass material  external to any given system of interest. Taking this exterior matter to begin at some scale $R_1$, the potential and its derivative in the intermediate $R_0<r<R_1$ region then take the form
\begin{eqnarray}
B(R_0<r<R_1)&=&w-\frac{2\beta^*}{r}+\gamma^* r +\gamma_0 r-\kappa r^2 -\frac{1}{2}\int_{R_1}^{\infty}
dr^{\prime}r^{\prime 3}f(r^{\prime})-Kr^2,
\nonumber\\
B^{\prime}(R_0<r<R_1)&=&\frac{2\beta^*}{r^2}+\gamma^* +\gamma_0 -2\kappa r -2Kr,
\label{7.12x}
\end{eqnarray}
where
\begin{eqnarray}
\kappa=\frac{1}{6}\int_{R_1}^{\infty}
dr^{\prime}r^{\prime }f(r^{\prime}).
\label{7.13x}
\end{eqnarray}
We have not merged the $-Kr^2$ and $-\kappa r^2$ terms in (\ref{7.12x}) since while the $-Kr^2$ term applies for all $r$, the  $-\kappa r^2$ term only has the form that it does for $r<R_1$, with the $-Kr^2$ term being due to the cosmological background (vanishing Weyl tensor) and the $-\kappa r^2$ term being due to the inhomogeneities in it (non-vanishing Weyl tensor). With new galactic data having subsequently come on line the data then went out far enough that the $\kappa r^2$ term was now relevant, and on ignoring the $-Kr^2$ term, through the use of (\ref{7.12x}) very good fitting to the rotation curves of the 138 galaxies was obtained in \cite{Mannheim2011,Mannheim2012,O'Brien2012} with fixed, universal (i.e., galaxy-independent) parameters
\begin{eqnarray}
\beta^*&=&1.48\times 10^5 {\rm cm},\quad \gamma^*=5.42\times 10^{-41} {\rm cm}^{-1},
\nonumber\\
\gamma_0&=&3.06\times
10^{-30} {\rm cm}^{-1},\quad \kappa = 9.54\times 10^{-54} {\rm cm}^{-2},
\label{7.14x}
\end{eqnarray} 
and with there being no need to introduce any dark matter. Since current dark matter fits require two free parameters per galactic halo, the galaxy-dependent 276 free dark matter halo parameters that are needed for the 138 galaxy sample are replaced by just the three universal parameters: $\gamma^*$, $\gamma_0$ and $\kappa$. (The luminous Newtonian $\beta^*$-dependent  contribution in (\ref{7.12x}) is common to both dark matter and conformal gravity fits and is included in both cases.) With $\gamma_0$ being fitted to be of order the inverse of the Hubble radius and with the fitted $\kappa$ being of order a typical cluster of galaxies scale, the values for $\gamma^0$ and $\kappa$ that are obtained show that they are indeed of the cosmological scales associated with the homogeneous Hubble flow and the inhomogeneities in it. We can thus use stars in galaxies to serve as test particles that measure the global geometry of the universe. From the perspective of a local $1/r$ Newtonian potential the fact that the measured velocities exceed the luminous Newtonian expectation is described as the missing mass problem, with undetected or dark matter within the galaxies  themselves being needed in order to be able to account for the shortfall. From the perspective of conformal gravity the shortfall is explained by the rest of the visible mass in the universe. The missing mass is thus not missing at all, it is the rest of the visible universe and it has been hiding in plain sight all along.

Since one cannot ignore the rest  of the Universe for bound orbits, one cannot do so for unbound photon trajectories either. As such, the presence of contributions coming from outside of any given galaxy of interest constitute an effect that is foreign to standard Newton-Einstein gravity, namely an external field effect. As well as occurring in conformal gravity, such effects are also present in other alternate gravity theories, and have been identified in a study \cite{Chae2020} of Milgrom's Modified Newtonian Dynamics (MOND)  and a study \cite{Moffat2021} of Moffat's Modified Gravity Theory (MOG), and are also discussed in \cite{Mannheim2021b}. These external field concerns are  of relevance for gravitational bending and lensing, with the intent of this paper being to provide a framework for future conformal gravity studies on the topic. In these studies one cannot restrict the range of $r$ as done in (\ref{7.12x}) since for a lens composed for instance of a cluster of galaxies the lens is an extended source that light rays both go around and through. Moreover, since light rays are coming in from a non-asymptotically flat background one has to include the background cosmological $\gamma_0r$ contribution to both the full metric and Ricci scalar, viz.
\begin{eqnarray}
B(r)&=&w-Kr^2-\frac{r}{2}\int_0^r
dr^{\prime}r^{\prime 2}f(r^{\prime})
-\frac{1}{6r}\int_0^r
dr^{\prime}r^{\prime 4}f(r^{\prime})
-\frac{1}{2}\int_r^{\infty}
dr^{\prime}r^{\prime 3}f(r^{\prime})
-\frac{r^2}{6}\int_r^{\infty}
dr^{\prime}r^{\prime }f(r^{\prime})+\gamma_0r,
\label{7.15x}
\end{eqnarray}                                 
\begin{align}
R^{\alpha}_{\phantom{\alpha}\alpha}&=-\frac{3}{r}\int_0^r
dr^{\prime}r^{\prime 2}f(r^{\prime})-\frac{1}{r^2}\int_r^{\infty}
dr^{\prime}r^{\prime 3}f(r^{\prime})-2\int_r^{\infty}
dr^{\prime}r^{\prime }f(r^{\prime})-12K+\frac{6\gamma_0}{r}+\frac{2}{r^2}(w-1),
\label{7.16x}
\end{align}
as integrated over all the visible material in the Universe. And in addition one has to adapt the analysis to incorporate the concerns described in \cite{Rindler2009} and \cite{Ishak2010}.

The issue of gravitational bending in  the conformal gravity theory has led to a spirited though not as  yet fully resolved discussion in the literature, with some quite  varying results being reported. Studies based on the pure null geodesic behavior associated with (\ref{5.6x}) may be found in  \cite{Walker1994,Edery1998,Edery2001}. However, since the linear potential leads to a non-asymptotically flat geometry one cannot actually use  (\ref{5.6x}). Studies that took the asymptotic behavior of the conformal gravity metric given in (\ref{7.10x}) into consideration may be found in  \cite{Sultana2010,Cattani2013,Villanueva2013,Hoseini2017,Lim2017,
Campigotto2019}, but for comments on \cite{Campigotto2019} and its relevance to \cite{Campigotto2015}, \cite{Jackiw2015},  \cite{Mannheim2021a} and \cite{Mannheim2017} see \cite{footnote12a}. The present work provides some additional insight into the issue of gravitational bending in the conformal theory, and its follow up could prove to be instructive. 
 
Finally, we should note that while there are potential departures from geodesic behavior for massless light rays there is no such departure for bound orbits of massive particles such as planets in the solar system or stars and gas in a galaxy, since, as noted in Sec. \ref{S3d} and \cite{footnote9}, the mass term typically dominates over curvature effects. (This is actually quite consequential for conformal gravity fits to galactic rotation curves since unlike the $-2MG/r$ term, the linear potential does contribute to the Ricci scalar, and  would be relevant if the Ricci scalar contribution were not suppressed by the mass term \cite{footnote13}.) For massless particles one has to compare the curvature scale not with a mass scale but with a wavelength scale, and for phenomena such as lensing one has to make the comparison over all radii $r$ including those far from the lens   \cite{footnote14}. In regard to curvature effects, we note that these are large in the merger region of two compact sources such as two massive black holes, and in this region curvature effects could impact on the trajectories of gravitational waves. In events such as the merger of two neutron stars high curvature could impact not just on gravitational waves  in the merger region but also on the light rays that are emitted in the merger region in these so-called multi-messenger events.

\section{Conformal Invariant Waves Propagating in a Conformal to Flat Geometry}
\label{S8}
\subsection{The General Situation}
\label{S8a}

The conformally coupled scalar field wave equation given in (\ref{3.1x}), viz.
\begin{eqnarray}
\nabla_{\mu}\nabla^{\mu}S
+\frac{1}{6}SR^{\alpha}_{\phantom{\alpha}\alpha}=0,
\label{8.1z}
\end{eqnarray}
is left invariant under $S(x)\rightarrow \Omega^{-1}(x)S(x)=\tilde{S}(x)$, $g_{\mu\nu}(x)\rightarrow \Omega^2(x)g_{\mu\nu}(x)=
\tilde{g}_{\mu\nu}(x)$, for a conformal factor $\Omega(x)$ that has a completely arbitrary dependence on $x^{\mu}$. Consequently, the physical  content of this wave equation does not change under a local conformal transformation. Now if the spacetime geometry in which the wave is propagating just happens to be conformal to flat (such as de Sitter or Robertson-Walker), viz. line element in polar coordinates of the form
\begin{eqnarray}
ds^2=\Omega^2(x)[dp^2-dr^2-r^2d\theta^2-r^2\sin^2\theta d\phi^2],
\label{8.2z}
\end{eqnarray}
then by a local conformal transformation one can bring the metric to a flat Minkowski form. The content of (\ref{8.1z}) and (\ref{8.2z}) is then the same as that of 
\begin{eqnarray}
\partial_{\mu}\partial^{\mu}\tilde{S}=0,\quad ds^2=dp^2-dr^2-r^2d\theta^2-r^2\sin^2\theta d\phi^2.
\label{8.3z}
\end{eqnarray}
But solutions to $\partial_{\mu}\partial^{\mu}\tilde{S}=0$ are plane waves that propagate on the flat space light cone $ds^2=dp^2-dr^2-r^2d\theta^2-r^2\sin^2\theta d\phi^2=0$. Hence solutions to (\ref{8.1z}) are of the form $S(x)=\Omega^{-1}\tilde{S}$, i.e., conformally transformed plane waves, transformed waves that propagate on the conformally transformed light cone $ds^2=\Omega^2(x)[dp^2-dr^2-r^2d\theta^2-r^2\sin^2\theta d\phi^2]=0$, with rays that travel on the standard null geodesic
\begin{eqnarray}
 \frac{d^2x^{\lambda} }{ dq^2}
+\Gamma^{\lambda}_{\mu \nu} 
\frac{dx^{\mu}}{dq}\frac{dx^{\nu } }{ dq} = 0 ,\quad R_{\lambda\mu\nu\kappa}\neq 0
\label{8.4z}
\end{eqnarray}
propagating in  a  space with non-vanishing Riemann tensor $R_{\lambda\mu\nu\kappa}$.

Now given that $ds^2=\Omega^2(x)[dp^2-dr^2-r^2d\theta^2-r^2\sin^2\theta d\phi^2]=0$ the rays must obey $ds^2=dp^2-dr^2-r^2d\theta^2-r^2\sin^2\theta d\phi^2=0$, so that for a ray propagating in the radial direction for instance the trajectory obeys $(p-p^{\prime})^2-(r-r^{\prime})^2=0$. However, these radial modes are associated with the flat space polar coordinate geodesic
\begin{eqnarray}
 \frac{d^2x^{\lambda} }{ dq^2} +\Gamma^{\lambda}_{\mu \nu} 
\frac{dx^{\mu}}{dq}\frac{dx^{\nu } }{ dq} = 0,\quad  R_{\lambda\mu\nu\kappa}=0
\label{8.5zz}
\end{eqnarray}
propagating in  a  space with vanishing $R_{\lambda\mu\nu\kappa}$. Now (\ref{8.4z}) and (\ref{8.5zz}) are quite different from each other. Nonetheless, in Sec. \ref{S3g} we showed that these two geodesics are equivalent. We now show this explicitly in a special case, viz. a Robertson-Walker geometry. Since the Maxwell equations are also locally conformal invariant, whatever outcome we obtain in the scalar field case will hold in the Maxwell case as well.

\subsection{Robertson-Walker Geometries}
\label{S8b}

To exhibit the required equivalence it suffices to consider a Robertson-Walker geometry with arbitrary expansion radius $a(t)$ and flat spatial three-curvature. In comoving time $t$ the line element is of the form 
\begin{eqnarray}
ds^2=dt^2-a^2(t)[dr^2+r^2d\theta^2+r^2\sin^2\theta d\phi^2],
\label{8.6zz}
\end{eqnarray}
and in conformal time with $p=\int dt/a(t)$ it is of the form
\begin{eqnarray}
ds^2=\Omega^2(p)[dp^2-dr^2-r^2d\theta^2-r^2\sin^2\theta d\phi^2],
\label{8.7zz}
\end{eqnarray}
where $\Omega(p)=a(t)$. (de Sitter is the special case in which $a(t)=e^{Ht}$, $\Omega=1/Hp$.) 

With the dot denoting the derivative with respect to $p$ and $0$ denoting $p$,  in the $(0,r)$ sector the connection and the Ricci scalar associated with (\ref{8.7zz}) are of the form
\begin{eqnarray}
\Gamma^0_{00}=\Gamma^0_{rr}=\Gamma^r_{0r}=\frac{\dot{\Omega}}{\Omega},\qquad \Gamma^0_{0r}=\Gamma^r_{00}=\Gamma^r_{rr}=0,\qquad R^{\alpha}_{\phantom{\alpha}\alpha}=-\frac{6\ddot{\Omega}}{\Omega^3}.
\label{8.8zz}
\end{eqnarray}
The null geodesic equation given in (\ref{8.4z}) yields
\begin{eqnarray}
\frac{d^2r}{dq^2}+2\frac{\dot{\Omega}}{\Omega}\frac{dr}{dq}\frac{dp}{dq}=0, \qquad
\frac{d^2p}{dq^2}+\frac{\dot{\Omega}}{\Omega}\frac{dp}{dq}\frac{dp}{dq}+\frac{\dot{\Omega}}{\Omega}\frac{dr}{dq}\frac{dr}{dq}
=0,
\label{8.9zz}
\end{eqnarray}
and with $d\Omega(p)/dq=(d\Omega(p)/dp)(dp/dq)$  we obtain
\begin{eqnarray}
\left(\frac{dr}{dq}\right)^2=\frac{e^2}{\Omega^4},\quad \left(\frac{dp}{dq}\right)^2=\frac{e^2}{\Omega^4}+\frac{f^2}{\Omega^2},
\label{8.10zz}
\end{eqnarray}
where $e^2$ and $f^2$ are integration constants. On requiring that the solutions to the massless geodesics be on the light cone yields
\begin{eqnarray}
-\Omega^2\left(\frac{dp}{dq}\right)^2+\Omega^2\left(\frac{dr}{dq}\right)^2=0,
\label{8.11zz}
\end{eqnarray}
to thus lead to $f^2=0$, and thereby give 
\begin{eqnarray}
\left(\frac{dr}{dq}\right)^2=\frac{e^2}{\Omega^4},\quad \left(\frac{dp}{dq}\right)^2=\frac{e^2}{\Omega^4}.
\label{8.12zz}
\end{eqnarray}
The trajectories are thus of the form
\begin{eqnarray}
\left(\frac{dr}{dp}\right)^2=1,
\label{8.13zz}
\end{eqnarray}
just as we would have directly obtained if we had set $ds^2=0$ in (\ref{8.7zz}). As we see, trajectories for conformally coupled scalar fields in a conformal to flat geometry are the same as those of a free flat space scalar field in a flat geometry. The conformal factor thus hides itself and is not observable. Also we recall that in Sec. \ref{S3g} null geodesics could be transformed into each other provided we used affine parameters for the two geodesics that were related according to $d/d\bar{q}=\Omega^{-2}(x)d/dq$. This is precisely the structure that we see in (\ref{8.11zz}) and (\ref{8.12zz}), since with  $\Omega^{2}(x)d/dq=d/d\bar{q}$ we can bring these equations to the flat spacetime form
\begin{eqnarray}
-\left(\frac{dp}{d\bar{q}}\right)^2+\left(\frac{dr}{d\bar{q}}\right)^2=0,
\label{8.14zzz}
\end{eqnarray}
\begin{eqnarray}
\left(\frac{dr}{d\bar{q}}\right)^2=e^2,\quad \left(\frac{dp}{d\bar{q}}\right)^2=e^2.
\label{8.15zzz}
\end{eqnarray}

For nonzero spatial curvature we set $r=\sinh \chi$ when $k=-1$ and $r=\sin\chi$ when $k=+1$. This then leads to metrics of the form
\begin{align}
&ds^2(k=-1)=\Omega^2(p)[dp^2-d\chi^2-\sinh^2\chi d\theta^2-\sinh^2\chi\sin^2\theta d\phi^2], 
\nonumber\\
&ds^2(k=+1)=\Omega^2(p)[dp^2-d\chi^2-\sin^2\chi d\theta^2-\sin^2\chi\sin^2\theta d\phi^2].
\label{8.16zzz}
\end{align}
These metrics are not explicitly in a conformal to flat form but could be brought to such a form by coordinate transformations involving $p$ and $\chi$ (see e.g. \cite{Amarasinghe2019}). Given the form of (\ref{8.16zzz}), the discussion follows the $k=0$ case, and leads to radial trajectories of the form
\begin{equation}
\left(\frac{d\chi}{dp}\right)^2=1. 
\label{8.17zzz}
\end{equation}

\section{Cosmological Fluctuations}
\label{S9}

In the presence of  inhomogeneities in a cosmological Robertson-Walker background the geometry is no longer homogeneous and isotropic and the Weyl tensor is no longer zero. There are two options for the  propagation of fields in such a situation, depending on whether the perturbation only affects the geometry (a fluctuating source emitting a light signal), or whether it affects the propagating scalar or Maxwell field equations as well (field already propagating in the absence of fluctuations).  For a conformally coupled scalar field one considers
\begin{eqnarray}
(g^{\mu\nu}+\delta g^{\mu\nu})[\partial_{\mu}\partial_{\nu}-\Gamma^{\lambda}_{\mu\nu}-\delta \Gamma^{\lambda}_{\mu\nu}]S+
\frac{1}{6}(R^{\alpha}_{\phantom{\alpha}\alpha}+\delta R^{\alpha}_{\phantom{\alpha}\alpha})S=0
\label{9.1a}
\end{eqnarray}
in the former case, while in the latter case one considers 
\begin{eqnarray}
(g^{\mu\nu}+\delta g^{\mu\nu})[\partial_{\mu}\partial_{\nu}-\Gamma^{\lambda}_{\mu\nu}-\delta \Gamma^{\lambda}_{\mu\nu}](S+\delta S)+
\frac{1}{6} (R^{\alpha}_{\phantom{\alpha}\alpha}+\delta R^{\alpha}_{\phantom{\alpha}\alpha})(S+\delta S)=0.
\label{9.2a}
\end{eqnarray}
In (\ref{9.2a}) the background scalar field obeys 
\begin{eqnarray}
\left[g^{\mu\nu}[\partial_{\mu}\partial_{\nu}-\Gamma^{\lambda}_{\mu\nu}]+
\frac{1}{6}R^{\alpha}_{\phantom{\alpha}\alpha}\right]S=0,
\label{9.3a}
\end{eqnarray}
and the first-order perturbation is of the form
\begin{eqnarray}
\left[g^{\mu\nu}[\partial_{\mu}\partial_{\nu}-\Gamma^{\lambda}_{\mu\nu}]+\frac{1}{6} R^{\alpha}_{\phantom{\alpha}\alpha}\right]\delta S=-\delta g^{\mu\nu}[\partial_{\mu}\partial_{\nu}-\Gamma^{\lambda}_{\mu\nu}]S
+g^{\mu\nu}\delta \Gamma^{\lambda}_{\mu\nu}S
-\frac{1}{6} \delta R^{\alpha}_{\phantom{\alpha}\alpha}S.
\label{9.4a}
\end{eqnarray}

In (\ref{9.1a}) a conformally coupled scalar field propagates in a background that is not conformal to flat, and as we had noted in Sec. \ref{S1e}, in such a case the propagator takes support off the light cone. Additionally, in (\ref{9.4a}) the action of the background wave operator on $\delta S$ (the left-hand side of (\ref{9.4a}))  is not zero (non-vanishing right-hand side). So unlike the unperturbed field $S$ that satisfies the unperturbed (\ref{9.3a}) and does propagate on the light cone (the cosmological background being conformal to flat), the fluctuating field $\delta S$ propagates off the light cone. Thus in neither (\ref{9.1a}) or (\ref{9.4a}) will the geodesics be null if curvature effects are sufficiently large.

Since the trajectories are null geodesics in the absence of the perturbation we need to explore what happens to such null geodesics when they are perturbed. To do this we recall that with the cosmological Robertson-Walker background being conformal to flat, in the background both the null geodesic equation and the light cone condition hold, viz. 
\begin{eqnarray}
 \frac{d^2x^{\lambda} }{ dq^2}
+\Gamma^{\lambda}_{\mu \nu} 
\frac{dx^{\mu}}{dq}\frac{dx^{\nu } }{ dq} = 0, \quad g_{\mu\nu}\frac{dx^{\mu}}{dq}\frac{dx^{\nu}}{dq}=0.
\label{9.5xx}
\end{eqnarray}
And in the simplest case where when $k=0$, the background radial solutions are given by (\ref{8.12zz}). We thus wish to see what happens to the light cone condition given in (\ref{9.5xx}) when the null geodesic equation is perturbed. On defining the perturbation according to 
\begin{eqnarray}
\frac{dx^{\lambda}}{dq}\rightarrow \frac{dx^{\lambda}}{dq}+\delta^{\lambda},\quad g_{\mu\nu}\rightarrow g_{\mu\nu}+h_{\mu\nu}, \quad g^{\mu\nu}\rightarrow g^{\mu\nu}-h^{\mu\nu},\quad \Gamma^{\lambda}_{\mu \nu} \rightarrow
\Gamma^{\lambda}_{\mu \nu} +\delta \Gamma^{\lambda}_{\mu \nu},
\label{9.6xx}
\end{eqnarray}
so that
\begin{eqnarray}
\frac{d\delta^{\lambda}}{dq}+\delta\Gamma^{\lambda}_{\mu \nu} \frac{dx^{\mu}}{dq}\frac{dx^{\nu } }{ dq} 
+\Gamma^{\lambda}_{\mu \nu}\left(\delta^{\mu}\frac{dx^{\nu } }{ dq} +\frac{dx^{\mu}}{dq}\delta^{\nu }\right)=0,
\label{9.7xx}
\end{eqnarray}
we look to see whether or not 
\begin{eqnarray}
A=g_{\mu\nu}\left(\delta^{\mu}\frac{dx^{\nu } }{ dq} +\frac{dx^{\mu } }{ dq} \delta^{\nu}\right)
+h_{\mu\nu}\frac{dx^{\mu } }{ dq} \frac{dx^{\nu } }{ dq} 
\label{9.8xx}
\end{eqnarray}
is zero. 

With $\delta(g^{\mu\nu})=-h^{\mu\nu}$,   we find that $\delta\Gamma^{\lambda}_{\mu \nu} $ evaluates to
\begin{eqnarray}
\delta\Gamma^{\lambda}_{\mu \nu}=-\frac{1}{2}h^{\lambda\alpha}\left(\partial_{\mu}g_{\alpha\nu}+\partial_{\nu}g_{\alpha\mu}
-\partial_{\alpha}g_{\mu\nu}\right)+\frac{1}{2}g^{\lambda\alpha}\left(\partial_{\mu}h_{\alpha\nu}+\partial_{\nu}h_{\alpha\mu}
-\partial_{\alpha}h_{\mu\nu}\right).
\label{9.9xx}
\end{eqnarray}
Given (\ref{8.8zz}) and taking the positive square roots in  (\ref{8.12zz}) so that $dp/dq=dr/dq=e/\Omega^2$, the radial sector of (\ref{9.7xx}) evaluates to 
\begin{eqnarray}
\frac{d\delta^{0}}{dq}+\left(\delta\Gamma^{0}_{00}+2\delta\Gamma^{0}_{0r}+\delta\Gamma^{0}_{rr} \right)\frac{e^2}{\Omega^4}
+\frac{2e\dot{\Omega}}{\Omega^3}(\delta^0+\delta^r)=0,
\label{9.10xx}
\end{eqnarray}
\begin{eqnarray}
\frac{d\delta^{r}}{dq}+\left(\delta\Gamma^{r}_{00}+2\delta\Gamma^{r}_{0r}+\delta\Gamma^{r}_{rr} \right)\frac{e^2}{\Omega^4}
+\frac{2e\dot{\Omega}}{\Omega^3}(\delta^0+\delta^r)=0.
\label{9.11xx}
\end{eqnarray}
Then with 
\begin{eqnarray}
\delta\Gamma^{0}_{00}+2\delta\Gamma^{0}_{0r}+\delta\Gamma^{0}_{rr}=2\Omega\dot{\Omega}(h^{00}-h^{0r})
-\frac{1}{2\Omega^2}\left(\partial_0h_{00}+2\partial_rh_{00}+2\partial_rh_{0r}-\partial_0h_{rr}\right),
\label{9.12xx}
\end{eqnarray}
\begin{eqnarray}
\delta\Gamma^{r}_{00}+2\delta\Gamma^{r}_{0r}+\delta\Gamma^{r}_{rr}=2\Omega\dot{\Omega}(h^{0r}-h^{rr})
+\frac{1}{2\Omega^2}\left(-\partial_rh_{00}+2\partial_0h_{rr}+2\partial_0h_{0r}+\partial_rh_{rr}\right),
\label{9.13xx}
\end{eqnarray}
(\ref{9.10xx}) and (\ref{9.11xx}) take the form
\begin{eqnarray}
\frac{d\delta^{0}}{dq}+\frac{2e\dot{\Omega}}{\Omega^3}(\delta^0+\delta^r)
+\frac{2e^2\dot{\Omega}}{\Omega^3}\left(h^{00}-h^{0r}\right)
+\frac{e^2}{2\Omega^6}\left(-\partial_0h_{00}-2\partial_rh_{00}-2\partial_rh_{0r}+\partial_0h_{rr}\right)
=0,
\label{9.14xx}
\end{eqnarray}
\begin{eqnarray}
\frac{d\delta^{r}}{dq}+\frac{2e\dot{\Omega}}{\Omega^3}(\delta^0+\delta^r)
+\frac{2e^2\dot{\Omega}}{\Omega^3}\left(h^{0r}-h^{rr}\right)
+\frac{e^2}{2\Omega^6}\left(-\partial_rh_{00}+2\partial_0h_{rr}+2\partial_0h_{0r}+\partial_rh_{rr}\right)
=0.
\label{9.15xx}
\end{eqnarray}
With $h^{\mu\nu}=g^{\mu\beta}g^{\nu\sigma}h_{\beta\sigma}$, we can rewrite (\ref{9.14xx}) and (\ref{9.15xx}) as 
\begin{eqnarray}
\frac{d\delta^{0}}{dq}+\frac{2e\dot{\Omega}}{\Omega^3}(\delta^0+\delta^r)
+\frac{2e^2\dot{\Omega}}{\Omega^7}\left(h_{00}+h_{0r}\right)
+\frac{e^2}{2\Omega^6}\left(-\partial_0h_{00}-2\partial_rh_{00}-2\partial_rh_{0r}+\partial_0h_{rr}\right)=0,
\label{9.16xx}
\end{eqnarray}
\begin{eqnarray}
\frac{d\delta^{r}}{dq}+\frac{2e\dot{\Omega}}{\Omega^3}(\delta^0+\delta^r)
+\frac{2e^2\dot{\Omega}}{\Omega^7}\left(-h_{0r}-h_{rr}\right)
+\frac{e^2}{2\Omega^6}\left(-\partial_rh_{00}+2\partial_0h_{rr}+2\partial_0h_{0r}+\partial_rh_{rr}\right)=0.
\label{9.17xx}
\end{eqnarray}
Subtracting (\ref{9.17xx}) from (\ref{9.16xx}) and using $d/dq=(dp/dq)d/dp+(dr/dq)d/dr$ gives
\begin{eqnarray}
\frac{d}{dq}(\delta^0-\delta^r)+\frac{2e^2\dot{\Omega}}{\Omega^7}B=\frac{e^2}{2\Omega^6}(\partial_0+\partial_r)B=\frac{e}{2\Omega^4}\frac{dB}{dq},
\label{9.18xx}
\end{eqnarray}
where $B=h_{00}+2h_{0r}+h_{rr}$. We can write (\ref{9.18xx}) very compactly as
\begin{eqnarray}
\frac{d}{dq}\left(\delta^0-\delta^r-\frac{eB}{2\Omega^4}\right)=0,
\label{9.19xx}
\end{eqnarray}
to thus obtain
\begin{eqnarray}
\delta^0-\delta^r-\frac{eB}{2\Omega^4}=C,
\label{9.20xx}
\end{eqnarray}
where $C$ is an integration constant.

From its definition in (\ref{9.8xx}) $A$ is given by 
\begin{eqnarray}
A=-2e(\delta^0-\delta^r)+\frac{e^2}{\Omega^4}B. 
\label{9.21xx}
\end{eqnarray}
Thus from (\ref{9.20xx}) it follows that 
\begin{eqnarray}
A=-2eC. 
\label{9.22xx}
\end{eqnarray}
We thus establish that $A$ is a constant, and recognize it as an integration constant. Since the solutions to (\ref{9.1a}) and (\ref{9.4a}) are not restricted to the light cone, we see that the integration constant $A$ must be nonzero. This situation completely parallels the analysis  that we gave in Sec. \ref{S5} of null geodesics in a background Schwarzschild de Sitter geometry, a geometry that also is not conformal to flat. Specifically, we found there that it is only the integration constant that determines whether the solution is or is not on the light cone. This information has to be provided separately as the null geodesic equation is a differential equation whose solution requires the supplying of boundary conditions that are not contained in the equation itself. 

While there does not appear to be a way to obtain a closed form expression for $\delta^0+\delta^r$, we can also  go back to (\ref{9.1a}) or (\ref{9.4a})  and try to eikonalize them directly. And in  the Appendix  we take a first step in this direction by evaluating 
$\delta R^{\alpha}_{\phantom{\alpha}\alpha}$ in the convenient scalar, vector, tensor basis that is commonly used in cosmological perturbation studies. However, the perturbation to the background $dr/dp=1$ given in (\ref{8.13zz}) can actually be evaluated in a closed form since it only involves the $\delta^0-\delta^r$ combination and not the $\delta^0+\delta^r$ one. Specifically, on recalling that in the background we have $dr/dq=dp/dq=e/\Omega^2$, through the use of (\ref{9.20xx}) we obtain 
\begin{eqnarray}
\frac{dr}{dp}\rightarrow \frac{dr/dq+\delta^r}{dp/dq+\delta^0}=1+\frac{\Omega^2}{e}(\delta^r-\delta^0)=1-\frac{\Omega^2C}{e} -\frac{B}{2\Omega^2}.
\label{9.23xx}   
\end{eqnarray}
On setting $A=-1$ so that $(dx_{\mu}/dq)(dx^{\mu}/dq)=-1$, we obtain $2eC=1$, and can thus rewrite (\ref{9.23xx}) as 
\begin{eqnarray}
\frac{dr}{dp}\rightarrow1-\frac{\Omega^2}{2e^2} -\frac{B}{2\Omega^2}.
\label{9.24xx}   
\end{eqnarray}
Finally, we note that given (\ref{8.16zzz}) and (\ref{8.17zzz}), fluctuations around background geometries with $k\neq 0$ follow identically, with (\ref{9.24xx}) being replaced by  
\begin{eqnarray}
\frac{d\chi}{dp}\rightarrow1-\frac{\Omega^2}{2e^2} -\frac{B}{2\Omega^2}, \quad B=h_{00}+2h_{0\chi}+h_{\chi\chi}.
\label{9.25qq}   
\end{eqnarray}
While we have now provided closed form expressions for both $dr/dp$ and $d\chi/dp$, to actually evaluate them we would need to know the relevant components of $h_{\mu\nu}$. These components are provided by the gravitational fluctuation equations themselves and are not sought here  \cite{footnote15}.

It is of interest to compare (\ref{9.24xx}) with the expression that would hold on the light cone, where with $k_{\mu\nu}=h_{\mu\nu}/\Omega^2$ the fluctuating radial sector line element is given by 
\begin{eqnarray}
ds^2=\Omega^2(dp^2-dr^2-k_{00}dp^2-2k_{0r}dpdr-k_{rr}dr^2).
\label{9.26qq}
\end{eqnarray}
On setting $ds^2=0$, we find that to lowest order in the perturbation
\begin{eqnarray}
\frac{dr}{dp}=1-\frac{1}{2}(k_{00}+2k_{0r}+k_{rr})=1-\frac{B}{2\Omega^2}.
\label{9.27qq}
\end{eqnarray}
We recognize (\ref{9.27qq}) as the form that (\ref{9.23xx}) would take if we set $C=0$, i.e., if we set $A=0$. Consequently, (\ref{9.23xx}) and (\ref{9.27qq}) only differ by an integration constant, an integration constant that determines whether or not the fluctuations are on the light cone. Eq. (\ref{9.23xx}) thus explicitly shows how the light cone condition given in (\ref{9.27qq}) is altered if the signal propagates off the light cone.

\section{Completely General Approach to Eikonalization}
\label{S10}

\subsection{Eikonalization, the Normal to the Wavefront and Embedding Theory}
\label{S10a}

We would like to note that it is possible to use the formalism we have developed even if the eikonal function $T$ is not larger than any other of the other terms in the wave equations. All of the minimally coupled scalar field (\ref{2.3y}), the conformally coupled scalar field (\ref{3.3z}), the real part of the mass generating (\ref{3.32qq})  and the real part of the Maxwell (\ref{4.6x}) as projected on to a polarization vector can be written without approximation in the generic  form $-\nabla_{\nu}T\nabla^{\nu}T=X$. (For convenience the associated forms for $X$ are listed in (\ref{1.62h}) to (\ref{1.64h}).) Thus in all cases as long as $X$ is nonzero we can introduce just one step, namely associating the eikonal function with a velocity along a one-dimensional trajectory according to 
\begin{eqnarray}
\frac{dx^{\mu}}{dq}=\frac{\nabla^{\mu}T}{(-\nabla_{\nu}T\nabla^{\nu}T)^{1/2}}=\frac{\nabla^{\mu}T}{X^{1/2}},\qquad \frac{dx_{\mu}}{dq}\frac{dx^{\mu}}{dq}=-1.
\label{10.1x}
\end{eqnarray}
Just as in (\ref{3.11z}), from (\ref{10.1x}) it follows that
\begin{eqnarray}
\frac{D^2x^{\lambda}}{Dq^2}=\frac{d^2x^{\lambda} }{ dq^2}
+\Gamma^{\lambda}_{\mu \nu} 
\frac{dx^{\mu}}{dq}\frac{dx^{\nu}}{dq}
=-\frac{1}{2X}
\left[g^{\lambda \mu}+\frac{dx^{\lambda}}{dq}\frac{dx^{\mu}}{dq}\right]\frac{\partial X}{\partial x^{\mu}}=0,
\label{10.2x}
\end{eqnarray}
where the covariant four-acceleration $D^2x^{\lambda}/Dq^2$ was defined in (\ref{1.2x}). From (\ref{10.2x}) we can construct a generalized four-acceleration
\begin{eqnarray}
\frac{\hat{D}^2x^{\lambda}}{ \hat{D}q^2}&=&\frac{d^2x^{\lambda} }{ dq^2}
+\Gamma^{\lambda}_{\mu \nu} 
\frac{dx^{\mu}}{dq}\frac{dx^{\nu}}{dq}
+\frac{1}{2X}
\left[g^{\lambda \mu}+\frac{dx^{\lambda}}{dq}\frac{dx^{\mu}}{dq}\right]\frac{\partial X}{\partial x^{\mu}}=0.
\label{10.3x}
\end{eqnarray}
Just as the acceleration four-vector defined in (\ref{10.2x}) is orthogonal  to $dx_{\lambda}/dq$, viz.   
\begin{eqnarray}
\frac{dx_{\lambda}}{dq}\frac{D^2x^{\lambda}}{Dq^2}=0, 
\label{10.4x}
\end{eqnarray}
the generalized acceleration four-vector obeys 
\begin{eqnarray}
\frac{dx_{\lambda}}{dq}\frac{\hat{D}^2x^{\lambda}}{\hat{D}q^2}=0,
\label{10.5x}
\end{eqnarray}
to thus also be orthogonal to the eikonal four-velocity $dx_{\lambda}/dq$ \cite{footnote15b}.   

The structure exhibited in (\ref{10.4x}) and (\ref{10.5x}) is familiar in a different context, namely in the theory of  geometric embeddings. The treatment is standard and we follow the typical discussion given in \cite{Mannheim2005}. We consider a space of $D$ dimensions embedded in a space of $D+1$ dimensions with $D+1$-dimensional metric $g_{AB}$. We introduce a normal $n^A$ in the  $D+1$-dimensional  space that is orthogonal to every vector in the $D$-dimensional space and obeys $g_{AB}n^An^B=\mp 1$, a vector that could be timelike or spacelike depending on the circumstances. (In the Appendix we consider both cases.) To characterize the 
$D$-dimensional space, as in Sec. \ref{S4}  we define an induced metric
\begin{eqnarray}
I_{AB}(\mp)=g_{AB}\pm n_An_B,
\label{10.6x}
\end{eqnarray}
and an associated extrinsic curvature 
\begin{eqnarray}
K_{MN}=I^A_{\phantom{A}M}(\mp)I^B_{\phantom{B}N}(\mp)\nabla_An_B,
\label{10.7x}
\end{eqnarray}
with $\nabla_An_B$ being evaluated with the full $D+1$-dimensional metric $g_{AB}$. As constructed, $K_{MN}$ is symmetric and obeys $n^{M}K_{MN}=0$, to thus lie in the $D$-dimensional tangent space orthogonal to $n^M$. Given (\ref{10.6x}) we can rewrite (\ref{10.7x}) in a manifestly symmetric form in which $K_{MN}$ is written as the Lie derivative of the induced metric, viz 
\begin{eqnarray}
K_{MN}=\frac{1}{2}\mathcal{L}_nI_{MN}(\mp)=\frac{1}{2}\left(\nabla_Mn_N\pm n_Mn^A\nabla_An_N+\nabla_Nn_M\pm n_Nn^A\nabla_An_M\right).
\label{10.8x}
\end{eqnarray}
In (\ref{10.8x}) we recognize the acceleration vector $a_N=n^A\nabla_An_N$. Since $n^An_A=\mp 1$ the acceleration vector obeys $n^Na_N=0$, to thus also lie entirely in the $D$-dimensional tangent space orthogonal to $n^M$. With $dx^{\lambda}/dq$ being normalized in (\ref{10.1x}), the identification of $dx^{\lambda}/dq$ with a normal $n^{\lambda}$ enables to us to recognize the condition given in (\ref{10.4x}) as $n^Na_N=0$. Moreover, according to (\ref{10.2x}) and (\ref{10.3x}) $\hat{D}^2x^{\lambda}/\hat{D}q^2$ is parallel to $D^2x^{\lambda}/Dq^2$, and according to (\ref{10.5x}) $\hat{D}^2x^{\lambda}/\hat{D}q^2$ is orthogonal to $dx_{\lambda}/dq$. Thus $\hat{D}^2x^{\lambda}/\hat{D}q^2$ also lies in the tangent space orthogonal to $n^M$. We can thus identify $dx^{\lambda}/dq$ as the normal to an advancing wavefront, with the generalized $\hat{D}^2x^{\lambda}/\hat{D}q^2$ lying in the wavefront. Thus use of the eikonal condition given in (\ref{10.1x}) enables us to interpret the eikonal velocity vector $dx^{\lambda}/dq$ as being normal to an advancing wavefront, with the wavefront lying in the space tangent to the normal, a space of one lower dimension than the entire space \cite{footnote15a}.

\subsection{Why there has to be an Eikonal Velocity Relation}
\label{S10b}

As just described in Sec. \ref{S10a}, the heart of the eikonal procedure is in associating the velocity along a trajectory with properties of the solution to the wave equation, just as in (\ref{10.1x}). To understand this in a very simple case we consider a free massive scalar field wave equation in flat spacetime, with equation of motion and solution 
\begin{eqnarray}
(-\partial_0^2+\vec{\nabla}^2-m^2)S=0,\qquad S=e^{i(k_ix^i-k_0x^0)},\quad k^0=(k_ik^i+m^2)^{1/2}.
\label{10.9x}
\end{eqnarray}
If this wave advances in the $x^3$ direction, points of constant phase will then travel with a  phase velocity $v^3_p=x^3/x^0=k^0/k^3=(k_3^2+m^2)^{1/2}/k_3$ that is greater than the speed of light. However since the wave is massive the energy and momentum transported by the wave are transported at a  velocity that is less than the speed of light (this velocity is essentially the group velocity). To identify this velocity we set $T=k_ix^i-k_0x^0$ and define an eikonal velocity vector
\begin{eqnarray}
\frac{dx^{\mu}}{dq}=\frac{\nabla^{\mu}T}{C}=\frac{k^{\mu}}{C},
\label{10.10x}
\end{eqnarray}
where $C$ is a constant with the dimension of inverse length. From (\ref{10.10x}) the eikonal velocity is given by
\begin{eqnarray}
v^3_e=\frac{dx^3}{dx^0}=\frac{k^3}{k^0}=\frac{k^3}{((k^3)^2+m^2)^{1/2}},
\label{10.11x}
\end{eqnarray}
with $C$ dropping out. This eikonal velocity is less than, not greater than, the velocity of light, and together with $v^3_p$ obeys  $v^3_ev^3_p=1$. For this wave the ray travels in the $x^3$ direction normal to the wavefront in the $(x^1,x^2)$ plane, and thus represents the velocity of transport of the energy and momentum carried by the wave. It is thus the eikonal velocity relation given in (\ref{10.1x}) that enables us to determine the velocity of transport of  the energy and momentum carried by a wave  \cite{footnote16}.

For this problem there are two quantities with the same dimension as $C$, viz, $(-\nabla_{\nu}T\nabla^{\nu}T)^{1/2}$ and $m$. But $\nabla_{\nu}T\nabla^{\nu}T=k_{\nu}k^{\nu}=-(k^0)^2+k_ik^i=-m^2$, so in fact there is only one independent quantity (setting $C$ equal to a dimensionless constant times $m$ would simply rescale $\eta_{\mu\nu}(dx^{\mu}/dq)(dx^{\nu}dq)=-1$). Thus we can set $C=(-\nabla_{\nu}T\nabla^{\nu}T)^{1/2}=m$, and rewrite (\ref{10.1x}) in the form
\begin{eqnarray}
\frac{dx^{\mu}}{dq}=\frac{\nabla^{\mu}T}{(-\nabla_{\nu}T\nabla^{\nu}T)^{1/2}}=\frac{\nabla^{\mu}T}{m},\qquad \frac{dx_{\mu}}{dq}\frac{dx^{\mu}}{dq}=-1.
\label{10.12x}
\end{eqnarray}
Similarly, since $\nabla_{\nu}T\nabla^{\nu}T=-m^2$, we can set $\nabla^{\nu}T\nabla_{\nu}\nabla^{\lambda}T=0$, from which it follows that the acceleration of the ray obeys $d^2x^{\lambda}/dq^2=0$. Then with $\eta_{\mu\nu}(dx^{\mu}/dq)(dx^{\nu}dq)=-1$ leading to  $(dx_{\lambda}/dq)(d^2x^{\lambda}/dq^2)=0$, the acceleration is perpendicular to the eikonal velocity and thus lies in the  wavefront.

The retarded propagator associated with the wave equation given in (\ref{10.9x}) is given by
\begin{eqnarray}
D(x^0,\vec{x};m,RET)=-\frac{1}{(2\pi)^4}\int d^4k \frac{e^{ik\cdot x}}{k_0^2-k_1^2-k_2^2-k_3^2- m^2+i\epsilon\theta(k^0)},
\label{10.13x}
\end{eqnarray}
and takes the form $D(x^0,\vec{x};m,RET)=\theta(x^0)\Delta(x)$ where $\Delta(x)$ can be constructed from the second-order differential equation $(-\partial_0^2+\vec{\nabla}^2-m^2)\Delta(x)=0$ as integrated with two initial conditions: $\Delta(0,\vec{x})=0$,  $\partial_0\Delta(x)|_{x^0=0}=-\delta^3(\vec{x})$. The resulting $\Delta(x)$ is found in many places with a recent evaluation being given in \cite{Mannheim2021}.  With this evaluation we obtain 
\begin{eqnarray}
D(x^0,\vec{x};m,RET)=\frac{1}{2\pi}\theta(x^0)\delta[-x^2]-\frac{m}{4\pi}\theta(x^0)\frac{J_1(m[-x^2]^{1/2})}{[-x^2]^{1/2}}.
\label{10.14x}
\end{eqnarray}
As we see, the propagator  takes support both on and within the light cone, just as required by the general analysis of \cite{DeWitt1960}. Thus with the propagator taking support off the light cone, the wavefront takes support off the light cone too, and the acceleration equation $d^2x^{\lambda}/dq^2=0$ for the normal to the wavefront has to be integrated under the timelike constraint $\eta_{\mu\nu}(dx^{\mu}/dq)(dx^{\nu}dq)=-1$. The connection between propagators, wavefronts and normals to wavefronts is thus provided by the eikonal velocity relation given in (\ref{10.1x}).

In the massless case the quantity $\nabla_{\nu}T\nabla^{\nu}T$ vanishes identically and (\ref{10.12x}) is replaced by the lightlike
\begin{eqnarray}
\frac{dx^{\mu}}{dq}=\nabla^{\mu}T,\qquad \frac{dx_{\mu}}{dq}\frac{dx^{\mu}}{dq}=0.
\label{10.15x}
\end{eqnarray}
The trajectory then follows a null geodesic, and only the delta function term appears in $D(x^0,\vec{x};m=0,RET)$. 

While the theta function term would be absent in the massless flat spacetime free scalar field case, in any situation  in which the theta function is to appear, then by dimensional analysis there has to be a parameter in the theory with the same dimension as $m$. If the wave equation has no mass term, then the required dimensionful term would have to be provided by the geometry itself. And indeed, as noted in Sec. \ref{S1e} for the curved space $\nabla_{\mu}\nabla^{\mu}S=0$ theory,  that is precisely what happens when the geometry is of the form $ds^2=\Omega^2(p)(dp^2-dx^2-dy^2-dz^2)$ and $\Omega(p)=\sin(Hp)$, with $H$ replacing $m$. Moreover, in this case the propagator given in (\ref{1.40h}) takes the form 
\begin{eqnarray}
D^+_{MIN}(t,\vec{x};H,RET)&=&\frac{1}{\Omega(p)\Omega(p^{\prime})}\left[\frac{1}{2\pi}\theta(x^0)\delta[-x^2]-\frac{H}{4\pi}\theta(x^0)\frac{J_1(H[-x^2]^{1/2})}{[-x^2]^{1/2}}\right],
\label{10.16x}
\end{eqnarray}
to take exactly the same support both on and within the light cone as $D(x^0,\vec{x};m,RET)$. Solutions to the wave equation are given by $e^{i(k_ix^i-k_0x^0)}/\sin(Hp)$ where  $k_0^2=k_1^2+k_2^2+k_3^2- H^2$. Normals to the wavefronts thus obey (\ref{10.1x}), and in the Appendix we evaluate (\ref{10.3x}) explicitly to show that this is in fact the case. To conclude then,  we see that (\ref{10.1x}) is generic for any situation in which normals to wavefronts travel off the light cone.

\section{Final Comments}
\label{S11}

In this paper we have shown that propagators associated with conformally invariant wave equations will take support off the light cone unless the waves are propagating in geometric backgrounds that are conformal to flat. In consequence, in geometries that are not conformal to flat departures from both null geodesics and the massless particle light cone are generally to be expected in an eikonalization of field theoretic wave equations, and are to be expected even if the wave equations themselves are locally conformal invariant. In particular, we have shown that such departures are to be expected in some of the most prominent geometries that are  used in astrophysics and cosmology, namely Schwarzschild de Sitter geometries and fluctuations around Robertson-Walker geometries. As such, our results are points of principle, and their phenomenological relevance will depend on how big the curvature contributions that we have identified might actually be in specific cases. In this paper we have integrated some specific modified geodesic equations in some simple cases. However, integrating these modified geodesic equations may not be so straightforward in more complicated cases. In those cases an alternate procedure is possible, one that bypasses the modified geodesics altogether. Specifically, we first solve the wave equation directly in order to determine the eikonal function $T$ either exactly or in the large $T$ limit. And only then do we set $dx^{\mu}/dq=\nabla^{\mu}T/(-\nabla_{\nu}T\nabla^{\nu}T)^{1/2}$ in order to then get the trajectories.  

In our work we have found an intricate connection between trajectories and the value of $(dx_{\mu}/dq)(dx^{\mu}/dq)$. If solutions are on the light cone we can obtain trajectories directly by setting $(dx_{\mu}/dq)(dx^{\mu}/dq)=0$, and  integrating the geodesic or modified geodesic equations only recovers this result. However if $(dx_{\mu}/dq)(dx^{\mu}/dq)=-1$ we do not find the trajectories from this condition. Rather, we use the geodesic or modified geodesic equations, with this condition then serving as a boundary condition that provides an integration constant for the integration of the geodesic or modified geodesic equations.

Beyond this, we have shown that,  just as we would want, the solution to the eikonal velocity relation  is precisely the trajectory followed by the normal to a wavefront, regardless in fact of whether the wave is propagating on or off the light cone, with the general, non-approximate (\ref{10.2x}) holding for the normal to the wavefront no matter how big or small $T$ might be. This provides quite strong support for the validity of both the eikonal approximation relation given in (\ref{10.2x}) and for the relevance of the normalized eikonal velocity condition $dx^{\mu}/dq=\nabla^{\mu}T/(-\nabla_{\nu}T\nabla^{\nu}T)^{1/2}$ from which it follows. Thus  when $\nabla_{\nu}T\nabla^{\nu}T=0$ we eikonalize according to $dx^{\mu}/dq=\nabla^{\mu}T$ and obtain null geodesics on the light cone, while when $\nabla_{\nu}T\nabla^{\nu}T\neq 0$ we eikonalize according to $dx^{\mu}/dq=\nabla^{\mu}T/(-\nabla_{\nu}T\nabla^{\nu}T)^{1/2}$ and obtain trajectories that take support off the light cone. Moreover, no matter which option is relevant we will finish up with $dx^i/dx^0=\nabla^i{T}/\nabla^0T$, so that the velocities of points on trajectories are always determinable solely from the eikonal function $T$, with this being so regardless of whether $T$ is large or small. We thus distinguish between identifying $dx^{\mu}/dq$ with an appropriate derivative of $T$ (either $\nabla^{\mu}T$ or $\nabla^{\mu}T/(-\nabla_{\nu}T\nabla^{\nu}T)^{1/2}$), and then actually solving for $T$. Solving for $T$ when $T$ is large is known as the eikonal approximation, but relating $dx^{\mu}/dq$ to $\nabla^{\mu}T$ or $\nabla^{\mu}T/(-\nabla_{\nu}T\nabla^{\nu}T)^{1/2}$ is not an approximation at all. Rather, it gives the direction of the normal to the wavefront according to $n^{\mu}=dx^{\mu}/dq$, and all wavefronts with either large or small wavelength have normals. 

We have shown that it is possible to obtain eikonal trajectories that are exact without approximation (i.e., large or small $T$), and have shown that normals to advancing wavefronts follow these exact eikonal trajectories, with these trajectories being the trajectories along which energy and momentum are transported. In general then, in going from flat space to curved space one does not generalize flat space geodesics to curved space geodesics. Rather, one generalizes flat space wavefront normals (normals that are geodesic in flat space) to curved space wavefront normals, and in curved space normals to wavefronts do not have to be geodesic or restricted to the light cone.

\begin{acknowledgments}
The author would like to thank Dr. R. J. Adler, Dr. R. A. Walton, Dr. T. Schucker, Dr. M. C. Campigotto, Dr. A. Diaferio and Dr. L. Fatibene for some helpful communications.
\end{acknowledgments}

\appendix
\numberwithin{equation}{section}
\setcounter{equation}{0}

\section{The Perturbed Scalar Field Wave Equation}
\label{SA}
For fluctuations around a Robertson-Walker background it is very convenient to use the scalar, vector, tensor decomposition of the fluctuations that was introduced in \cite{Lifshitz1946,Bardeen1980}. Some recent reviews may be found in \cite{Kodama1984,Mukhanov1992,Stewart1990,Ma1995,Bertschinger1996,Zaldarriaga1998} and \cite{Dodelson2003,Mukhanov2005,Weinberg2008,Lyth2009,Ellis2012}. With such a decomposition the metric is written as 
\begin{align}
ds^2&=-(g_{\mu\nu}+h_{\mu\nu})dx^{\mu}dx^{\nu}=\Omega^2(p)\left[dp^2-\frac{dr^2}{1-kr^2}-r^2d\theta^2-r^2\sin^2\theta d\phi^2\right]
\nonumber\\
&+\Omega^2(p)\left[2\phi dp^2 -2(\tilde{\nabla}_i B +B_i)dp dx^i - [-2\psi\tilde{\gamma}_{ij} +2\tilde{\nabla}_i\tilde{\nabla}_j E + \tilde{\nabla}_i E_j + \tilde{\nabla}_j E_i + 2E_{ij}]dx^i dx^j\right].
\label{A.1xx}
\end{align}
In (\ref{A.1xx})  $\tilde{\nabla}_i=\partial/\partial x^i$ and  $\tilde{\nabla}^i=\tilde{\gamma}^{ij}\tilde{\nabla}_j$  (with Latin indices) are defined with respect to the background three-space metric $\tilde{\gamma}_{ij}$, and $(1,2,3)=(r,\theta,\phi)$. And with
\begin{eqnarray}
\tilde{\gamma}^{ij}\tilde{\nabla}_j V_i=\tilde{\gamma}^{ij}[\partial_j V_i-\tilde{\Gamma}^{k}_{ij}V_k]
\label{A.2xx}
\end{eqnarray}
for any three-vector $V_i$ in a three-space with three-space connection $\tilde{\Gamma}^{k}_{ij}$, the elements of (\ref{A.1xx}) are required to obey
\begin{eqnarray}
\tilde{\gamma}^{ij}\tilde{\nabla}_j B_i = 0,\quad \tilde{\gamma}^{ij}\tilde{\nabla}_j E_i = 0, \quad E_{ij}=E_{ji},\quad \tilde{\gamma}^{jk}\tilde{\nabla}_kE_{ij} = 0, \quad\tilde{\gamma}^{ij}E_{ij} = 0.
\label{A.3xx}
\end{eqnarray}
With the  three-space sector of the background geometry being maximally three-symmetric, it is described by a Riemann tensor of the form
\begin{eqnarray}
\tilde{R}_{ijk\ell}=k[\tilde{\gamma}_{jk}\tilde{\gamma}_{i\ell}-\tilde{\gamma}_{ik}\tilde{\gamma}_{j\ell}].
\label{A.4z}
\end{eqnarray}
The perturbed conformally coupled scalar field wave equation is of the form given in Sec. \ref{S9}, viz.

\begin{eqnarray}
-h^{\mu\nu}[\partial_{\mu}\partial_{\nu}-\Gamma^{\lambda}_{\mu\nu}]S
-g^{\mu\nu}\delta \Gamma^{\lambda}_{\mu\nu}S
+g^{\mu\nu}[\partial_{\mu}\partial_{\nu}-\Gamma^{\lambda}_{\mu\nu}]\delta S
+\frac{1}{6} \delta S R^{\alpha}_{\phantom{\alpha}\alpha}
+\frac{1}{6} S\delta R^{\alpha}_{\phantom{\alpha}\alpha}=0,
\label{A.5z}
\end{eqnarray}
and thus we evaluate the most complicated term in it, viz. $\delta R^{\alpha}_{\phantom{\alpha}\alpha}$.

To this end we note that with $R^{\alpha}_{\phantom{\alpha}\alpha}=g^{\alpha\beta}R_{\alpha\beta}$, the fluctuation in the Ricci scalar is given by
\begin{eqnarray}
\delta R^{\alpha}_{\phantom{\alpha}\alpha}=-h^{\alpha\beta}R_{\alpha\beta}+g^{\alpha\beta}\delta R_{\alpha\beta}.
\label{A.6z}
\end{eqnarray}
With the Einstein tensor being of the form $G_{\mu\nu}=R_{\mu\nu}-(1/2)g_{\mu\nu}R^{\alpha}_{\phantom{\alpha}\alpha}$, we can rewrite (\ref{A.6z}) in the form 
\begin{eqnarray}
\delta R^{\alpha}_{\phantom{\alpha}\alpha}=h^{\alpha\beta}G_{\alpha\beta}-g^{\alpha\beta}\delta G_{\alpha\beta}.
\label{A.7z}
\end{eqnarray}
This form is more convenient since for the metric given in  (\ref{A.1xx})  $G_{\alpha\beta}$ and $\delta G_{\alpha\beta}$ have been tabulated in many studies. Here we follow the presentation given in \cite{Phelps2019,Mannheim2020,Amarasinghe2020,Amarasinghe2021}, where it was shown that
\begin{align}
G_{00}&= -3k- 3 \dot{\Omega}^2\Omega^{-2},\quad G_{0i} =0,
\quad G_{ij} = \tilde{\gamma}_{ij}(k - \dot\Omega^2\Omega^{-2}+ 2\ddot\Omega \Omega^{-1}),\quad R^{\alpha}_{\phantom{\alpha}\alpha}=-6\Omega^{-2}k-6\ddot\Omega \Omega^{-3},
\label{A.8z}
\end{align}
\begin{eqnarray}
\delta G_{00}&=& -6 k \phi - 6 k \psi + 6 \dot{\psi} \dot{\Omega} \Omega^{-1} + 2 \dot{\Omega} \Omega^{-1} \tilde{\nabla}_{a}\tilde{\nabla}^{a}B - 2 \dot{\Omega} \Omega^{-1} \tilde{\nabla}_{a}\tilde{\nabla}^{a}\dot{E} - 2 \tilde{\nabla}_{a}\tilde{\nabla}^{a}\psi, 
 \nonumber\\ 
\delta G_{0i}&=& 3 k \tilde{\nabla}_{i}B -  \dot{\Omega}^2 \Omega^{-2} \tilde{\nabla}_{i}B + 2 \overset{..}{\Omega} \Omega^{-1} \tilde{\nabla}_{i}B - 2 k \tilde{\nabla}_{i}\dot{E} - 2 \tilde{\nabla}_{i}\dot{\psi} - 2 \dot{\Omega} \Omega^{-1} \tilde{\nabla}_{i}\phi +2 k B_{i} -  k \dot{E}_{i} \nonumber \\ 
&& -  B_{i} \dot{\Omega}^2 \Omega^{-2} + 2 B_{i} \overset{..}{\Omega} \Omega^{-1} + \frac{1}{2} \tilde{\nabla}_{a}\tilde{\nabla}^{a}B_{i} -  \frac{1}{2} \tilde{\nabla}_{a}\tilde{\nabla}^{a}\dot{E}_{i},
 \nonumber\\ 
\delta G_{ij}&=& -2 \overset{..}{\psi}\tilde{\gamma}_{ij} + 2 \dot{\Omega}^2\tilde{\gamma}_{ij} \phi \Omega^{-2} + 2 \dot{\Omega}^2\tilde{\gamma}_{ij} \psi \Omega^{-2} - 2 \dot{\phi} \dot{\Omega}\tilde{\gamma}_{ij} \Omega^{-1} - 4 \dot{\psi} \dot{\Omega}\tilde{\gamma}_{ij} \Omega^{-1} - 4 \overset{..}{\Omega}\tilde{\gamma}_{ij} \phi \Omega^{-1} \nonumber \\ 
&& - 4 \overset{..}{\Omega}\tilde{\gamma}_{ij} \psi \Omega^{-1} - 2 \dot{\Omega}\tilde{\gamma}_{ij} \Omega^{-1} \tilde{\nabla}_{a}\tilde{\nabla}^{a}B - \tilde{\gamma}_{ij} \tilde{\nabla}_{a}\tilde{\nabla}^{a}\dot{B} +\tilde{\gamma}_{ij} \tilde{\nabla}_{a}\tilde{\nabla}^{a}\overset{..}{E} + 2 \dot{\Omega}\tilde{\gamma}_{ij} \Omega^{-1} \tilde{\nabla}_{a}\tilde{\nabla}^{a}\dot{E} 
\nonumber \\ 
&& - \tilde{\gamma}_{ij} \tilde{\nabla}_{a}\tilde{\nabla}^{a}\phi +\tilde{\gamma}_{ij} \tilde{\nabla}_{a}\tilde{\nabla}^{a}\psi + 2 \dot{\Omega} \Omega^{-1} \tilde{\nabla}_{j}\tilde{\nabla}_{i}B + \tilde{\nabla}_{j}\tilde{\nabla}_{i}\dot{B} -  \tilde{\nabla}_{j}\tilde{\nabla}_{i}\overset{..}{E} - 2 \dot{\Omega} \Omega^{-1} \tilde{\nabla}_{j}\tilde{\nabla}_{i}\dot{E} \nonumber \\ 
&& + 2 k \tilde{\nabla}_{j}\tilde{\nabla}_{i}E - 2 \dot{\Omega}^2 \Omega^{-2} \tilde{\nabla}_{j}\tilde{\nabla}_{i}E + 4 \overset{..}{\Omega} \Omega^{-1} \tilde{\nabla}_{j}\tilde{\nabla}_{i}E + \tilde{\nabla}_{j}\tilde{\nabla}_{i}\phi -  \tilde{\nabla}_{j}\tilde{\nabla}_{i}\psi +\dot{\Omega} \Omega^{-1} \tilde{\nabla}_{i}B_{j} + \frac{1}{2} \tilde{\nabla}_{i}\dot{B}_{j}
\nonumber \\ 
&& -  \frac{1}{2} \tilde{\nabla}_{i}\overset{..}{E}_{j} -  \dot{\Omega} \Omega^{-1} \tilde{\nabla}_{i}\dot{E}_{j} + k \tilde{\nabla}_{i}E_{j} -  \dot{\Omega}^2 \Omega^{-2} \tilde{\nabla}_{i}E_{j} + 2 \overset{..}{\Omega} \Omega^{-1} \tilde{\nabla}_{i}E_{j} + \dot{\Omega} \Omega^{-1} \tilde{\nabla}_{j}B_{i} + \frac{1}{2} \tilde{\nabla}_{j}\dot{B}_{i} \nonumber \\ 
&& -  \frac{1}{2} \tilde{\nabla}_{j}\overset{..}{E}_{i} -  \dot{\Omega} \Omega^{-1} \tilde{\nabla}_{j}\dot{E}_{i} + k \tilde{\nabla}_{j}E_{i} -  \dot{\Omega}^2 \Omega^{-2} \tilde{\nabla}_{j}E_{i} + 2 \overset{..}{\Omega} \Omega^{-1} \tilde{\nabla}_{j}E_{i}- \overset{..}{E}_{ij} - 2 \dot{\Omega}^2 E_{ij} \Omega^{-2} \nonumber \\ 
&& - 2 \dot{E}_{ij} \dot{\Omega} \Omega^{-1} + 4 \overset{..}{\Omega} E_{ij} \Omega^{-1} + \tilde{\nabla}_{a}\tilde{\nabla}^{a}E_{ij},
 \nonumber\\
g^{\mu\nu}\delta G_{\mu\nu} &=& 6 \dot{\Omega}^2 \phi \Omega^{-4} + 6 \dot{\Omega}^2 \psi \Omega^{-4} - 6 \dot{\phi} \dot{\Omega} \Omega^{-3} - 18 \dot{\psi} \dot{\Omega} \Omega^{-3} - 12 \overset{..}{\Omega} \phi \Omega^{-3} - 12 \overset{..}{\Omega} \psi \Omega^{-3} - 6 \overset{..}{\psi} \Omega^{-2} + 6 k \phi \Omega^{-2} \nonumber \\ 
&& + 6 k \psi \Omega^{-2} - 6 \dot{\Omega} \Omega^{-3} \tilde{\nabla}_{a}\tilde{\nabla}^{a}B - 2 \Omega^{-2} \tilde{\nabla}_{a}\tilde{\nabla}^{a}\dot{B} + 2 \Omega^{-2} \tilde{\nabla}_{a}\tilde{\nabla}^{a}\overset{..}{E} + 6 \dot{\Omega} \Omega^{-3} \tilde{\nabla}_{a}\tilde{\nabla}^{a}\dot{E} \nonumber \\ 
&& - 2 \dot{\Omega}^2 \Omega^{-4} \tilde{\nabla}_{a}\tilde{\nabla}^{a}E + 4 \overset{..}{\Omega} \Omega^{-3} \tilde{\nabla}_{a}\tilde{\nabla}^{a}E + 2 k \Omega^{-2} \tilde{\nabla}_{a}\tilde{\nabla}^{a}E - 2 \Omega^{-2} \tilde{\nabla}_{a}\tilde{\nabla}^{a}\phi + 4 \Omega^{-2} \tilde{\nabla}_{a}\tilde{\nabla}^{a}\psi. 
\label{A.9z}
\end{eqnarray}
In  \cite{Phelps2019,Mannheim2020,Amarasinghe2020,Amarasinghe2021} it was shown that for any Robertson-Walker background  with any $k$ or $\Omega(p)$ the particular combinations 
\begin{align}
\alpha=\phi + \psi + \dot B - \ddot E,\quad  \gamma= - \dot\Omega^{-1}\Omega \psi + B - \dot E,\quad  B_i-\dot{E}_i,  \quad E_{ij}
\label{A.10z}
\end{align}
of the fluctuations  are gauge invariant. Interestingly, these combinations are independent of $k$ even if $k$ is nonzero.

Given (\ref{A.10z}), we thus look to reexpress (\ref{A.7z}) in terms of gauge invariant combinations, and following some algebra, we obtain
\begin{eqnarray}
g^{\alpha\beta}\delta G_{\alpha\beta}=&-&12 \ddot{\Omega}  \Omega^{-3}(\alpha - \dot\gamma) -6 \dot{\Omega} \Omega^{-3}(\dot{\alpha} -\ddot\gamma)-2 \Omega^{-2} \tilde{\nabla}_{a}\tilde{\nabla}^{a}(\alpha +3\dot\Omega\Omega^{-1}\gamma)
+(2k\Omega^{-2}+4\ddot{\Omega}\Omega^{-3}-2\dot{\Omega}^2\Omega^{-4}) \tilde{\nabla}_{a}\tilde{\nabla}^{a}E
\nonumber\\
&+&6\dot{\Omega}^2\Omega^{-4}\phi+6\dot{\Omega}^2\Omega^{-4}\psi+6\ddot{\Omega}\Omega^{-3}\psi+6k\Omega^{-2}\phi+6k\Omega^{-2}\psi-6\overset{...}{\Omega}\Omega^{-2}\dot{\Omega}^{-1}\psi,
\label{A.11y}
\end{eqnarray}
\begin{eqnarray}
-h^{\alpha\beta}G_{\alpha\beta}=&-&(2k\Omega^{-2}+4\ddot{\Omega}\Omega^{-3}-2\dot{\Omega}^2\Omega^{-4}) \tilde{\nabla}_{a}\tilde{\nabla}^{a}E
\nonumber\\
&-&6k\Omega^{-2}\phi-6\dot{\Omega}^2\Omega^{-4}\phi+6k\Omega^{-2}\psi-6\dot{\Omega}^2\Omega^{-4}\psi+12\ddot{\Omega}\Omega^{-3}\psi.
\label{A.12y}
\end{eqnarray}
Combining these relations we obtain 
\begin{eqnarray}
\delta R^{\alpha}_{\phantom{\alpha}\alpha}
=12 \ddot{\Omega}  \Omega^{-3}(\alpha - \dot\gamma) +6 \dot{\Omega} \Omega^{-3}(\dot{\alpha} -\ddot\gamma)
+2 \Omega^{-2} \tilde{\nabla}_{a}\tilde{\nabla}^{a}(\alpha +3\dot\Omega\Omega^{-1}\gamma)+6\Omega^{-3}\dot{\Omega}^{-1}\dot{A}\psi,
\label{A.13z}
\end{eqnarray}
where 
\begin{eqnarray}
A=\Omega\ddot{\Omega}-2 \dot{\Omega}^2-k\Omega^2,\quad \dot{A}=\Omega\overset{...}{\Omega}-3\dot{\Omega}\ddot{\Omega}-2k\Omega\dot{\Omega}.
\label{A.14z}
\end{eqnarray}
We should thus use (\ref{A.13z}) in (\ref{A.5z}). We note that in (\ref{A.13z}) only the $\psi$-dependent  term is not gauge invariant.

Since $\delta R^{\alpha}_{\phantom{\alpha}\alpha}$ is associated with fluctuations in the spin zero scalar field sector, only the scalar fluctuations given in (\ref{A.1xx}) appear in (\ref{A.13z}). The other gauge invariant combinations given in (\ref{A.10z}) can appear in fluctuations involving vector modes or even those involving the propagation of gravitational modes themselves. For spin one and spin two modes we would replace $\delta R^{\alpha}_{\phantom{\alpha}\alpha}$ by $\delta R^{\alpha\beta}\epsilon_{\alpha (a)}\epsilon_{\beta (a)}$ and $\delta R^{\alpha\beta}\epsilon_{\alpha\beta (a)}$ as appropriately summed on $(a)$, where $\epsilon_{\alpha (a)}$ and $\epsilon_{\alpha\beta (a)}$ are polarization vectors and tensors. These polarization vectors and tensors would project onto the various components of $\delta G_{\mu\nu}$ as listed in (\ref{A.9z}) and not just onto its trace.

While in general we would have to deal with the function  $A(p)$ in any application of (\ref{A.5z})  and (\ref{A.13z}), it was noted in  \cite{Mannheim2020,Amarasinghe2021} that if the background perfect fluid is just a cosmological constant term (such as in the inflationary universe), then for standard gravity or conformal gravity fluctuations $A(p)$ just happens to vanish identically (with $\delta R^{\alpha}_{\phantom{\alpha}\alpha}$ then being gauge invariant). For $k=0$ this entails that $A(p)=1/p$, for $k=-1$ we find that $\Omega(p)=1/\sinh p$, while for $k=+1$ we find that $\Omega(p)=1/\sin p$. 

As noted in \cite{Weinberg2008}, the standard use of the Boltzmann equation in cosmological fluctuation theory can be recast as a study of the propagation of light on a perturbed Robertson-Walker light cone. Since we are arguing in this paper that gravity actually takes massless particles off the light cone we need to ask what then happens to the Boltzmann equation. In equilibrium the distribution function $f(q,x^{\alpha},U^{\beta})$ schematically obeys 
\begin{eqnarray}
\left[\frac{\partial}{\partial q}+U^{\lambda}\frac{\partial}{\partial x^{\lambda}}+\frac{\partial U^{\lambda}}{\partial q}\frac{\partial}{\partial U^{\lambda}}\right]f(q,x^{\alpha},U^{\beta})=0, 
\label{A.15xx}
\end{eqnarray}
where $U^{\lambda}=dx^{\lambda}/dq$. For particles that obey the standard massless particle geodesic equation (\ref{1.6x})  we can rewrite (\ref{A.15xx}) as 
\begin{eqnarray}
\left[\frac{\partial}{\partial q}+U^{\lambda}\frac{\partial}{\partial x^{\lambda}}-\Gamma^{\lambda}_{\mu\nu} U^{\mu}U^{\nu}\frac{\partial}{\partial U^{\lambda}}\right]f(q,x^{\alpha},U^{\beta})=0. 
\label{A.16xx}
\end{eqnarray}
However, given (\ref{9.7xx}), we  replace (\ref{A.16xx}) by 
\begin{eqnarray}
&&\left[\frac{\partial}{\partial q}+U^{\lambda}\frac{\partial}{\partial x^{\lambda}}-\Gamma^{\lambda}_{\mu\nu} U^{\mu}U^{\nu}\frac{\partial}{\partial U^{\lambda}}\right]f(q,x^{\alpha},U^{\beta})
\nonumber\\
&&=-\delta^{\lambda}\frac{\partial f(q,x^{\alpha},U^{\beta})}{\partial x^{\lambda}}
+\left[\frac{d\delta^{\lambda}}{dq}+\left[\delta\Gamma^{\lambda}_{\mu \nu} U^{\mu}U^{\nu}+\Gamma^{\lambda}_{\mu \nu}(\delta^{\mu}U^{\nu}+U^{\mu}\delta^{\nu})\right]\right]
\frac{\partial f(q,x^{\alpha},U^{\beta})}{\partial U^{\lambda}}.
\label{A.17xx}
\end{eqnarray}
In the language of transport theory the right-hand side of (\ref{A.17xx}) may be thought of as being a gravitationally-induced viscosity.

\section{Equivalence of the Eikonal Trajectory to the Normal to the Wavefront -- Timelike Case}
\label{SB}

When a massless wave equation has solutions that take support  off the light cone the normals to the wavefronts follow trajectories that are also off the light cone. We now show in a simple solvable example that the modified trajectory that we obtained in the eikonal approximation in such a case coincides with none other than the normal to the wavefront. The model that we consider is the massless minimally coupled scalar field we introduced in Sec. \ref{S1e} in which 
\begin{eqnarray}
ds^2=\Omega^2(p)(dp^2-dx^2-dy^2-dz^2),
\label{B.1x}
\end{eqnarray}
with $\Omega(p)=\sin(Hp)$. Solutions to $\Omega^{-2}(-\partial_0^2-2\Omega^{-1}\dot{\Omega}\partial_0+\vec{\nabla}^2)S=0$ are of the form
\begin{eqnarray}
S(x)=\frac{e^{ik\cdot x}}{\sin(Hp)},
\label{B.2x}
\end{eqnarray}
where $k_0^2=k_1^2+k_2^2+k_3^2+H^2$. If we now set $S(x)=Ae^{iT}$ we obtain
\begin{eqnarray}
A=\frac{1}{\sin(Hp)},\quad T=k\cdot x,
\label{B.3x}
\end{eqnarray}
with the wave equation taking the form
\begin{eqnarray}
\nabla_{\mu}T\nabla^{\mu}T=A^{-1}\nabla_{\mu}\nabla^{\mu}A=\Omega^{-3}\ddot{\Omega}=-\frac{H^2}{\sin^2(Hp)}.
\label{B.4x}
\end{eqnarray}
The  utility of this model is that it is solvable analytical, and can thus provide some general insight.

For the eikonal procedure, we take the propagation to be the timelike $(dx_{\mu}/dq)(dx^{\mu}/dq)=-1$, and following (\ref{10.1x}) and  (\ref{10.2x}) we set  
\begin{eqnarray}
\frac{dx^{\mu}}{dq}=\frac{\nabla^{\mu}T}{(-\nabla_{\nu}T\nabla^{\nu}T)^{1/2}}=\frac{\nabla^{\mu}T}{(-A^{-1}\nabla_{\nu}\nabla^{\nu}A)^{1/2}},
\quad \frac{dx_{\mu}}{dq}\frac{dx^{\mu}}{dq}=-1.
\label{B.5x}
\end{eqnarray}
However unlike the short wavelength limit that is usually required for the eikonal approximation, in this particular case the relation $\nabla_{\mu}T\nabla^{\mu}T=A^{-1}\nabla_{\mu}\nabla^{\mu}A$ is exact. Thus the only thing that we need to test is the validity of  the identification of $dx^{\mu}/dq$ with  $\nabla^{\mu}T/(-\nabla_{\nu}T\nabla^{\nu}T)^{1/2}$.

With the steps leading to (\ref{3.12z}) being identical we obtain 
\begin{eqnarray}
\frac{d^2x^{\lambda} }{ dq^2}
+\Gamma^{\lambda}_{\mu \nu} 
\frac{dx^{\mu}}{ dq}\frac{dx^{\nu } }{dq}=
-\frac{1}{2(-A^{-1}\nabla_{\nu}\nabla^{\nu}A)}
\left[g^{\lambda \mu}+\frac{dx^{\lambda}}{dq}\frac{dx^{\mu}}{dq}\right]\frac{\partial (-A^{-1}\nabla_{\nu}\nabla^{\nu}A)}{\partial x^{\mu}},
\label{B.6x}
\end{eqnarray}
Since in this model $-R^{\alpha}_{\phantom{\alpha}\alpha}/6=\ddot{\Omega}/\Omega^3=-H^2/\Omega^2=A^{-1}\nabla_{\mu}\nabla^{\mu}A$, we could  write (\ref{B.6x}) in the form equivalent to that given in (\ref{3.19qq}).

For a wave propagating in the $x^3$ direction the relevant components of the connection are
\begin{eqnarray}
\Gamma^0_{00}=\Gamma^0_{33}=\Gamma^3_{03}=\frac{\dot{\Omega}}{\Omega}=\frac{H\cos(Hp)}{\sin(Hp)},\qquad \Gamma^0_{03}=\Gamma^3_{00}=\Gamma^3_{33}=0.
\label{B.7x}
\end{eqnarray}
For this mode (\ref{B.6x}) yields
\begin{eqnarray}
\frac{d^2x^3}{dq^2}+\frac{\dot{\Omega}}{\Omega}\frac{dx^3}{dq}\frac{dx^0}{dq}=0, 
\label{B.8x}
\end{eqnarray}
\begin{eqnarray}
\frac{d^2x^0}{dq^2}+\frac{\dot{\Omega}}{\Omega}\left(\frac{dx^3}{dq}\frac{dx^3}{dq}+\frac{1}{\Omega^2}\right)=0.
\label{B.9x}
\end{eqnarray}
For (\ref{B.8x}) the solution is 
\begin{eqnarray}
\frac{dx^3}{dq}=\frac{f}{\Omega}, 
\label{B.10x}
\end{eqnarray}
where $f$ is an integration constant. To find the solution to (\ref{B.9x}) we set $(dx_{\mu}/dq)(dx^{\mu}/dq)=-B$ where initially $B$ is a general constant. This yields
\begin{eqnarray}
-\Omega^2\left(\frac{dx^0}{dq}\right)^2+\Omega^2\left(\frac{dx^3}{dq}\right)^2=-B,
\label{B.11x}
\end{eqnarray}
so that
\begin{eqnarray}
\left(\frac{dx^0}{dq}\right)^2-\frac{(B+f^2)}{\Omega^2}=0.
\label{B.12x}
\end{eqnarray}
Differentiating (\ref{B.12x}) with respect to $q$ gives 
\begin{eqnarray}
\frac{d^2x^0}{dq^2}+\frac{\dot{\Omega}}{\Omega}\frac{(f^2+B)}{\Omega^2}=0.
\label{B.13x}
\end{eqnarray}
As we see, (\ref{B.13x}) is consistent with (\ref{B.9x}), and entails that $B=1$, just as required. With this value for $B$ the general solution to (\ref{B.6x}) is then of the form
\begin{eqnarray}
\frac{dx^0}{dq}=\frac{(f^2+1)^{1/2}}{\Omega}, \qquad \frac{dx^3}{dq}=\frac{f}{\Omega}, 
\label{B.14x}
\end{eqnarray}
and leads to a slower than light velocity in conformal and comoving time of the form
\begin{eqnarray}
\frac{dx^3}{dp}=\frac{1}{(1+f^{-2})^{1/2}},\qquad a(t)\frac{dx^3}{dt}=\frac{1}{(1+f^{-2})^{1/2}}, 
\label{B.15x}
\end{eqnarray}
where $a(t)=(1-H^2t^2)^{1/2}$. We can thus interpret $(1+f^{-2})^{1/2}$ as a refractive index $n$, one that in this case obeys $n>1$, just as it would need to. Our model is thus of interest since it shows how gravity can produce a refractive index that acts in complete analog to the refractive index of geometrical optics.

With (\ref{B.5x}) leading to 
\begin{eqnarray}
\frac{dx_{\mu}}{dq}=\frac{ k_{\mu}}{(-k_{\nu}k^{\nu})^{1/2}}=\frac{\Omega k_{\mu}}{H},
\label{B.16x}
\end{eqnarray}
and (\ref{B.14x}) leading to 
\begin{eqnarray}
\frac{dx_0}{dq}=-\Omega(f^2+1)^{1/2}, \qquad \frac{dx_3}{dq}=\Omega f, 
\label{B.17x}
\end{eqnarray}
we thus identify
\begin{eqnarray}
k_0=-H(f^2+1)^{1/2}, \qquad k_3=H f, 
\label{B.18x}
\end{eqnarray}
and this brings us right back to 
\begin{eqnarray}
k_0^3- k_3^2=H^2 
\label{B.19x}
\end{eqnarray}
just as it should. 

We thus see that solutions to the eikonal relation given in (\ref{B.6x}) indeed represent rays propagating in the $x^3$ direction.  However, the wave equation solution given in (\ref{B.2x}) describes a wave propagating in the $x^3$ direction with the transverse $x^1$ and $x^2$ being in the wavefront. The solution to the eikonal approximation is thus precisely the trajectory followed by the normal to the wavefront, just as we would want it to be. This provides quite strong support for the validity of (\ref{B.6x}) and for the normalized eikonal condition given in (\ref{B.5x}) from which (\ref{B.6x}) follows.

\section{Equivalence of the Eikonal Trajectory to the Normal to the Wavefront -- Spacelike Case}
\label{SC}

Having solved the massless minimally coupled scalar field model in the timelike case, for completeness we now solve  the  model in the spacelike case. As before, we take the background line element to be of the form  
\begin{eqnarray}
ds^2=\Omega^2(p)(dp^2-dx^2-dy^2-dz^2),
\label{C.1x}
\end{eqnarray}
and this time set $\Omega(p)=e^{Hp}$. Solutions to $\Omega^{-2}(-\partial_0^2-2\Omega^{-1}\dot{\Omega}\partial_0+\vec{\nabla}^2)S=0$ are of the form
\begin{eqnarray}
S(x)=e^{-Hp}e^{ik\cdot x},
\label{C.2x}
\end{eqnarray}
where $k_0^2=k_1^2+k_2^2+k_3^2-H^2$. If we now set $S(x)=Ae^{iT}$ we obtain
\begin{eqnarray}
A=e^{-Hp},\quad T=k\cdot x,
\label{C.3x}
\end{eqnarray}
with the wave equation taking the form
\begin{eqnarray}
\nabla_{\mu}T\nabla^{\mu}T=A^{-1}\nabla_{\mu}\nabla^{\mu}A=\Omega^{-3}\ddot{\Omega}=H^2e^{-2Hp}.
\label{C.4x}
\end{eqnarray}
Even though this model is not physical since the propagator takes support outside the light cone, its utility is that it is solvable analytical, and can thus provide some general insight.

For the eikonal procedure, we take the propagation to be the spacelike $(dx_{\mu}/dq)(dx^{\mu}/dq)=+1$, and following (\ref{10.1x}) and  (\ref{10.2x}) we set  
\begin{eqnarray}
\frac{dx^{\mu}}{dq}=\frac{\nabla^{\mu}T}{(\nabla_{\nu}T\nabla^{\nu}T)^{1/2}}=\frac{\nabla^{\mu}T}{(A^{-1}\nabla_{\nu}\nabla^{\nu}A)^{1/2}},
\quad \frac{dx_{\mu}}{dq}\frac{dx^{\mu}}{dq}=+1.
\label{C.5x}
\end{eqnarray}
In this particular case the relation $\nabla_{\mu}T\nabla^{\mu}T=A^{-1}\nabla_{\mu}\nabla^{\mu}A$ is exact. Thus the only thing that we need to test is the validity of  the identification of $dx^{\mu}/dq$ with  $\nabla^{\mu}T/(+\nabla_{\nu}T\nabla^{\nu}T)^{1/2}$.

With the steps leading to (\ref{3.12z}) being identical we obtain 
\begin{eqnarray}
\frac{d^2x^{\lambda} }{ dq^2}
+\Gamma^{\lambda}_{\mu \nu} 
\frac{dx^{\mu}}{ dq}\frac{dx^{\nu } }{dq}=
\frac{1}{2(A^{-1}\nabla_{\nu}\nabla^{\nu}A)}
\left[g^{\lambda \mu}-\frac{dx^{\lambda}}{dq}\frac{dx^{\mu}}{dq}\right]\frac{\partial (A^{-1}\nabla_{\nu}\nabla^{\nu}A)}{\partial x^{\mu}},
\label{C.6x}
\end{eqnarray}
and note the sign change on the right-hand side of (\ref{C.6x}) compared to the right-hand side of (\ref{B.6x}). (Since in this model $-R^{\alpha}_{\phantom{\alpha}\alpha}/6=\ddot{\Omega}/\Omega^3=H^2/\Omega^2=A^{-1}\nabla_{\mu}\nabla^{\mu}A$, we could  write (\ref{C.6x}) in the form equivalent to that given in (\ref{3.19qq}) as adjusted for the spacelike $(dx_{\mu}/dq)(dx^{\mu}/dq)=+1$.)

For a wave propagating in the $x^3$ direction the relevant components of the connection are
\begin{eqnarray}
\Gamma^0_{00}=\Gamma^0_{33}=\Gamma^3_{03}=\frac{\dot{\Omega}}{\Omega}=H,\qquad \Gamma^0_{03}=\Gamma^3_{00}=\Gamma^3_{33}=0.
\label{C.7x}
\end{eqnarray}
For this mode (\ref{C.6x}) yields
\begin{eqnarray}
\frac{d^2x^3}{dq^2}+H\frac{dx^3}{dq}\frac{dx^0}{dq}=0, 
\label{C.8x}
\end{eqnarray}
\begin{eqnarray}
\frac{d^2x^0}{dq^2}+H\frac{dx^3}{dq}\frac{dx^3}{dq}-\frac{H}{\Omega^2}=0.
\label{C.9x}
\end{eqnarray}
For (\ref{C.8x}) the solution is 
\begin{eqnarray}
\frac{dx^3}{dq}=\frac{f}{\Omega}, 
\label{C.10x}
\end{eqnarray}
where $f$ is an integration constant. To find the solution to (\ref{C.9x}) we set $(dx_{\mu}/dq)(dx^{\mu}/dq)=B$ where initially $B$ is a general constant. This yields
\begin{eqnarray}
-\Omega^2\left(\frac{dx^0}{dq}\right)^2+\Omega^2\left(\frac{dx^3}{dq}\right)^2=B,
\label{C.11x}
\end{eqnarray}
so that
\begin{eqnarray}
\left(\frac{dx^0}{dq}\right)^2+\frac{(B-f^2)}{\Omega^2}=0.
\label{C.12x}
\end{eqnarray}
Differentiating (\ref{C.12x}) with respect to $q$ gives 
\begin{eqnarray}
\frac{d^2x^0}{dq^2}+\frac{Hf^2}{\Omega^2}-\frac{HB}{\Omega^2}=0.
\label{C.13x}
\end{eqnarray}
As we see, (\ref{C.13x}) is consistent with (\ref{C.9x}), and entails that $B=1$, just as required. Then with $B=1$, from (\ref{C.12x}) it follows that $f^2>1$. With this value for $B$ the general solution to (\ref{C.6x}) is then of the form
\begin{eqnarray}
\frac{dx^0}{dq}=\frac{(f^2-1)^{1/2}}{\Omega}, \qquad \frac{dx^3}{dq}=\frac{f}{\Omega}, 
\label{C.14x}
\end{eqnarray}
and leads to superluminal velocity in conformal and comoving time of the form
\begin{eqnarray}
\frac{dx^3}{dp}=\frac{1}{(1-f^{-2})^{1/2}},\qquad a(t)\frac{dx^3}{dt}=\frac{1}{(1-f^{-2})^{1/2}}, 
\label{C.15x}
\end{eqnarray}
where $a(t)=Ht$. We can thus interpret $(1-f^{-2})^{1/2}$ as a refractive index $n$, one that in this case obeys $n<1$, just as desired.

With (\ref{C.5x}) leading to 
\begin{eqnarray}
\frac{dx_{\mu}}{dq}=\frac{\Omega k_{\mu}}{H},
\label{C.16x}
\end{eqnarray}
and (\ref{C.14x}) leading to 
\begin{eqnarray}
\frac{dx_0}{dq}=-\Omega(f^2-1)^{1/2}, \qquad \frac{dx_3}{dq}=\Omega f, 
\label{C.17x}
\end{eqnarray}
we thus identify
\begin{eqnarray}
k_0=-H(f^2-1)^{1/2}, \qquad k_3=H f, 
\label{C.18x}
\end{eqnarray}
and this brings us right back to the tachyonic
\begin{eqnarray}
k_0^3- k_3^2=-H^2 
\label{C.19x}
\end{eqnarray}
just as it should. 

We thus see that solutions to the eikonal relation given in (\ref{C.6x}) indeed represent rays propagating in the $x^3$ direction.  However, the wave equation solution given in (\ref{C.2x}) describes a wave propagating in the $x^3$ direction with the transverse $x^1$ and $x^2$ being in the wavefront. The solution to the eikonal approximation is thus precisely the trajectory followed by the normal to the wavefront, just as we would want it to be. This provides quite strong support for the validity of (\ref{C.6x}) and for the normalized eikonal condition given in (\ref{C.5x}) from which (\ref{C.6x}) follows.


\begin{thebibliography}{99}
\bibitem{footnote1} Under the transformation $x^{\prime \lambda} =
x^{\lambda}+ \tfrac{1}{2}x^{\mu}x^{\nu}(\Gamma^{\lambda}_{\mu \nu})_{P}$, the
primed  coordinate  $(\Gamma^{\prime \lambda}_{\mu
\nu})_{P}$ will vanish at the point $P$, regardless in fact
of how large the Riemann tensor in the neighborhood of the point $P$
might actually be.
\bibitem{footnote1a} Any given four-vector $V_{\mu}$ can be written as $V_{\mu}=A_{\mu} +\nabla _{\mu}B$, where $A_{\mu}$ is transverse, viz. $\nabla_{\mu}A^{\mu}=0$, and $B$ is a scalar. In consequence, $\nabla_{\mu}V^{\mu}=\nabla_{\mu}\nabla^{\mu}B$. Thus with $\nabla_{\mu}\nabla^{\mu}D(x-x^{\prime})=\delta^4(x-x^{\prime})$, we can write $V_{\mu}=A_{\mu}+\nabla_{\mu}\int d^4x^{\prime}D(x-x^{\prime})\nabla^{\prime}_{\mu}V^{\mu}(x^{\prime})$. The content of being able to write a four-vector as a total derivative is that the transverse $A_{\mu}$ type component is absent.
\bibitem{footnote1d} While this example is only meant to be illustrative,  for $\Omega(p)=e^{Hp}$ it does lead to propagation outside the light cone for modes with $|\vec{k}|^2<H^2$. To  ensure that $(k^0)^2$ could not go negative would require that $\ddot{\Omega}/\Omega$ be negative (cf. $\Omega(p)=\sin(Hp)$) or zero. For a power-behaved  $\Omega(p)=p^{n}$ this would require that $0\leq n\leq 1$. In turn, we would have $t=p^{n+1}/(n+1)$ and $a(t)=t^{n/(n+1)}$. Thus the associated $a(t)$ would also be power behaved with a power $a(t)=t^{\alpha}$ where $\alpha$ would be required to obey $0\leq \alpha \leq 1/2$. In the standard gravity cosmological zero three-curvature radiation era the expansion radius is given by  $a=t^{1/2}$ ($\Omega(p)=p$). Consequently, propagators associated with $\nabla_{\mu}\nabla^{\mu}S=0$ would be causal. However, propagators associated with the standard gravity cosmological matter era where $a(t)=t^{2/3}$ ($\Omega(p)=p^2$) would not be. Moreover, propagators associated with the standard gravity de Sitter era where $a(t)=e^{Ht}$ ($\Omega(p)=1/Hp$) would also not be causal.
\bibitem{footnote1b} There is still the issue of whether the associated geodesics obey a geodesic equation based on the wave equation given in (\ref{1.41h}) or on that given in (\ref{1.28h}). In Sec. \ref{S3g} we show that the two are equivalent.
\bibitem{footnote1c} While we can always conformally transform the Ricci scalar away, if the underlying geometry is conformal to flat the transformation that transforms it away will at the same time bring the entire geometry to flat, and bring the wave equation to (\ref{1.28h}) and not to the second relation in (\ref{1.48h}).
\bibitem{footnote2} For $T$ possessing a term of the form $k_0x^0$, $\nabla_{\mu}T\nabla^{\mu}T$ behaves like $k_0^2$ for large $k_0$ while $\nabla_{\mu}\nabla^{\mu}T$ is of only order $k_0$.
\bibitem{footnote3b} By first transforming to the barred basis in which $\bar{\nabla}_{\mu}\bar{\nabla}^{\mu}\bar{S}=0$, we transformed to a basis in which we knew that the associated propagator took support off the light cone. On conformally transforming back we obtain (\ref{3.19qq}), which then automatically takes support off the light cone too.
\bibitem{footnote4} Choosing to take the integrand to be $(-R^{\alpha}_{\phantom{\alpha}\alpha})^{1/2}$ in the action $I_T=- \kappa\int ds(-R^{\alpha}_{\phantom{\alpha}\alpha})^{1/2}$ renders the constant $\kappa$ dimensionless.
\bibitem{footnote5} While beyond the scope of the present paper, this result does raise the question of whether any eikonalized trajectory that can be obtained form a wave equation could also correspond to one that could be obtained from some appropriate test particle action.
\bibitem{footnote7} Similarly, if we drop the $\nabla_{\mu}\nabla^{\mu}A$ term from (\ref{3.3z}) and set $T=\int^x k_{\mu}dx^{\mu}$, we would obtain $k_{\mu}k^{\mu}-R^{\alpha}_{\phantom{\alpha}\alpha}/6=0$. Thus unless $R^{\alpha}_{\phantom{\alpha}\alpha}=0$, $k_{\mu}k^{\mu}$ could not be zero. Heuristically, we could regard $k_{\mu}k^{\mu}-R^{\alpha}_{\phantom{\alpha}\alpha}/6=0$ as being the ``light cone" in this case as conformal modes would satisfy this condition.
\bibitem{footnote8} The curved space massless fermion  Dirac action $I_D=-\int d^4x(-g)^{1/2}i\bar{\psi}\gamma^{c}V^{\mu}_c(x)[\partial_\mu+\Gamma_\mu(x)]\psi$  is locally conformal invariant under $g_{\mu\nu}(x)\rightarrow \Omega^2(x)g_{\mu\nu}(x)$, $V^a_{\mu}(x)\rightarrow \Omega(x)V^a_{\mu}(x)$, $\psi(x)\rightarrow \Omega^{-3/2}(x)\psi(x)$, where $V^{\mu}_c(x)$ is a vierbein and $\Gamma_{\mu}=-(1/8)[\gamma_a,\gamma_b](V^b_{\nu}\partial_{\mu}V^{a\nu}+V^b_{\lambda}\Gamma^{\lambda}_{\phantom{\lambda}\nu\mu}V^{a\nu})$ is the spin connection. On ``squaring" the Dirac equation $i\gamma^{c}V^{\mu}_c(x)[\partial_\mu+\Gamma_\mu(x)]\psi(x)=0$ we obtain \cite{Mannheim2006} $[\partial_{\mu}+\Gamma_{\mu}][\partial^{\mu}+\Gamma^{\mu}]\psi(x)+\tfrac{1}{4}R^{\alpha}_{\phantom{\alpha}\alpha}\psi(x)=0$, to thus generate a dependence on the Ricci scalar similar to that found in the scalar field case. 
On defining $D_{\mu}=\partial_{\mu}+\Gamma_{\mu}$ and setting $\psi=\psi_0\exp(iT)$ where $\psi_0$ is a Dirac spinor, eikonalization  for fermions is of the form $\bar{\psi}_0(dx_{\mu}/dq)\psi_0=\bar{\psi}_0D_{\mu}T\psi_0$. In passing we note that both the scalar and the fermion wave equation operators can generically be written in the form $\nabla_{\mu}\nabla^{\mu}+(d/6)R^{\alpha}_{\phantom{\alpha}\alpha}$, $D_{\mu}D^{\mu}+(d/6)R^{\alpha}_{\phantom{\alpha}\alpha}$, where $d$ is the dimension of the field ($d=1$ for scalars and $d=3/2$ for fermions).
\bibitem{footnote9} For fermions we would take the action to be of the form $I_D=-\int d^4x(-g)^{1/2}\bar{\psi }[i\gamma^{c}V^{\mu}_c(x)[\partial_\mu+\Gamma_\mu(x)]+Q]\psi$, and on setting $Q=Q_0=m$ would obtain $[\partial_{\mu}+\Gamma_{\mu}][\partial^{\mu}+\Gamma^{\mu}]\psi(x)+[\tfrac{1}{4}R^{\alpha}_{\phantom{\alpha}\alpha}-m^2]\psi(x)=0$. On setting $D_{\mu}=\partial_{\mu}+\Gamma_{\mu}$ as per \cite{footnote8}, eikonalization in a form analogous to (\ref{3.35qq}) would follow if we set $\bar{\psi}_0(dx_{\mu}/dq)\psi_0=\bar{\psi}_0D_{\mu}T\psi_0/(m^2-R^{\alpha}_{\phantom{\alpha}\alpha}/4)^{1/2}$. While this would in principle actually lead to a departure from the standard massive particle geodesics given in (\ref{1.2x}), in practice this would not be the case since $m^2$ is typically much greater than curvature scales such as $R^{\alpha}_{\phantom{\alpha}\alpha}$.
\bibitem{footnote10a} To appreciate the distinction, we note that each of the $\int d^4x(-g)^{1/2}R_{\mu\nu}R^{\mu\nu}$ and $\int d^4x(-g)^{1/2}(R^{\alpha}_{\phantom{\alpha}\alpha})^2$ actions is separately globally conformal invariant under $g_{\mu\nu}\rightarrow e^{2\alpha}g_{\mu\nu}$ when $\alpha$ is as constant, but only the particular linear combination $\int d^4x(-g)^{1/2}[R_{\mu\nu}R^{\mu\nu}-(1/3)(R^{\alpha}_{\phantom{\alpha}\alpha})^2]$ is invariant (up to a total divergence Gauss-Bonnet term) when $\alpha$ is an arbitrary function of the spacetime coordinates.
\bibitem{footnote10} Even though $g_{\mu\nu}(x)$ does transform non-trivially under a local conformal transformation, and even though $A_{\mu}(x)$ has  engineering dimension one, under a local conformal transformation $A_{\mu}(x)$ is left invariant. This behavior is complementary to the behavior found in a gauge theory, since there it is $A_{\mu}(x)$ that transforms non-trivially under a local gauge transformation (viz.  $A_{\mu}(x)\rightarrow A_{\mu}(x)-\nabla_{\mu}\chi$) while it is $g_{\mu\nu}(x)$ that is left invariant.
\bibitem{footnote9b} Since a pure de Sitter geometry is on its own conformal to flat, for pure de Sitter the trajectories are null geodesics, $(dx_{\mu}/dq)(dx^{\mu}/dq)$ is zero, and (\ref{4.12x}) does not apply.
\bibitem{footnote11}  As had been noted in Sec. \ref{S1e}, obtaining contributions inside the light cone is actually quite familiar in conformal theories.
\bibitem{footnote11a} While the $Kr^2$ term does drop out of (\ref{5.6x}), we should note that not all trace of this term is lost since the point of closest approach, viz. $r_0$,  is the point at which  $dr/d\phi$ is zero, i.e. the point at which $r_0^2=J^2(1-2MG/r_0-Kr_0^2)$, with $r_0$ thus depending on $K$.
\bibitem{footnote11b} Technically in \cite{Kraniotis2011,Sultana2013,Kraniotis2014,Arakida2021} it was only shown that the null geodesic equation contained a cross term between the de Sitter term and the angular momentum of the source. To actually show that this term is present in the relevant bending formula we need first to introduce a closet approach point $r_0$ just as we did with (\ref{5.6x}), as it was that step which enabled the $-Kr^2$ term to decouple from the $r_0$-dependent bending formula given in (\ref{5.6x}). The present author has checked that the de Sitter term does remain in the bending formula when this is done in the rotating source case.
\bibitem{footnote12} In going from (\ref{5.2x}) to (\ref{5.3x})  we multiply the $d^2r/dq^2$ equation by $Adr/dq$  and then encounter the integration constant $E$. However, we cannot do this when $dr/dq$ is zero, and thus for circular orbits we use the $d^2r/dq^2$ equation as is.
\bibitem{footnote13} In a typical $10^{11}$ solar mass galaxy the square of the inverse of the Schwarzschild radius  is $10^{-32}~cm^{-2}$. In the kiloparsec distance range of relevance to galactic rotation curves the $\gamma_0/r$ term in (\ref{7.16x}) is of order the altogether smaller $10^{-51}~cm^{-2}$.
\bibitem{footnote14} These asymptotic considerations are not of concern for the gravitational bending of light by the Sun, since what is relevant  is not a single measurement but a difference between two, one when the Sun is along a line of sight to the target stars and the other half a year later when  it is not. In this difference the background contribution drops out. Moreover, since only contributions at radii close to the Sun remain, its $\gamma^*r$ contribution is not  significant. However, on the much larger distances associated with the lensing of quasar light by an intervening cluster of galaxies one can only make one measurement, and then the asymptotic behavior of the background and the linear potentials of the galaxies in the cluster cannot be ignored.
\bibitem{footnote15b} Once $T$ is not dominant we also need to consider the imaginary parts given in (\ref{2.4y}), (\ref{3.4z}),  (\ref{3.32qq}) and (\ref{4.6x}). This does not affect the formalism associated with the real parts that is presented here, but does restrict the allowed solutions to the theory. However, we should also note that the examples considered in the Appendix are ones in which the wave equations are solved exactly. In those cases both the real and imaginary parts of the wave equations are automatically satisfied.
\bibitem{footnote15a} In the event that the wave equation is conformal invariant and the background geometry is conformal to flat, given (\ref{1.43h}) we only need to consider the flat space case. Then we can set $dx^{\mu}/dq=\nabla^{\mu}T$ and $\nabla_{\nu}T\nabla^{\nu}T=0$,  and the null geodesic eikonal trajectory automatically coincides with the normal to the wavefront.
\bibitem{footnote16} A similar analysis holds for (\ref{1.9g}). When $n$ is constant the solution to (\ref{1.9g}) is of the form $\alpha=e^{i(k_ix^i-nk x^0)}$ where $k=|k_i|$. Thus with $T=k_ix^i-nk x^0$, $dx^i/dx^0=\partial^iT/\partial^0T=-k^i/nk$, to give a  velocity of magnitude $c/n$. In contrast, we should note that the phase velocity is given by $x^i/x^0=nk/k_i$, i.e., by $cn$.  The eikonal velocity $dx^i/dx^0=\partial^iT/\partial^0T$ is thus like a group velocity rather than a phase velocity, with the product of the eikonal and phase velocities being $c^2$.
\end{thebibliography}
\end{document}